# Light, the universe, and everything – 12 Herculean tasks for quantum cowboys and black diamond skiers

Girish Agarwal[a], Roland E. Allen[a], Iva Bezděková[b,c], Robert W. Boyd[d,e,f], Goong Chen[g,h,i], Ronald Hanson[j], Dean L. Hawthorne[k,l], Philip Hemmer[m,i], Moochan B. Kim[a,i], Olga Kocharovskaya[a,i], David M. Lee[a,i,l], Sebastian K. Lidström[n], Suzy Lidström[o,a], Harald Losert[p], Helmut Maier[q], John W. Neuberger[r], Miles J. Padgett[f], Mark Raizen[s], Surjeet Rajendran[t], Ernst Rasel[u], Wolfgang P. Schleich[a,i,p,v], Marlan O. Scully[a,i], Gavriil Shchedrin[a,w], Gennady Shvets[x], Alexei Sokolov[a,i,y], Anatoly Svidzinsky[a], Ronald L. Walsworth[z,27], Rainer Weiss[28,29], Frank Wilczek[30,31,32,33], Alan E. Willner[34], Eli Yablonovich[35] and Nikolay Zheludev[36,37]

[a]Department of Physics and Astronomy, Texas A&M University, College Station, Texas 77843, USA; [b]Department of Physics, Faculty of Nuclear Sciences and Physical Engineering, Czech Technical University in Prague, Brehova 7, 115 19 Praha 1 - Staré město, Czech Republic; [c]Department of Physics, Technical University of Ostrava 17, Listopadu 15, Ostrava -Poruba, 708 33, Czech Republic; [d]Department of Physics and School of Electrical Engineering and Computer Science, University of Ottawa, Ottawa, Ontario K1N 6N5, Canada; [e]Institute of Optics and Department of Physics and Astronomy, University of Rochester, Rochester, New York 14627, USA; [f]School of Physics and Astronomy, University of Glasgow, Glasgow, G12 8QQ, UK; [g]Department of Mathematics, Texas A&M University, College Station, Texas 77843, USA; [h]Science Program, Texas A&M University at Qatar, Education City, Doha, Qatar; [i]Institute for Quantum Science and Engineering (IQSE), Texas A&M University, College Station, Texas 77843, USA; [j]QuTech and Kavli Institute of Nanoscience, Delft University of Technology, P.O. Box 5046, 2600 GA Delft, The Netherlands; [k]Cornell Laboratory of Ornithology, Cornell University, Ithaca, New York 14850, USA; [l]Laboratory of Atomic and Solid State Physics, Cornell University, Ithaca, New York 14853, USA; [m]Department of Electrical & Computer Engineering, Texas A&M Energy Institute, Texas A&M University, Texas 77843, USA; [n]Department of Physics, Strathclyde University, Glasgow G1 1XJ, UK; [o]Physica Scripta, Royal Swedish Academy of Sciences, Stockholm, SE-104 05, Sweden (on submission); [p]Institut für Quantenphysik and Center for Integrated Quantum Science and Technology (IQ[ST]), Universität Ulm, Albert-Einstein-Allee 11, 89081 Ulm, Germany; [q]Institut für Zahlentheorie und Wahrscheinlichkeitstheorie, Universität Ulm, Albert-Einstein-Allee 11, 89081 Ulm, Germany; [r]Department of Mathematics, University of North Texas, Denton, Texas 76203-5017, USA; [s]College of Natural Sciences, University of Texas, Austin, Texas 78712, USA; [t]Department of Physics, University of California, Berkeley, California 94720-7300, USA; [u]Institut fuer Quantenoptik and QUEST-LFS, Welfengarten 1, 30167 Hannover, Germany; [v]Hagler Institute for Advanced Study at Texas A&M University, Texas A&M AgriLife Research, Texas A&M University, College Station 77843, TX, USA; [w]Colorado School of Mines, Golden, Colorado 80401, USA; [x]Applied and Engineering Physics, Cornell University, Ithaca, New York 14853, USA; [y]Quantum Optics Laboratory, Baylor Research Collaborative, Baylor University, Waco, Texas 76798, USA; [z]Department of Physics, Harvard University, Cambridge, Massachusetts 02138, USA; [27]Center for Brain Science, Harvard University, Cambridge, Massachusetts 02138, USA; [28]Department of Physics, Massachusetts Institute of Technology, Cambridge, Massachusetts 02139-4307, USA; [29]LIGO Group, MIT Kavli Institute for Astrophysics and Space Research, MIT, Cambridge, Massachusetts 02139, USA; [30]Center for Theoretical Physics, MIT, Cambridge, Massachusetts 02139, USA; [31]Wilczek Quantum Center, Shanghai Jiao Tong University, Shanghai 200240, China; [32]Arizona State University, Tempe, Arizona 85287, USA; [33]Department of Physics, Stockholm University, Stockholm, Sweden; [34]Department of Electrical Engineering, University of Southern California, Los Angeles, California 90089, USA; [35]Electrical Engineering and Computer Sciences Department, University of California, Berkeley, USA [36]Optoelectronics Research Centre, Institute for Life Sciences, University of Southampton, SO17 1BJ, UK; [37]Centre for Disruptive Photonic Technologies, Nanyang Technological University, Singapore 637371

**ABSTRACT**

The Winter Colloquium on the Physics of Quantum Electronics (PQE) has been a seminal force in quantum optics and related areas since 1971. It is rather mindboggling to recognize how the concepts presented at these conferences have transformed scientific understanding and human society. In January, 2017, the participants of PQE were asked to consider the equally important prospects for the future, and to formulate a set of questions representing some of the greatest aspirations in this broad field. The result is this multi-authored paper, in which many of the world's leading experts address the following fundamental questions: (1) What is the future of gravitational wave astronomy? (2) Are there new quantum phases of matter away from equilibrium that can be found and exploited – such as the time crystal? (3) Quantum theory in uncharted territory: What can we learn? (4) What are the ultimate limits for laser photon energies? (5) What are the ultimate limits to temporal, spatial, and optical resolution? (6) What novel roles will atoms play in technology? (7) What applications lie ahead for nitrogen-vacancy centers in diamond? (8) What is the future of quantum coherence, squeezing, and entanglement for enhanced superresolution and sensing? (9) How can we solve (some of) humanity's biggest problems through new quantum technologies? (10) What new understanding of materials and biological molecules will result from their dynamical characterization with free electron lasers? (11) What new technologies and fundamental discoveries might quantum optics achieve by the end of this century? (12) What novel topological structures can be created and employed in quantum optics?

E-mail: girish.agarwal@tamu.edu

E-mail: allen@tamu.edu

E-mail: bezdekova.iva@gmail.com

E-mail: boydrw@mac.com
E-mail: gchen@math.tamu.edu
E-mail: r.hanson@tudelft.nl
E-mail: dh27@cornell.edu
E-mail: prhemmer@ece.tamu.edu
E-mail: moochankim@tamu.edu
E-mail: kochar@physics.tamu.edu
E-mail: dmlee@physics.tamu.edu
E-mail: seb@lidstrom.fr
E-mail: suzy@intonate.com
E-mail: harald.losert@uni-ulm.de
E-mail: helmut.maier@uni-ulm.de
E-mail: jwn@unt.edu
E-mail: Miles.Padgett@glasgow.ac.uk
E-mail: raizen@physics.utexas.edu
E-mail: surjeet@berkeley.edu
E-mail: rasel@iqo.uni-hannover.de
E-mail: wolfgang.schleich@uni-ulm.de

E-mail: scully@tamu.edu
E-mail: shchedrin@mines.edu
E-mail: gs656@cornell.edu
E-mail: sokol@physics.tamu.edu
E-mail: asvid@physics.tamu.edu
E-mail: rwalsworth@cfa.harvard.edu
E-mail: weiss@ligo.mit.edu
E-mail: wilczek@mit.edu
E-mail: willner@usc.edu
E-mail: eliy@eecs.berkeley.edu
E-mail: niz@orc.soton.ac.uk

## 1. Introduction[1]

The precipitous slopes surrounding the Snowbird ski resort (see Figure 1) are not for the faint-hearted, and neither are the scientific challenges confronted within its meeting rooms below, as participants during each annual PQE conference explore the limits of 21st century quantum technologies.

Those who aspire to great science implicitly have the same guiding principle that steers powder skiers descending mountains or cowboys driving their herds across uncharted territory (or, in the original context, Hannibal crossing the Alps with elephants): "I will find a way or make one."

Some leading figures in the grand history of quantum optics are shown in Figures 2, 3 and 4: Max Planck, who started quantum mechanics to understand the behavior of radiation; Albert Einstein, who essentially introduced the photon (in 1905) and the understanding of absorption and emission (in 1916); Willis Lamb, whose experimental and theoretical work started modern quantum electrodynamics; and Charles Townes, the most central figure in the development of the maser and laser. The history continues with the major discoveries of those who are still active, many of whom are among the authors of this paper.

At PQE 2017, the participants were asked for suggestions regarding the future of quantum optics and related areas, and many ideas were forthcoming. These were consolidated into a final list of twelve, which far transcend the twelve tasks set to Hercules:

### (1) What is the future of gravitational wave astronomy?

Rainer Weiss (shown in Figure 5 with Marlan Scully) invented the techniques which ultimately led to the famous and spectacular double success of the Laser Interferometer Gravitational-Wave Observatory (LIGO), represented by Figure 6: after a century, Einstein's prediction of gravitational waves finally received direct confirmation, and gravitational wave astronomy was discovered to be a technique with tremendous potential for new discoveries.

This potential is explored in detail by the first contribution in this article, as Rai Weiss leads us through the varied phenomena that can be probed with gravitational radiation.

Since amplitude rather than power is observed, the signal strength falls as $1/R$ rather than $1/R^2$ (or worse), where R is the distance from source to detector. This means that an increase of 2 in sensitivity implies an increase of 8 in accessible volume. In addition, gravitational waves are the most penetrating radiation in the universe, capable of revealing phenomena that are hidden from observation with all of the many forms of electromagnetic radiation. The existence of black holes which form binaries and have unexpectedly large masses is already a major surprise and success. In the future, it may be possible to probe the properties of neutron stars – including their structures and equations of state – as they are deformed during a binary merger that emits gravitational waves. Prof. Weiss's very substantial contribution also discusses the prospects for deeper understanding of supernovas, supermassive black holes, cosmological features, and other sources, at least some of which will surely be unanticipated.

We mention that theorist Kip Thorne was also a major driving force behind LIGO as well as other enterprises such as the one represented in Figure 7.

The prospects for gravitational wave astronomy are further discussed by Surjeet Rajendran and by Ernst Rasel, with emphasis on atom interferometry to reach lower frequencies in terrestrial detectors, as a complement to the planned Laser Interferometer Space Antenna (LISA).

---

[1] Roland Allen and Suzy Lidström

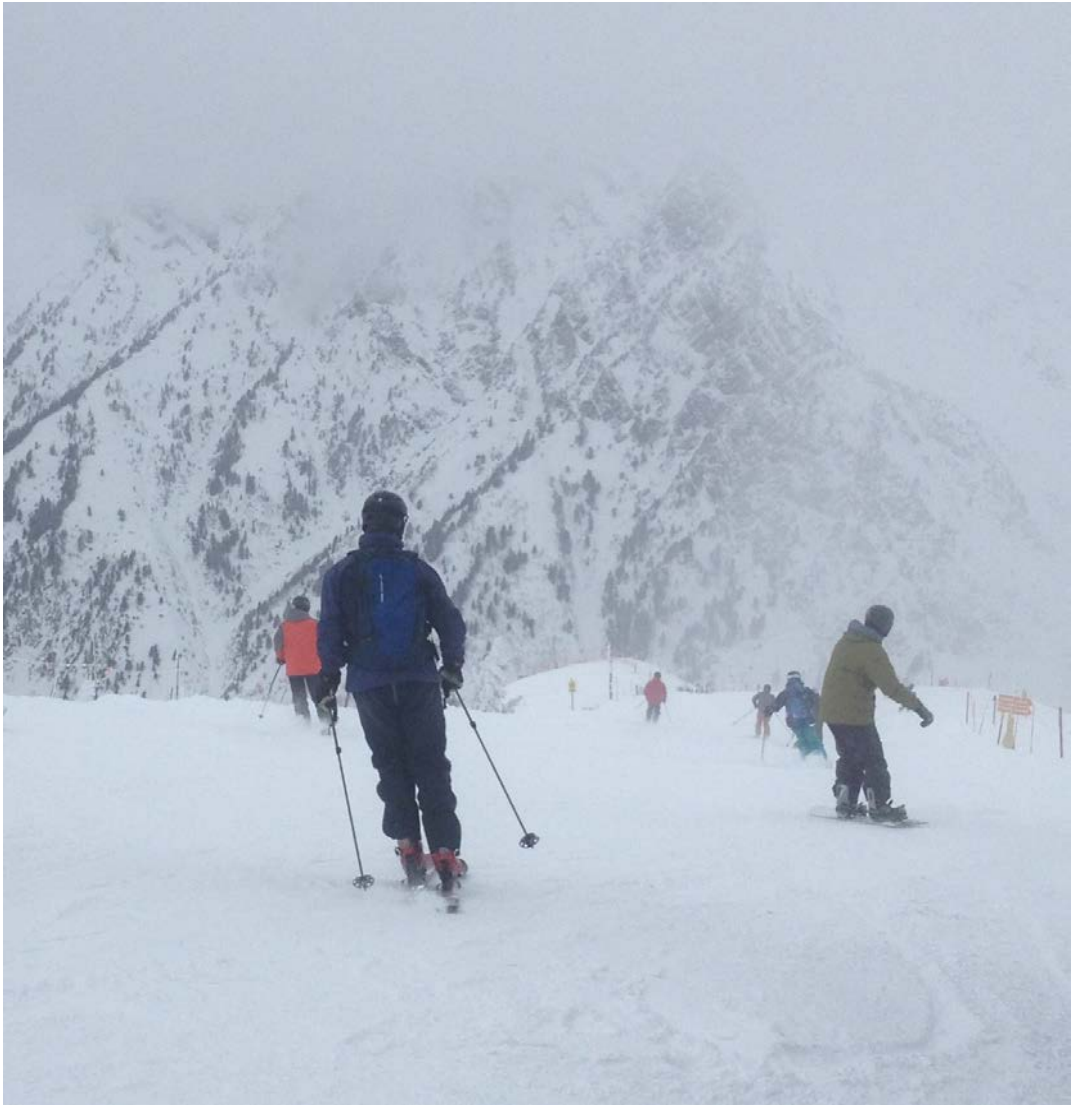

**Figure 1.** Preparing to descend a black diamond slope at Snowbird, January, 2017. A metaphor for quantum optics research: What challenges lie ahead in the depths and distant mists? Photograph: Roland Allen.

## (2) Are there new quantum phases of matter away from equilibrium that can be found and exploited – such as the time crystal?

Frank Wilczek, shown in Figure 8, surveys recent ideas and experiments related to a new form of symmetry breaking he has proposed: either spontaneous or driven breaking of time translation symmetry. Examples include Floquet time crystals and prethermal time crystals, with potential for developments in many different directions.

## (3) Quantum theory in uncharted territory: What can we learn?

In this contribution, Wolfgang Schleich – shown in Figures 9, 10 and 11 – provides a taste of the vast range of fundamental projects that an imaginative physicist can formulate in this broad context.

Each question involves a feasible approach to a deep issue: (a) Can we create fractal radiation? (b) What is the elasticity of spins? (c) Is the event horizon of a black hole a beam splitter? (d) Do continental divides of the Newton flow offer a path towards the Riemann Hypothesis?

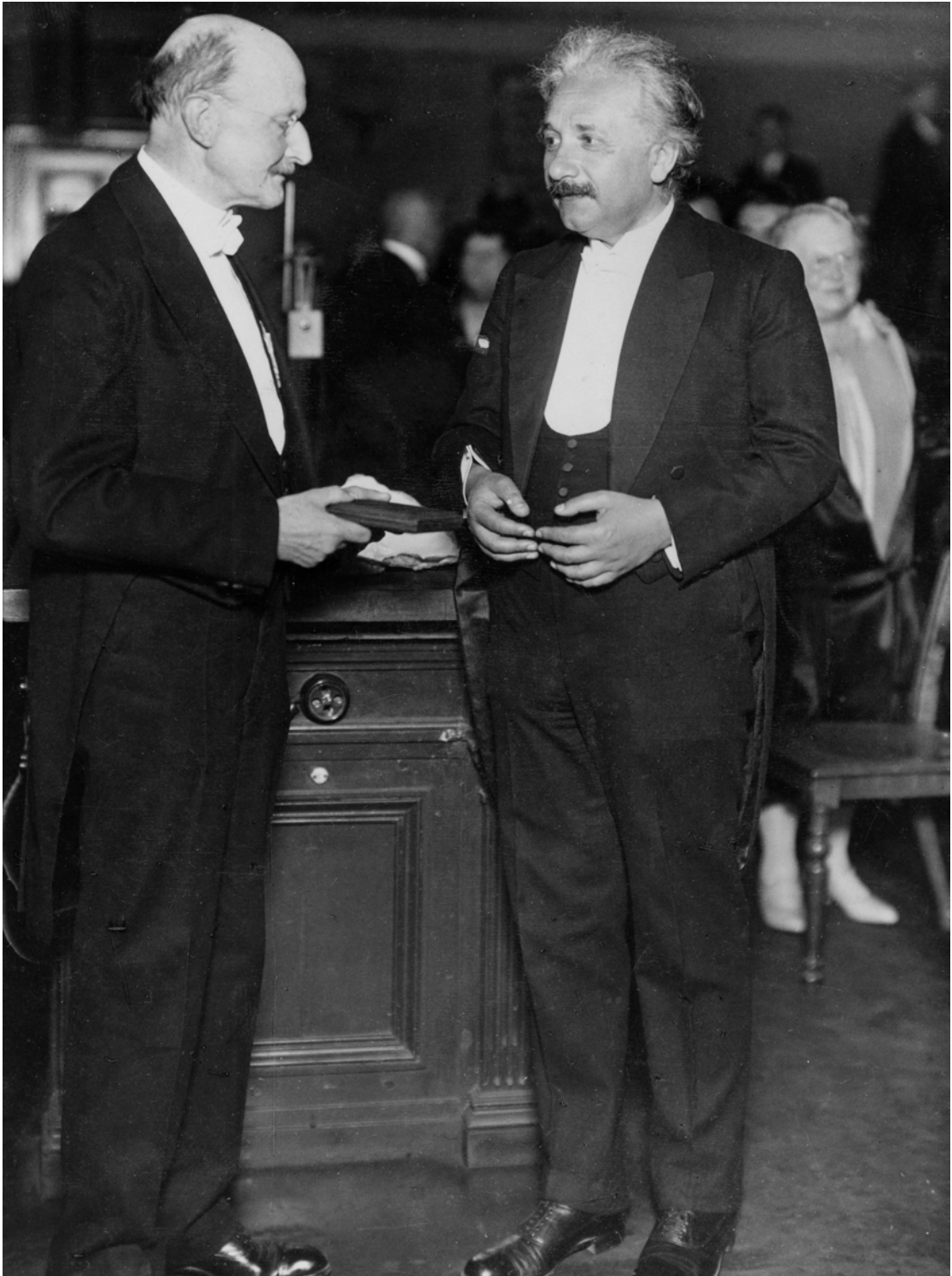

**Figure 2.** Quantum optics pioneers: Max Planck and Albert Einstein in 1929. Einstein is receiving the Planck medal. Credit: AIP Emilio Segrè Visual Archives, Fritz Reiche Collection.

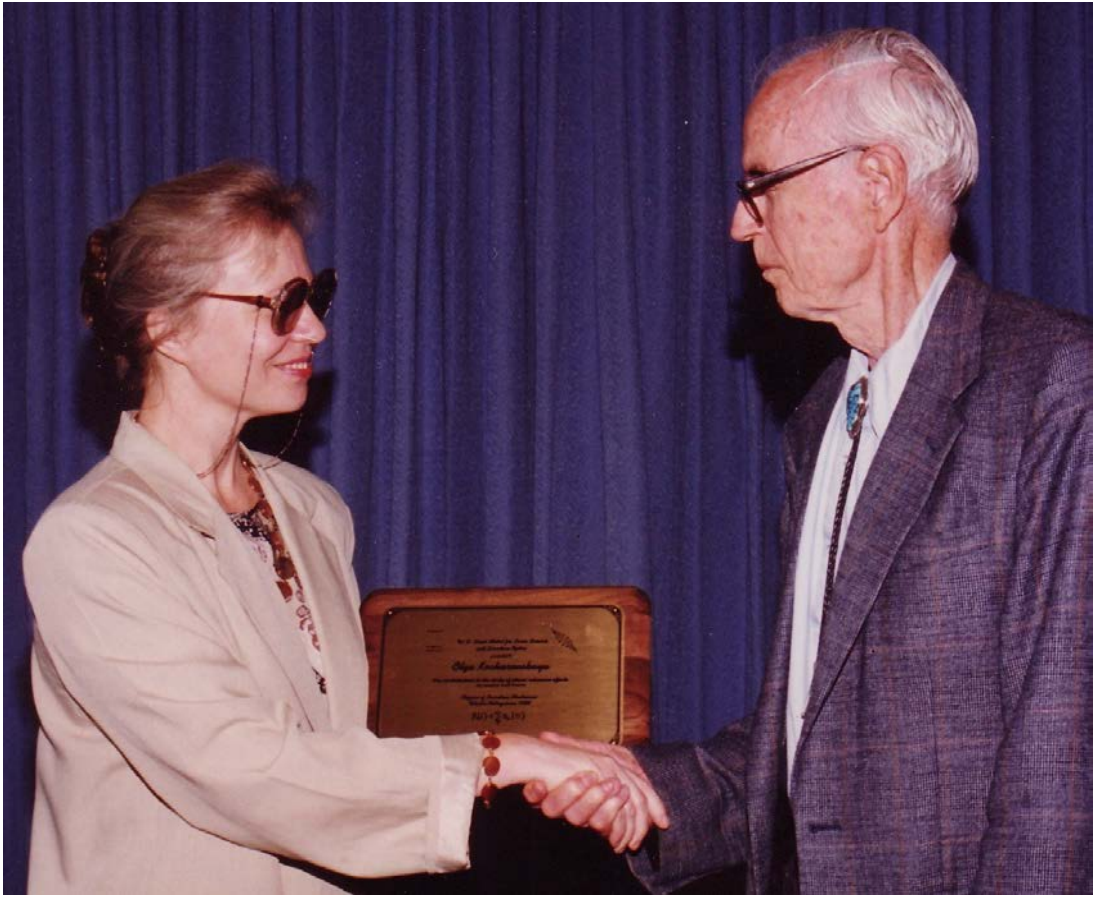

**Figure 3.** Willis Lamb and Olga Kocharovskaya in 1998. Kocharovskaya is receiving the Lamb Award, for achievements in laser science and quantum electronics. Credit: Institute for Quantum Science and Engineering.

Together with his collaborators Ernst Rasel, Harald Losert, Dean Hawthorne, Gavriil Shchedrin, Iva Bezděková, Moochan Kim, Helmut Maier, John Neuberger, Marlan Scully, and David Lee (shown in Figure 10), Prof. Schleich presents a stimulating analysis of each of these questions, which are currently under investigation by the groups credited in this section and its four subsections.

Ernst Rasel discusses the prospects for quantum tests of the equivalence principle with unprecedented sensitivity, using interferometry with Bose-Einstein condensates in microgravity, with reference to experiments demonstrating the high potential of this technique for probing the fundamental aspects of quantum mechanics and general relativity.

### (4) What are the ultimate limits for laser photon energies?

Olga Kocharovskaya leads us through the evolution of thinking and achievements in the section addressing the ultimate limits for laser photon energies. After a detailed discussion of the seemingly insurmountable barriers to lasing with very short wavelengths, she describes in equally clear detail the lasers which have been made to operate in the extreme ultraviolet and even X-ray ranges, with the shortest laser wavelength achieved being 0.6 Å, corresponding to a photon energy of 19.6 keV. Prof. Kocharovskaya's section ends with truly exotic proposals for achieving gamma ray lasers, a supreme goal included in the famous list of 30 important problems that was composed by the Nobel Laureate Vitaly Ginzburg.

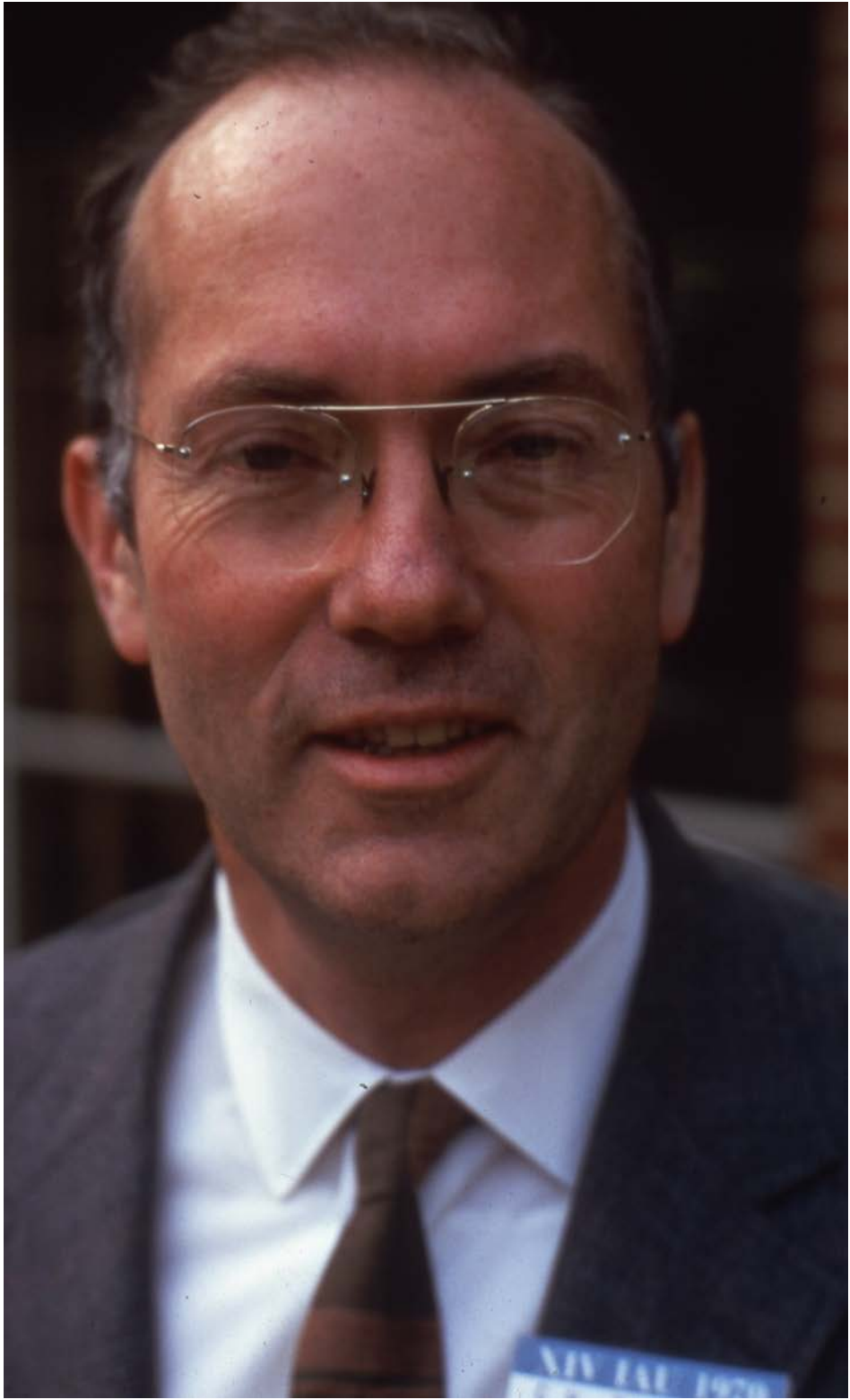

**Figure 4.** Quantum optics pioneer Charles Townes at the International Astronomical Union meeting at the University of Sussex, 1970. Credit: AIP Emilio Segrè Visual Archives, John Irwin Slide Collection.

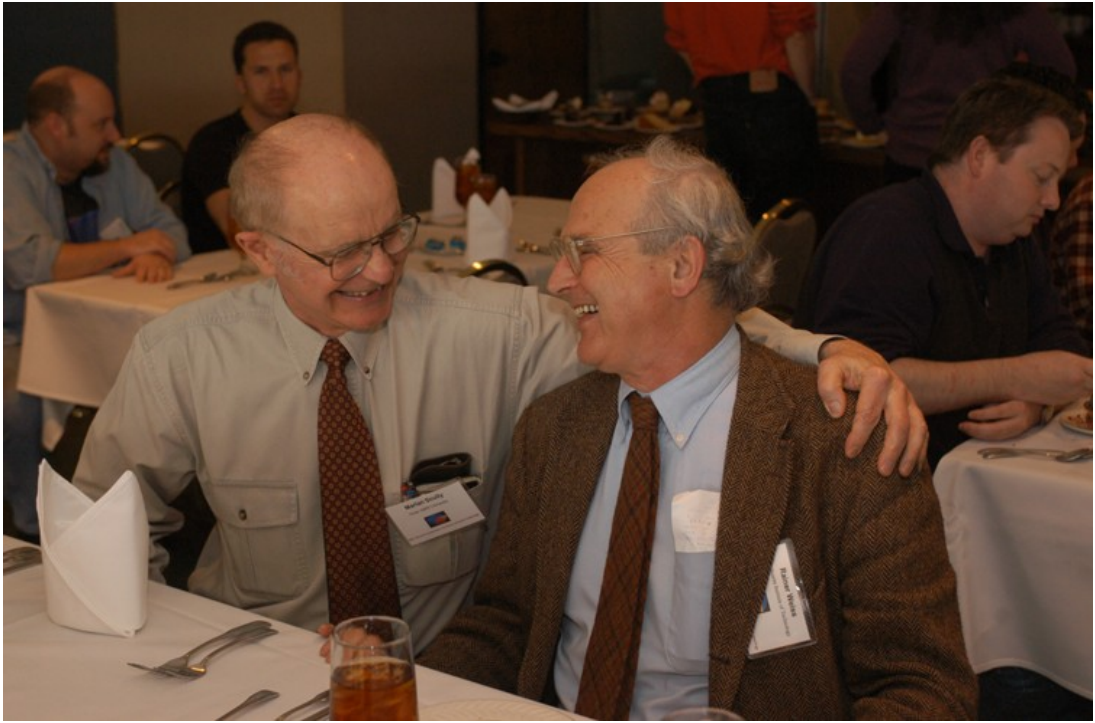

**Figure 5.** Two old M.I.T. friends: Rainer Weiss and Marlan Scully in 2004. Rainer Weiss invented the sophisticated laser interferometric techniques which are the basis for LIGO (the Laser Interferometer Gravitational Wave Observatory). He was also the chairman of the science working group for the COBE (Cosmic Background Explorer) satellite mission, which observed the inhomogeneities in the cosmic background radiation that seeded initial structure formation in the early Universe, as well as an originator of COBE with John Mather. He has thus been at the center of two of the greatest discoveries in modern physics and astronomy. Photograph: Roland Allen.

## (5) What are the ultimate limits to temporal, spatial, and optical resolution?

In their contribution, based partly on their own work, Robert Boyd, Miles Padgett, and Alan Willner consider very sophisticated approaches to the question "exactly how much information can reliably be carried by a single photon?" For encoding information, they first consider orbital angular momentum states of light, and then more broadly the various degrees of freedom of the photon: polarization, wavelength, time bins, and transverse spatial structure.

Nikolay Zheludev discusses the prospects for superresolution imaging technology based on superoscillation, a phenomenon first described by Berry and Popescu, and inspired by an earlier analysis of Aharonov et al., which allows optical waves to form arbitrarily small spatial energy localizations that propagate far from a source. The author's group has demonstrated resolution up to one sixth of the wavelength, with the technique recently applied in biological imaging.

## (6) What novel roles will atoms play in technology?

Mark Raizen describes several ways in which atoms can be employed in important technologies, including one which has no parallel with electrons or photons and which was developed by his group: Magnetically-Activated and Guided Isotope Separation (MAGIS), which relies on optical pumping of atomic beams and separation by magnetic-moment-to-mass ratios, and which will soon be employed in obtaining isotopes for medicine.

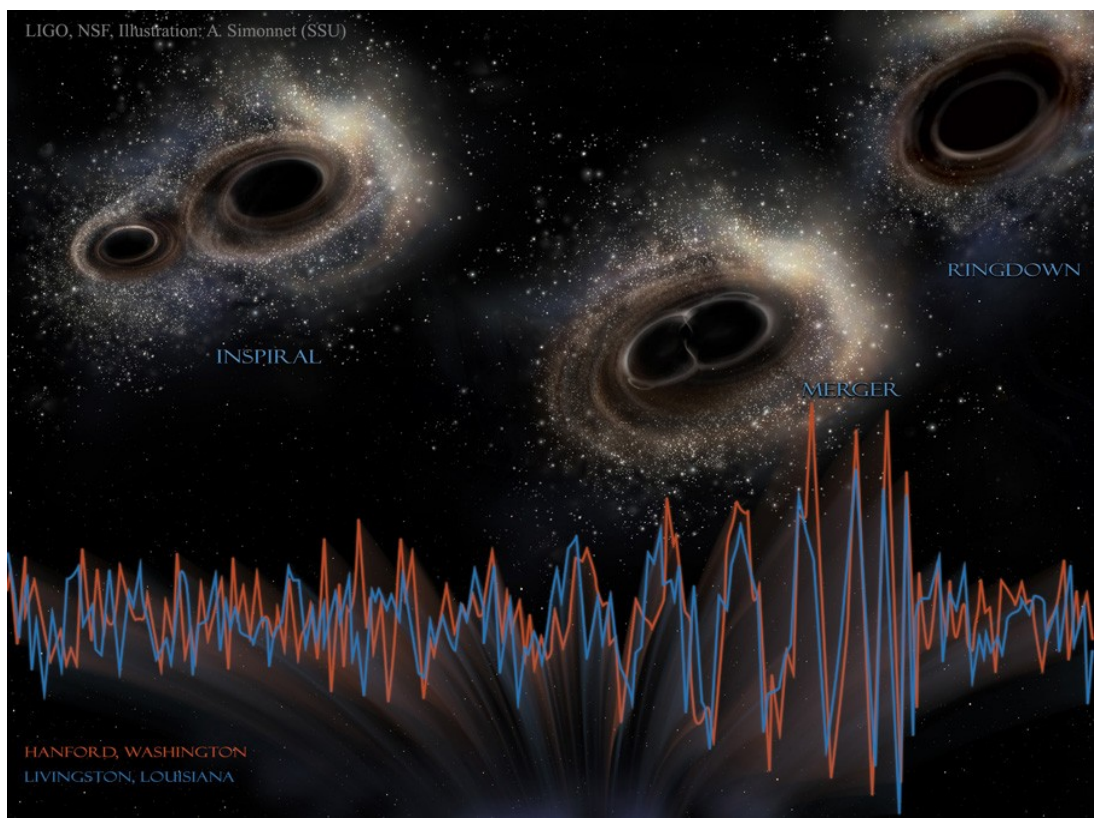

**Figure 6.** The LIGO collaboration has a spectacular double success: the first direct observation of gravitational waves—predicted by Einstein a century earlier—and the unexpected observation of the merger of two very massive black holes. As indicated in the figure, the waves were simultaneously detected at the two separate LIGO observatories in Hanford, Washington and Livingston, Louisiana, with amazingly consistent signals for the spiralling in and merger of the black holes, and the ringdown of the single remnant afterward. This observation marks the beginning of gravitational wave astronomy. Credit: LIGO, NSF, Aurore Simonnet (Sonoma State U.).

### (7) What applications lie ahead for nitrogen-vacancy centers in diamond?

Philip Hemmer describes the favorable attributes of these complexes which has made them attractive for diverse applications like sensing, imaging, and quantum computing. He goes further in describing the search for related complexes that may have even greater promise, with the SiV and GeV centers in diamond each having specific properties that are superior to those of the NV center.

Ronald Walsworth then provides a very detailed and stimulating description of the wide-ranging applications (already achieved or envisioned) for diamond NV color centers as precision quantum sensors in both the physical and life sciences. They have already been used to study important aspects of proteins, biomagnetism, living human cells, and advanced materials, and they may play a role in a wide variety of other sensing and imaging applications for which they are perhaps uniquely qualified.

Surjeet Rajendran explains how crystal defects such as nitrogen vacancy centers in diamond, paramagnetic F-centers in metal halides, and defects in silicon carbide could assist in the search for weakly interacting massive particles (WIMPs) as dark matter candidates, by helping to determine the direction of WIMP induced nuclear recoils.

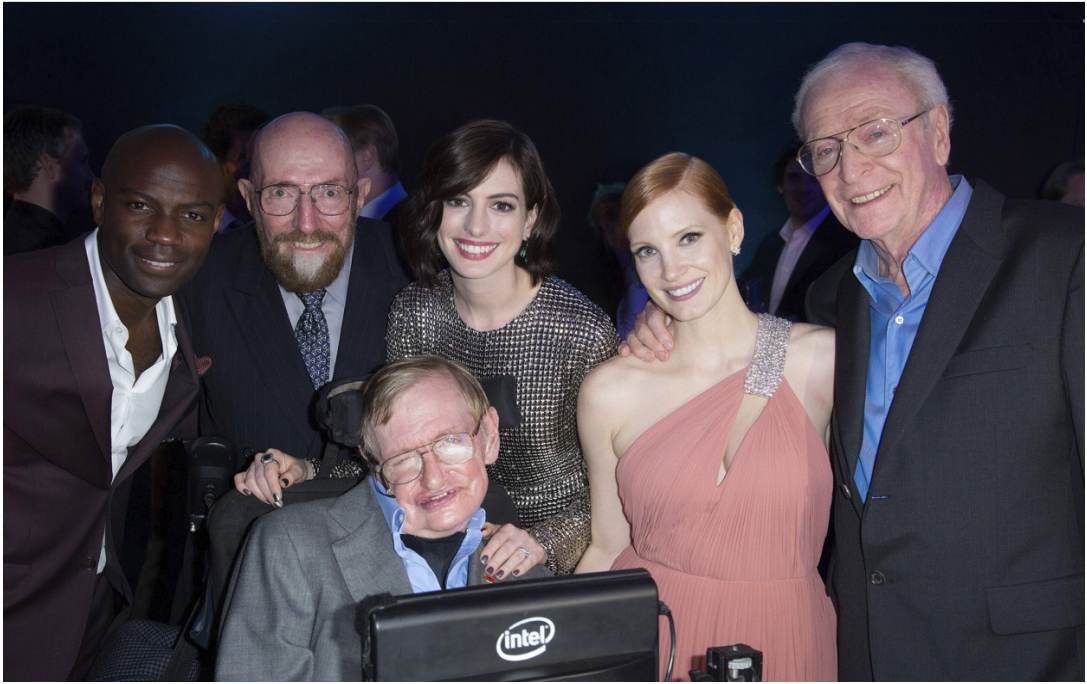

**Figure 7.** Kip Thorne and Stephen Hawking with actors of the movie Interstellar: David Gyasi, Anne Hathaway, Jessica Chastain, and Michael Caine. Kip Thorne has also been a principal driving force on LIGO, following a 1975 all-night discussion with Rai Weiss, when they shared a room at a NASA meeting on cosmology and relativity. (Weiss, an experimentalist, had reserved a room, and Thorne, a theorist, had not.) Credit: Kip Thorne.

## (8) What is the future of quantum coherence, squeezing, and entanglement for enhanced superresolution and sensing?

In his contribution, Girish Agarwal emphasizes enhanced resolution in imaging and sensing to study the fundamental physics of biological systems on the molecular and atomic scale. He considers three principal issues: (a) How can intensity-intensity correlations and structured illumination be employed far beyond the diffraction limit? (b) Quantum metrology: excitation with quantum light (entangled photons, squeezed light,…). (c) Quantum coherence for deep sub-wavelength localization and tracking.

## (9) How can we solve (some of) humanity's biggest problems through new quantum technologies?

Eli Yablonovich discusses the issue of high efficiency solar cells, with the central principle "a great solar cell needs to also be a great light emitting diode", and the additional point that the ability to create fuels would increase the size of the photovoltaic panel industry at least 10-fold.

Goong Chen considers the many potential applications of nonlinear science in this context, including those relevant to aircraft and spacecraft and to climate change and global warming.

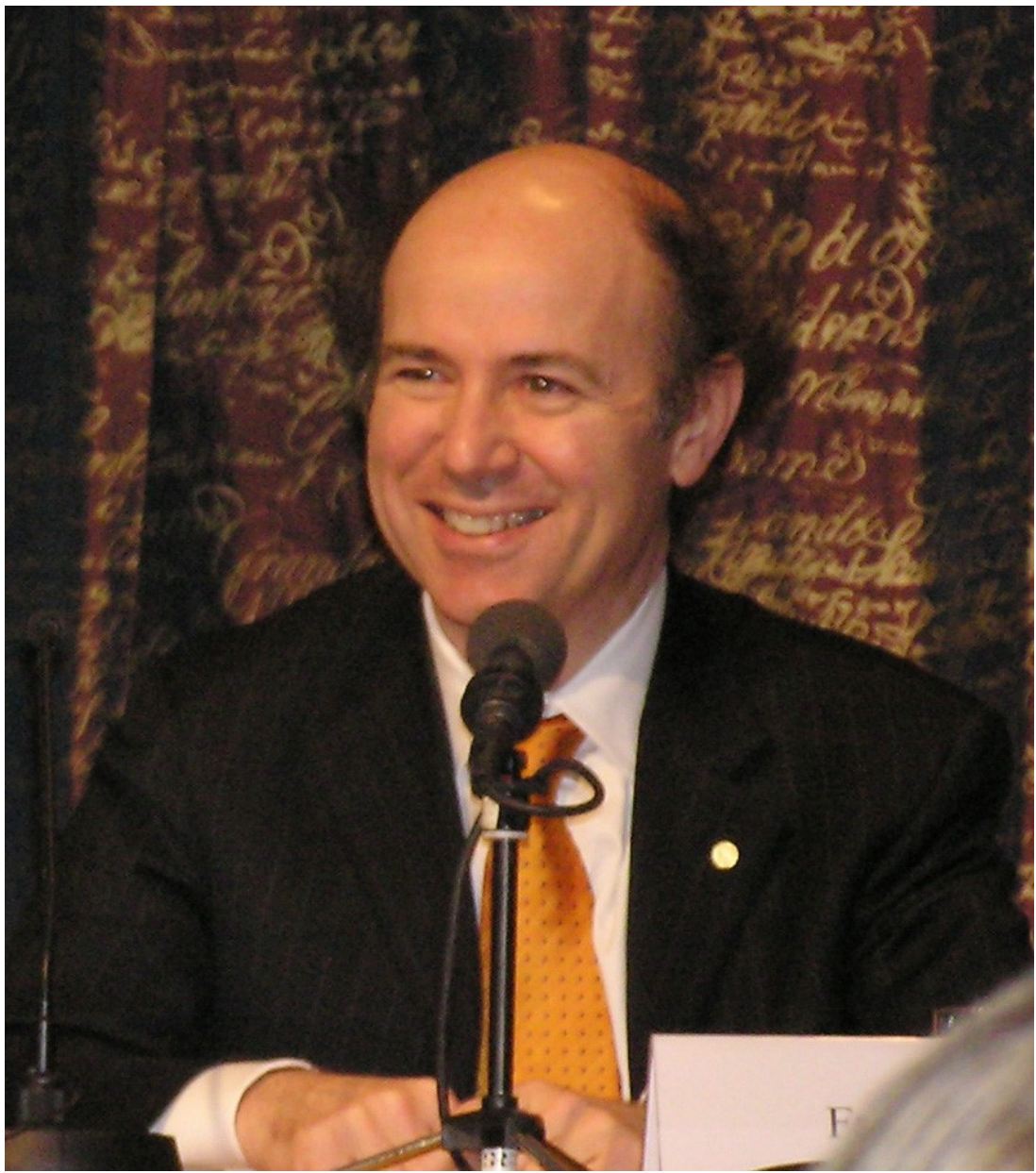

**Figure 8.** Frank Wilczek (Nobel Prize in Physics, 2004) first proposed the idea of a time crystal. Credit: Betsy Devine.

**(10) What new understanding of materials and biological molecules will result from their dynamical characterization with free electron lasers?**

In this section, the emphasis is on the current spectacular capabilities of the X-ray free electron lasers at Stanford, DESY, RIKEN, and elsewhere. Diffraction before destruction permits the structure of a single molecule or nanoparticle to become experimentally accessible, and the dream is that "movies" can be made by analyzing the behavior of thousands of individual molecules that are successively observed.

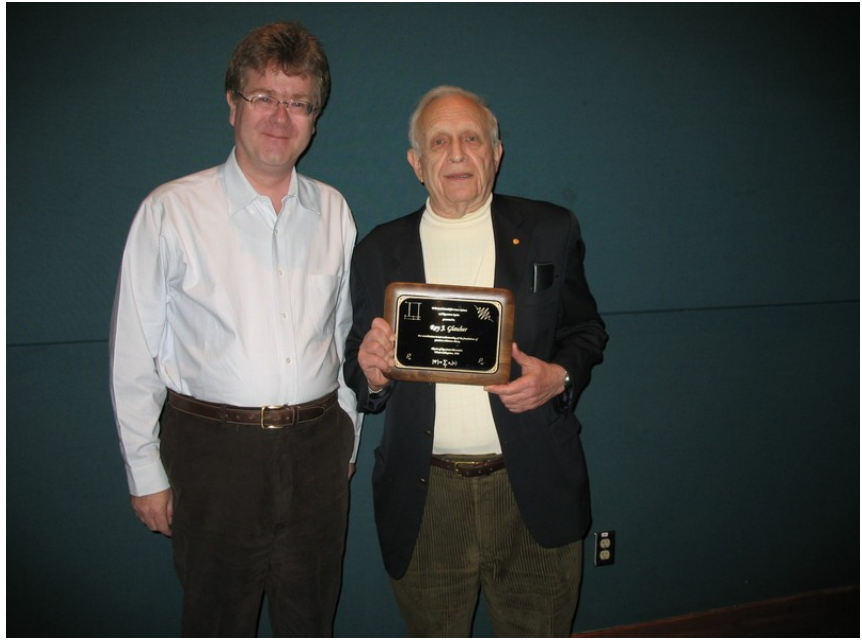

**Figure 9.** The Lamb Award: Wolfgang Schleich (left) pictured with Roy Glauber, recipient of the Lamb award in 2006 (and the Nobel Prize in Physics, 2005). Credit: Institute for Quantum Science and Engineering.

## (11) What new technologies and fundamental discoveries might quantum optics achieve by the end of this century?

Ronald Hanson describes the visionary possibility of a quantum internet – a network that enables the generation of entanglement between qubits at any two places on Earth, and which would have many far-reaching applications, including private communication secured by the laws of physics, synchronization protocols for enhanced time-keeping, and a means to link future quantum computers.

Alexei Sokolov discusses in substantial detail his proposal for a "Maxwell's demon for light", which would provide a means for determining the entropy of light – a nontrivial issue.

Marlan Scully (shown in Figures 12 and 13) poses a series of very fundamental questions based on earlier work of his and other groups: (i) Can amplification be achieved by (cooperative) spontaneous emission? (ii) Will tomorrow's particle accelerators be based on lasers? (iii) Will new optical techniques enable us to go beyond the Rayleigh limit? (iv) How is quantum coherence in lasers conceptually related to the Higgs? (v) Why is the many particle Lamb shift divergence free? (vi) How does Bayesian logic impact quantum thinking? (vii) Can we use quantum noise to improve on biological efficiency and information processing? Inspiration for the last question comes from the detected quantum beating in plant photosystem reaction center II by researchers in the Netherlands, Sweden, and Russia.

## (12) What novel topological structures can be created and employed in quantum optics?

Nikolay Zheludev discusses "flying doughnuts" (electromagnetic doughnut pulses), which have not yet been observed experimentally but can be generated from short transverse oscillations in a singular metamaterial converter. Their nonseparable space-time dependence, which distinguishes doughnut pulses from the vast majority of electromagnetic waveforms, allows for novel schemes of information transfer and spectroscopy.

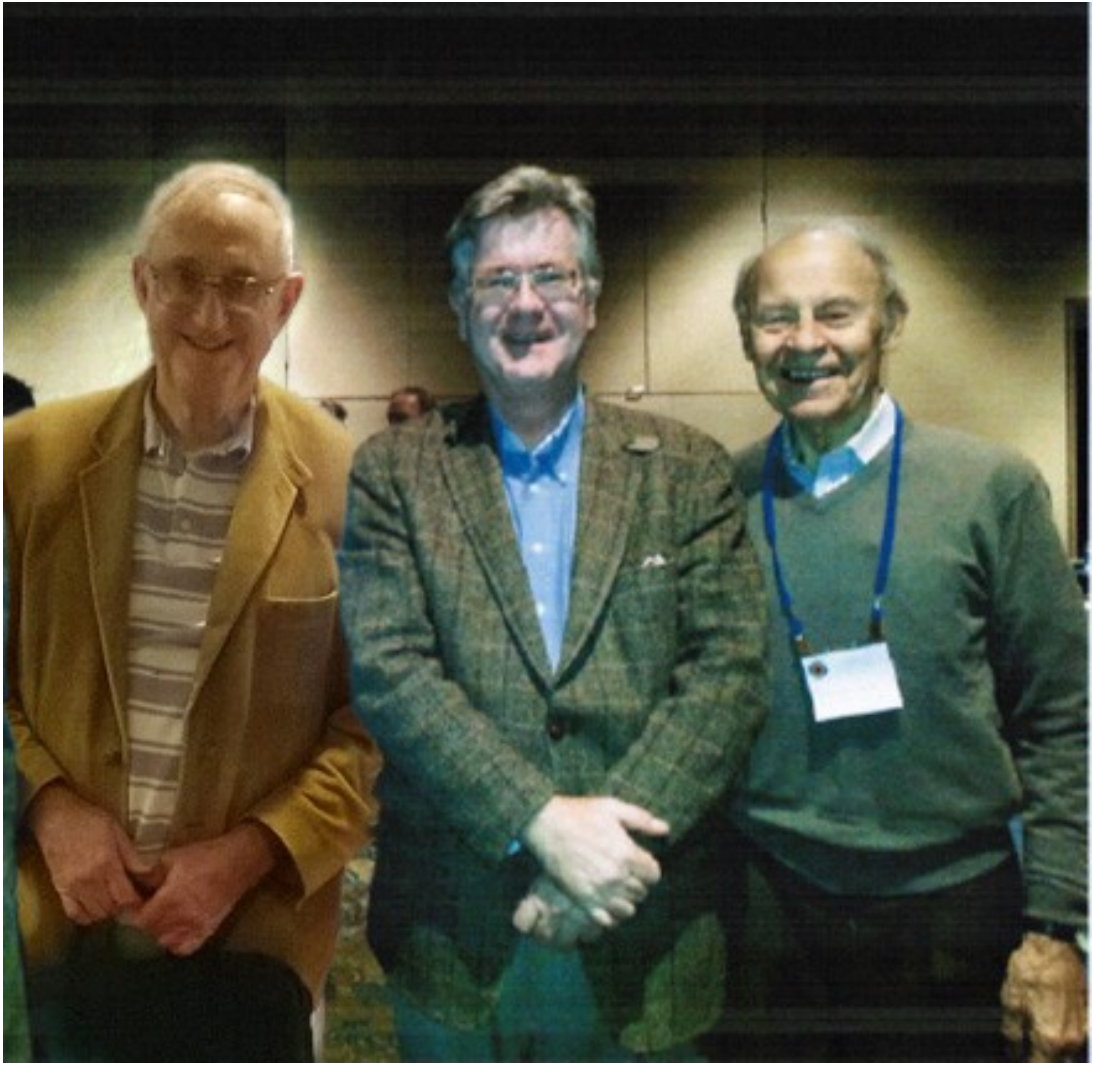

**Figure 10.** Quantum optics scientists imported from condensed matter physics, general relativity, and physical chemistry: David Lee (Nobel Prize in Physics, 1996), Wolfgang Schleich (Lamb Award for Laser Science and Quantum Optics, 2008), and Dudley Herschbach (Nobel Prize in Chemistry, 1986). Credit: Institute for Quantum Science and Engineering.

Finally, drawing largely on the work of his own group, Gennady Shvets gives a broad overview of topological photonics, providing context through a comparison with recent understanding of the role of topology in condensed matter physics. He points out that the three basic condensed matter systems supporting topological insulating phases – quantum Hall, quantum spin-Hall, and quantum valley-Hall topological insulators – have all been emulated in photonics, with potential for application in novel devices.

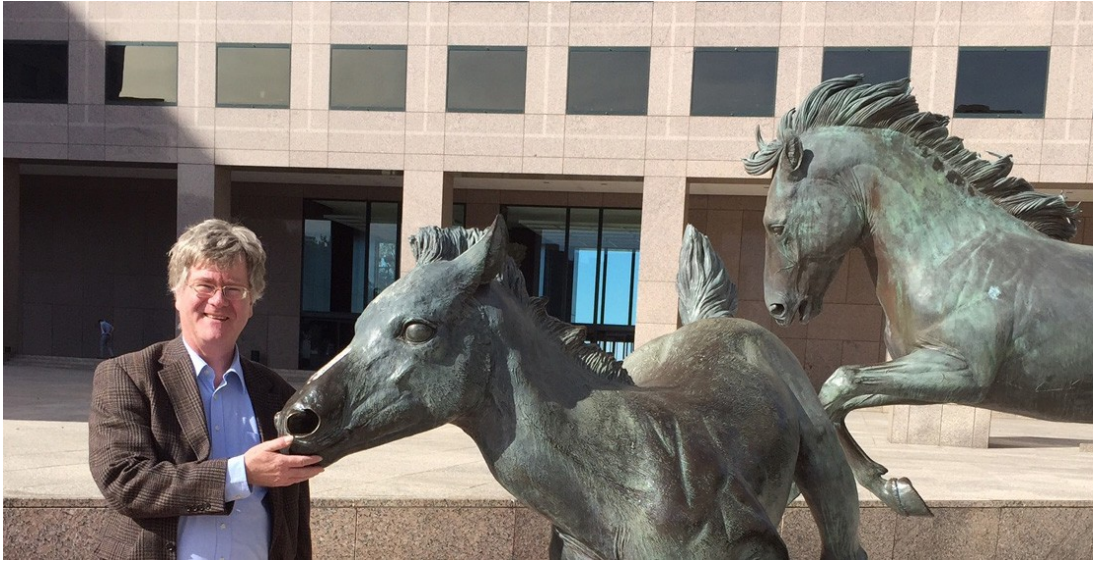

**Figure 11.** The relativistic quantum cowboy contemplates: what would you see if you rode a horse at the speed of light? Credit: Ernst Rasel.

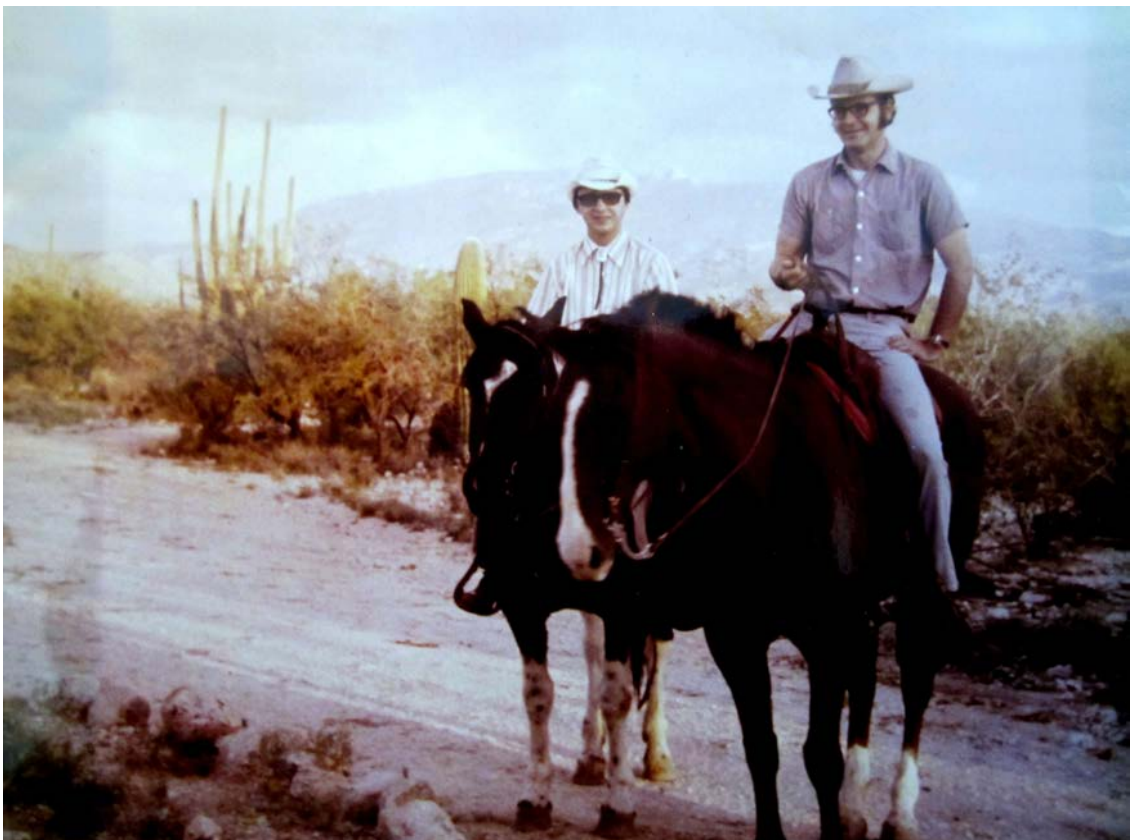

**Figure 12.** Cowboy physicists. Marlan Scully with Vladilen Letokhov in the late 1970s, at the Scully "hbar" ranch in Arizona. Credit: Institute for Quantum Science and Engineering.

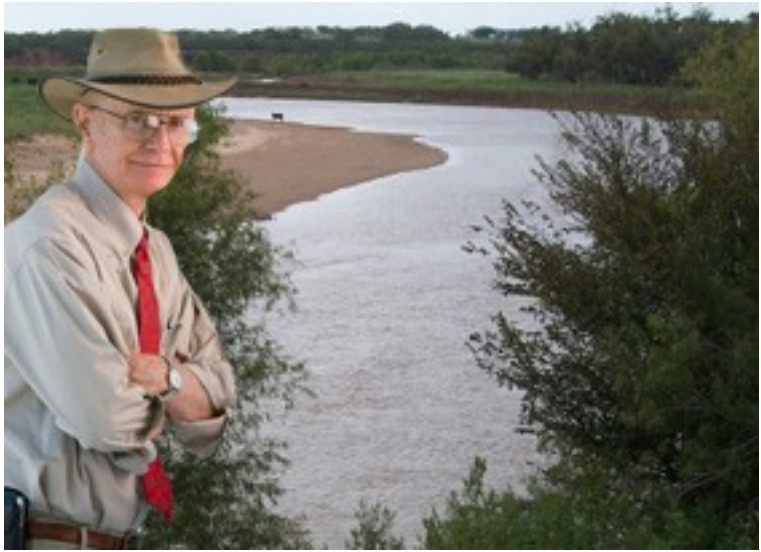

**Figure 13.** Marlan Scully in cowboy country, Robertson County, Texas, on the banks of the Brazos River. This 1000 acre ranch near Texas A&M University is "quantum + cowboy", with many research oriented activities. Credit: Institute for Quantum Science and Engineering.

## 2. (1.) What is the future of gravitational wave astronomy?[2]
### 2.1. Introduction

The recent observation of gravitational waves from the merger of binary black holes has several significant consequences. It has established that black holes can come in pairs and that they most likely dominate the gravitational wave sky. The direct measurement of the waveforms (Figure 6) – the time series of the wave amplitude – has established that the Einstein field equations discovered in 1915 work over a huge dynamic range from the weak field of Cavendish experiments to the dynamics around a pair of colliding black holes where stellar masses are moving close to the speed of light. They are a confirmation that the concepts developed in the middle of the 20th Century to directly measure gravitational waves are indeed valid. Most important, they are the first steps in what we hope will be a new way to explore the world around us – gravitational wave astronomy.

We expect that as with every other advance in observing the universe, much as was the case with the ability to look for the first time with radio waves and x-rays, gravitational waves will provide new insights into astrophysical phenomena we know about, but also will lead to discoveries of entirely new things, some of which we may not have imagined. The waves are emitted by accelerating masses and do not get scattered significantly by matter. In fact, gravitational waves are the most penetrating radiation in nature; they flow out of stellar systems unimpaired and can cross the universe without being blocked or altered.

### 2.2. What will gravitational wave astronomy reveal about black holes?

Start with the current discovery of binary black hole systems. We are not sure where they are made. Are they a result of the collapse of a heavy star after the nuclear fuel that maintains it has been exhausted, or are they produced by collisions in highly dense regions of stars called globular clusters that reside in many galaxies? Could they be the end point of the very first stars in the universe made primarily of hydrogen and helium? Or were they created in the very beginning of the universe – primordial black holes? The answer will come from observations with improved detectors. The binary black holes, since they are now moderately easy to see, could become standard astrophysical systems to map the geometry of the universe with systematic problems that are quite different from those of the electromagnetic detection of supernova.

---

[2] Rainer Weiss (2.1-2.11)

Black holes are the simplest compact systems we know about in nature. They are pure geometric objects explained by General Relativity.

## *2.3. What will gravitational wave astronomy reveal about neutron stars?* [3]

Neutron stars are enormously dense; with the mass of the sun in a radius of a typical large city, they are composed primarily of neutrons. Although the space in which they reside is significantly curved, their dynamics is still influenced by matter, which makes them more difficult to understand than black holes and interesting in their own way.

Binary neutron star systems were known to exist after the major discovery by Hulse and Taylor of the radio pulses emitted by a binary pulsar system. The system gave the first measurement of the energy carried away by gravitational waves, thereby settling the issue of the reality of gravitational waves. It also provided the first firm evidence for a source that might be observed by a gravitational wave detector; the coalescence of a binary neutron star system into a black hole. These are also possible models for short gamma-ray bursts which are detected at a rate of a few per month from all over the universe. We had expected to measure the gravitational waves from them as our first likely detections. To date we still have not detected any. [After this paper was written and submitted for publication, LIGO and Virgo announced the observation of a dramatic neutron star merger, as well as additional black hole mergers.]

When we measure the waveforms associated with the coalescence of neutron stars, we will obtain new information about the properties of nuclear matter. One of the most interesting will be the stiffness of nuclear matter, its equation of state, which will be measured through the tidal gravitational deformation of the two neutron stars as they closely approach each other during the coalescence. The deformation will leave a signature in the coalescence gravitational wave time series. In the process we may also learn more about how the heaviest nuclei in nature (such as gold and tungsten) are formed. Current modeling of the collision of neutron stars shows a spectrum of oscillations of the nuclear material before it becomes a black hole. The oscillation frequencies occur in the 1 to 4 kHz band with the exact frequency depending on the nuclear equation of state. Special high frequency gravitational wave detectors will need to be designed to sense these oscillations.

Another possible source of gravitational waves associated with neutron stars is the emission of a continuous gravitational wave from the rotation of the star. The star needs to have some kind of non-spherical character to do this; for example, a little mountain just a few centimeters high on the otherwise spherical star would suffice, or a large magnetic field which has deformed the star into an ellipsoid for which one of the principal axes is not along the rotation axis (similar to the earth whose magnetic dipole is not along the spin axis). This would be a wonderful source to test the basic ideas of gravitational waves such as the polarization and the propagation speed; it would also provide new information about neutron stars themselves.

---

[3] Rainer Weiss (2.1-2.11)

## 2.4. What will gravitational wave astronomy reveal about supernovas? [4]

Supernovas are the explosions of stars when they begin to collapse due to the exhaustion of their nuclear fuel. A milkyway galaxy such as ours experiences a supernova about every 30 years. Depending on how non-spherical the explosion and resulting collapse of the star is, a supernova may be a significant source of gravitational waves. Since gravitational waves are so penetrating they would be an excellent way to find out what is actually going on inside the explosion. The observation of gravitational waves from a supernova would be as important as the discovery of neutrinos from supernova 1987a (see Figure 14) as it will provide details of the bulk motion of the mass in the explosion. In order to make a scientific program, however, it would be necessary to observe many supernovas, which means being able to sense many galaxies, and to achieve that would require a considerable improvement in detector sensitivity over present capabilities.

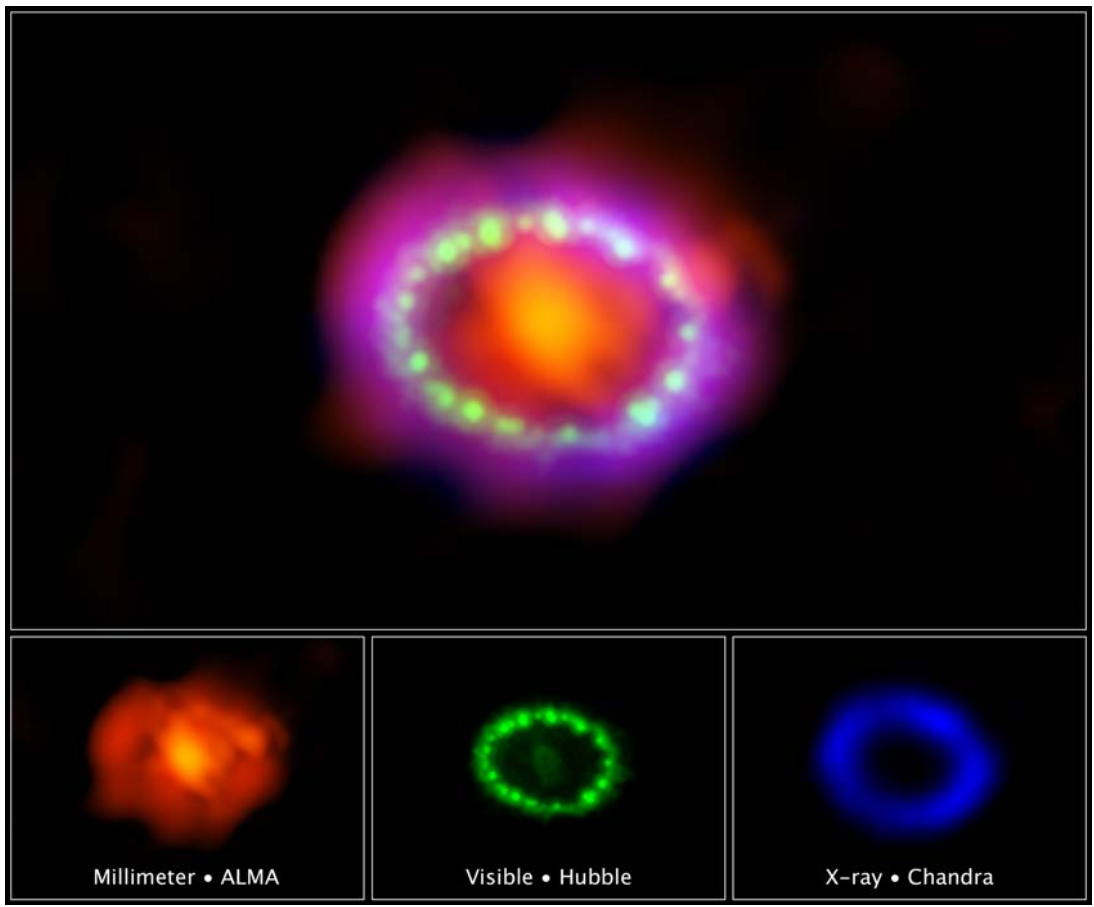

**Figure 14.** The remains of Supernova 1987A obtained by combining three observations from different telescopes/observatories (Credit: NASA, ESA and A. Angelich (NRAO/AUI/NSF)); the individual images are displayed in the lower part of the figure. On the left, in the millimetre range (Credit: ALMA; ESO/NAOJ/NRAO and R. Indebetouw (NRAO/AUI/NSF), in the centre, the Hubble Telescope measurements in the visible range (Credit: NASA, ESA and R. Kirshner (Harvard-Smithsonian Centre for Astrophysics and Gordon and Betty Moore Foundation)) and on the right, an image in the x-ray spectrum from Chandra. Credit: NASA/CXC/Penn State/K. Frank et al. https://www.nasa.gov/sites/default/files/thumbnails/image/stsci-h- p1708b-f-3000x2500.jpg

---

[4] Rainer Weiss (2.1-2.11)

## 2.5. What information can be obtained from the stochastic background of gravitational waves? [5]

A source certain to be detected is the incoherent emission of many unresolved gravitational wave sources as one observes further into the universe – a background noise of gravitational waves. Studies of such backgrounds will give information about the population of the sources back in time as well as whether there was an evolution of the sources with cosmic time. The most dramatic of these background sources would be the gravitational radiation that might have accompanied the primeval explosion that generated the 3K cosmic microwave background and the cosmological recession of the galaxies. If the current theories of the initial formation of the universe are correct – inflation or other concepts where there is a rapid initial expansion of the universe – most likely the direct detection of the gravitational radiation from this early epoch is too small to be sensed by instruments and techniques currently being considered. One idea for very much later in the present century would be a space mission, a big bang observer, which might be configured to make such a measurement.

## 2.6. Can anything be said about unanticipated sources?

It is worth highlighting what was mentioned in the introduction to this section: Given the radical difference between the sources of gravitational waves, accelerated masses, and the sources of electromagnetic ones, accelerated charges, as well as the deeply penetrating nature of the gravitational waves, it would be remarkable if there were no new sources for gravitational waves – sources not triggered by our knowledge of the electromagnetic sky.

## 2.7. What developments can be envisaged in the near and long term?

The long baseline interferometric gravitational wave detectors will advance in several ways in the short term. There are ideas by which the detectors will be improved in sensitivity by factors of 3 to 4 in the next decade and by possibly as much as another factor of 10 with new facilities (3rd generation detectors) in the next two decades. With such improvements, gravitational wave astronomy will extend into cosmological studies. In the nearer term is the prospect of more large-baseline detectors allowing better definition of the position of the gravitational wave sources on the sky and enabling the coupling of gravitational wave astronomy to the older and more mature field of electromagnetic observations. This will put the gravitational wave sources into their astronomical settings and improve the science (multi-messenger astronomy).

## 2.8. How important is it to continue to improve the sensitivity?

Now that we have come to a threshold marking the beginning of gravitational wave astronomy, it is critical to continue the development of detector technology. The evolving science that can be accomplished depends almost entirely on the improvement in sensitivity of the detectors. This is the case for all the sources of gravitational waves. The detectors measure the amplitude of the gravitational waves and not the power carried by the waves. As a consequence the detectable signal from a specific type of source varies as 1/distance and not as 1/distance$^2$. That means that if one improves the sensitivity, S, of the detector by say a factor of 2, being able to look twice as deep into the universe, the number of sources one might expect to measure grows as the new volume of the universe open for the detection, a factor $S^3 = 2^3 = 8$.

Simply running the detector for longer, for a time T, increases the number of events one expects linearly with time, much more slowly than making improvements in the detector.

---

[5] Rainer Weiss (2.1-2.11)

For periodic sources where one integrates for long periods, the ratio of signal to noise grows as $T^{1/2}$ while it grows directly with S. For stochastic backgrounds of gravitational waves the signal to noise improves as $T^{1/4}$ while it grows directly with S.

Besides the ground based large baseline detectors, a set of other detectors and ideas are being tried out to extend the spectral range of the field.

The ground based systems measure phenomena with periods ranging from $10^{-4}$ to about $10^{-1}$ seconds. The short period detections are limited by quantum noise in the light, while limits on the longer period are set by the noise from fluctuating gravitational forces due to seismic compressional waves and atmospheric density fluctuations which perturb the mirrors (and, in atom interferometers, the motion of the atoms). The sources in this band have been discussed above.

## 2.9. What will LISA reveal?[6]

The band from a period of several hours to a few minutes is going to be measured by the Laser Interferometer Space Antenna (LISA, Figure 15). The project is a joint effort of the ESA and NASA to detect the gravitational radiation using three spacecraft ranged by laser light configured as an equilateral triangle with sides 5 million kilometers in length. The system orbits the sun at the same distance as the Earth, but behind it at a Lagrange point (no gravitational gradients) of the sun/earth system. LISA will detect the gravitational waves from the collision of $10^5$ to $10^6$ solar mass black holes throughout the universe, the radiation of smaller black holes falling into the large ones, and the radiation caused by white dwarf binary systems in our own galaxy. Some of the ground based binary neutron and binary black hole sources will be observed by LISA long before they come to coalescence. As a consequence of which, it has important application to anticipating when the collisions occur, and in preparation for electromagnetic observations of the actual coalescence by both telescopes and ground based detectors.

## 2.10. Will pulsar asymmetries be observable through gravitational radiation?

The gravitational waves with periods of years to fractions of a year will be observed through radio observations of deviations in pulse frequency of galactic millisecond pulsars. With a good selection of millisecond pulsars distributed throughout the galaxy, gravitational waves passing through the galaxy will cause the pulsar rates to vary with a quadrupolar pattern on the sky. For example, millisecond pulsars in the northern and southern sky will change pulse period in the same manner, say speed up, while those in the eastern and western sky will slow down. The technique, called millisecond pulsar timing detection of gravitational waves, will observe gravitational waves from the collision of monster black holes with masses as large as $10^{11}$ to $10^{12}$ solar masses when galaxies collide, as has been seen in Hubble telescope images. The technique may also uncover a stochastic background of gravitational waves with such long periods.

## 2.11. Will inflation be confirmed through its signature in the polarization of primordial background radiation?[7]

There are current measurements of the patterns of the polarization of the cosmic microwave background (CMB) which may produce evidence for primeval gravitational waves, generated by the rapid accelerations driven by quantum fluctuations at the time of the initial formation of the universe (inflationary epoch), a period more than $10^{10}$ years ago. The idea is that these gravitational waves cause density variations in the cosmic plasma with quadrupolar symmetry. The density variations are associated with temperature fluctuations in the plasma. The hotter parts radiate electromagnetically into the colder parts, which then scatter the radiation in a polarized manner. (This Thomson scattering is much like the polarization of sunlight when it is scattered by density fluctuations in the Earth's atmosphere.) When one superposes gravitational waves going in all directions through the plasma, a pinwheel like pattern of the CMB polarization is seen by a distant observer. The patterns are called B modes. Density fluctuations in the plasma due to the simple adiabatic temperature fluctuations of the plasma at equilibrium make CMB polarization patterns called E modes, which have the polarization going around in circles or in radial lines around a point. The B modes are delicate and can be aliased by polarized emission in objects between the earth and the surface of last scattering, such as dust and electrons in our own galaxy. Furthermore, B mode patterns can be generated by E mode patterns that have been gravitationally lensed by interceding galaxies. Nevertheless, with care and enough ancillary measurements of the foreground emission, it is hoped that B modes induced by primeval gravitational waves will be found in the next decade. Should this occur, it will be a landmark discovery for cosmology as it allows us to look back to conditions in the universe at the moment of creation.

---

[6] Rainer Weiss (2.1-2.11)
[7] Rainer Weiss (2.1-2.11)

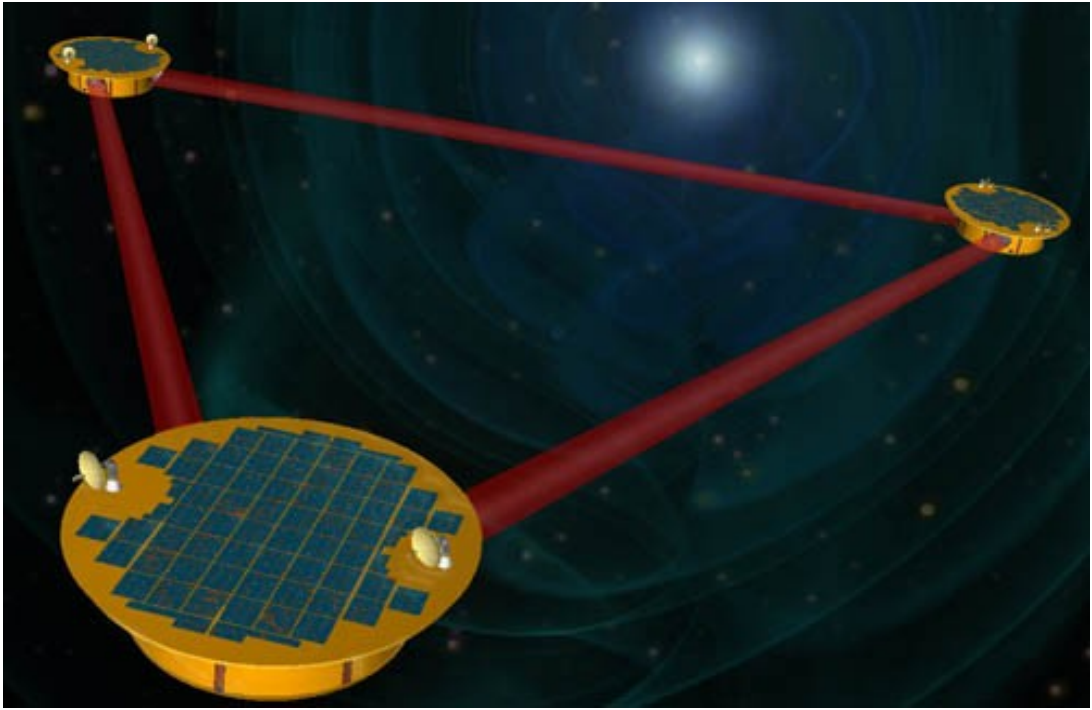

**Figure 15.** The Laser Interferometer Space Antenna (LISA) is a joint NASA-ESA project to develop and operate a space-based gravitational wave detector. LISA can detect gravitational-wave induced strains in spacetime by measuring changes of the separation between masses in three spacecraft 5 million kilometers apart. Credit: NASA.

## 2.12. Could atom interferometers be useful in searching for gravitational waves between 10 mHz and 10 Hz, bridging the gap between LIGO and LISA?[8]

The strain caused by gravitational waves can be observed by measuring oscillations in the distance between two spatially separated objects. This requires inertial proof masses whose positions are decoupled from the environment and a clock that is accurate enough to monitor these oscillations. In laser interferometers such as LIGO, the proof masses are suspended from well engineered vibration isolation systems, while the laser provides the clock necessary for the measurement. However, current lasers do not have the accuracy to measure the tiny strain caused by gravitational waves in order to mitigate this problem, and laser interferometers operate using two non-parallel baselines. In this scheme, the noise from the laser is common to both baselines while the gravitational wave strain is different. A differential measurement results in significant cancellation of noise from the laser while preserving the gravitational wave strain.

Recent advances in optical atomic clocks and atom interferometry may permit a new class of gravitational wave sensors that only require a single baseline. In these interferometers, the atoms can act as inertial proof masses and their internal energy levels can also be used to measure time. In this scheme, two atom interferometers separated by a distance L (see Figure 16) are operated by a common set of lasers. The atoms are exposed to the laser light at times 0, T and 2T. The atom-light interaction causes the atom to change its internal state, resulting in the development of a measurable phase-shift. This phase-shift is sensitive to the arrival time between the laser pulses. In the absence of a gravitational wave, the relative distance between atom interferometers is constant and thus the arrival times of the laser pulses do not change. In the presence of a gravitational wave, the distance between the atom interferometers modulates, resulting in a differential phase between the two atom interferometers. Since the interferometers are operated by the same laser, the noise from the laser is common and is cancelled to a high degree in the differential phase (*1*), while the gravitational wave signal is retained (*2*) . In this scheme, the cancellation of noise from the laser solely relies on the constancy of the speed of light. The atom clouds themselves need to be decoupled from environmental activity; this could be achieved by simply dropping them in free-fall (ballistic interferometers) (*2*) or by confining them to an optical lattice whose position is well engineered to be decoupled from the environment (*3*). While the atom technology is not presently as mature as optical interferometry, rapid technological developments in this field may make single baseline gravitational wave detectors suitable to search for gravitational waves in the frequency band 100 mHz to 10 Hz between LIGO and LISA.

## 2.13. Terrestrial detectors for infrasound gravitational waves[9]

As discussed in the previous sections, the direct observation of gravitational waves heralded a new form of astronomy. Gravitational waves carry information about a large number of exciting phenomena, which range from mergers of massive binary systems comprising neutron stars, white dwarfs, and black holes to the evolution of the early universe. These events create signals at rather diverse frequencies covering a broad spectrum.

---

[8] Surjeet Rajendran
[9] Ernst Rasel

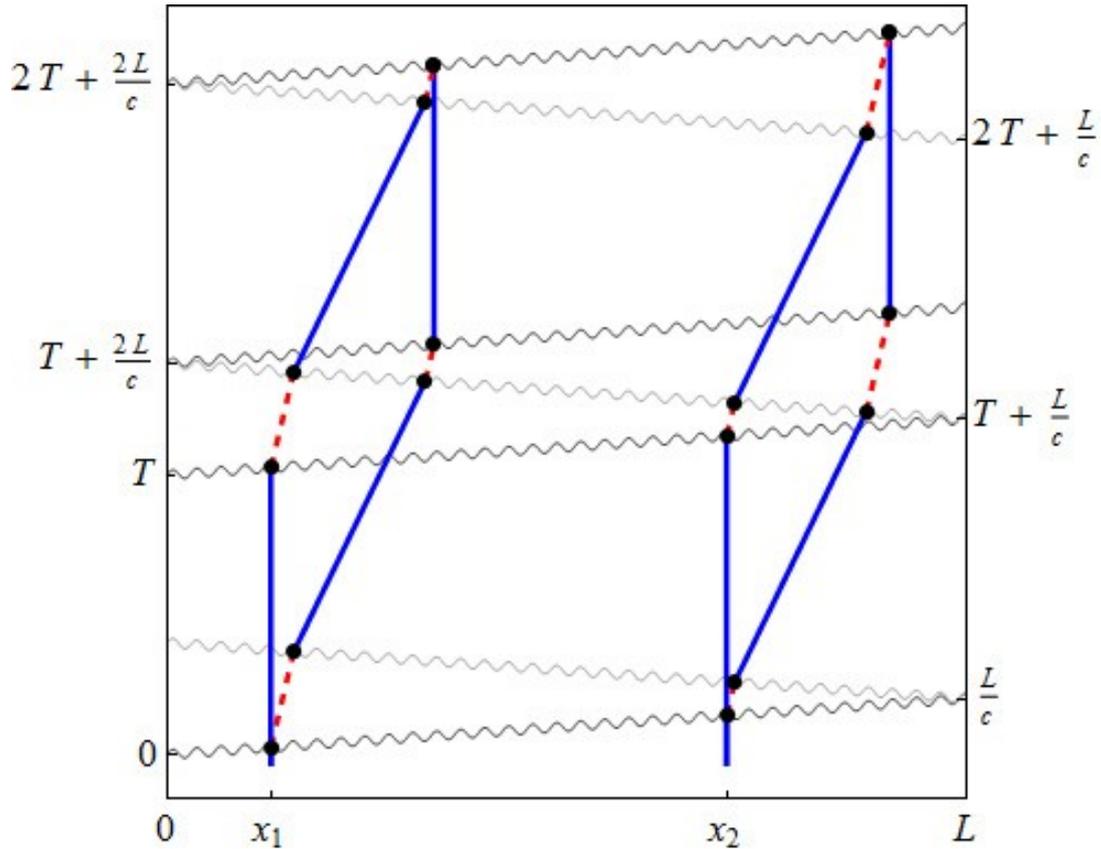

**Figure 16.** Laser interferometers. Credit: Surjeet Rajendran.

The incredible resolution required for the detection of tiny ripples in space-time caused by gravitational waves was first attained with laser interferometers. Such interferometers measure changes in the distance between mirrors employed as proof masses, which are shielded by extraordinarily sophisticated pendula from the noisy environment. Today's interferometers measure signals with frequencies comparable to acoustic waves of several tens to hundreds of Hertz (*4*), (*5*), (*6*), (*7*). These signals, persisting for only about a tenth of a second, stem from the very last second of binary mergers. The events are of such short duration that they make synchronous observation by standard astronomical instruments a challenge.

However, before the collapse, these objects emit signals in the infrasound domain, which last much longer (*8*), (*9*). Generally, many other fascinating astronomical phenomena create gravitational waves at much lower frequencies. For this reason, long-standing efforts are continuing to build detectors for this frequency range. The biggest hopes are for space-born laser interferometers (*10*) (*11*). Indeed, only recently, the first laser interferometer was successfully operated onboard the satellite Lisa Pathfinder (*12*). Lowering the frequency response of terrestrial antennas requires, however, new ways to overcome the environmental noise masking the gravitational-wave signals in present detectors. An extension of the frequency band of ground-based laser interferometers to the infrasound domain, as low as a tenth of a Hertz, would be spectacular indeed. This may be obtained by replacing the proof masses in laser interferometers by atomic de Broglie waves (*13*), (*14*), (*2*), (*15*), (*16*), (*17*), (*18*). Progress in light pulse interferometry with quantum degenerate matter has inspired scientists to explore this novel path (*19)*, (*20)*, (*21)*.

## 3. (2.) Are there new quantum phases of matter away from equilibrium that can be found and exploited – such as the time crystal?[10]

Symmetry and its breaking have been fertile themes in modern physics. In the context of relativistic quantum field theory, they have guided us to formulating fundamental laws. They have also helped us to understand the possible states of matter, and to analyze behavior within those states.

The classification of possible regular arrangements of molecules into crystals, and the recognition that physical properties of materials – including cleavage patterns, optical and elastic response, and above all band structure and quasiparticle behavior – is an especially impressive application. Here the governing symmetry is spatial translation. In the formation of a crystal lattice, the complete group of spatial translations is broken down to some discrete subgroup. (A full analysis should also include rotation symmetries, spatial reflection, and, when magnetic structure is involved, time reversal.) Perhaps the most fundamental symmetry of all is time translation symmetry. It is the statement that the laws of physics are unchanging and eternal. Strangely enough, there does not seem to be a convenient shorthand for the seven-syllable phrase "time translation symmetry"; here I will call it $\tau$ (tau). $\tau$ is related, through Emmy Noether's fundamental theorem, to the conservation of energy.

By analogy, it is natural to consider the possibility of states of matter wherein $\tau$ is broken down to a discrete subgroup. In that case, we may refer to a time crystal (*22*), (*23*). Whereas an ordinary (spatial) crystal contains an orderly pattern of molecules, a time crystal contains an orderly pattern of events (*22*), (*23*).

A beating heart is a time crystal in the broadest, purely mathematical sense. But a heart is complicated to construct, delicate, imprecise, and needy of nourishment. It is an interesting question for physics, whether there are simple (i.e., well-characterized and reproducible), robust, precise, and autonomous time crystals. Ideally, one would like to have systems that exhibit typical hallmarks of spontaneous symmetry breaking, such as long-range order, sharp phase transitions, and soft modes, wherein $\tau$ is the relevant (broken) symmetry.

Although spontaneous symmetry breaking is an established and mature topic in modern physics, it is not entirely straightforward to extend that concept to $\tau$ (*24*). Indeed, the usual heuristic to motivate spontaneous symmetry breaking is that a system will reduce its symmetry in order to minimize its energy (or, at finite temperature, free energy), but if $\tau$ is broken, then energy is no longer a useful, conserved quantity.

Nevertheless, there are physical systems which exhibit several of the hallmarks of spontaneous $\tau$ breaking, and thus deserve to be called physical time crystals. One class is related to the AC Josephson effect (*25*). In that effect, a constant voltage, applied across a superconducting junction, produces an oscillatory response. The AC Josephson effect in its usual form is an imperfect time crystal (*25*), since the current is degraded by radiation (when the circuit is closed) and by resistive dissipation, but those limitations can be overcome in systems inspired by similar concepts.

Recently a new and very interesting class of time crystals, the so-called Floquet time crystals, was predicted, and then demonstrated experimentally (*26*) (*27*). These are driven systems, subject to a time-dependent Hamiltonian with period T, so H(t+T) = H(t). In different examples, they exhibit response which is periodic only with the longer periods 2T or 3T. Thus, the equations of a Floquet time crystal exhibit a discrete version of $\tau$, but its response exhibits only a smaller (discrete) symmetry.

---

[10] Frank Wilczek

Other classes of time crystals have been predicted, but not yet observed. Prethermal time crystals (*28*) are closed systems which exhibit $\tau$ breaking for arbitrarily long, but finite, times; open time crystals are a variant of those, allowed (*28*) access to a heat bath, which can display permanent $\tau$ breaking.

Exploration of time crystals is in its infancy. Many open questions suggest themselves. Are there time liquids, glasses, and quasicrystals? Are there space-time crystals, exhibiting dynamical patterns? Can one classify the possibilities? What are the thermal, and electromagnetic properties of these new states of matter? Indeed, all the questions that arise when analyzing conventional states of matter, and more, must be addressed anew for these new, dynamical states.

## 4. (3.) Quantum theory in uncharted territory: What can we learn?[11]

We often gain insight into the foundations and the limitations of a theoretical framework by considering its inner workings at the interface with another one. Quantum theory is a perfect example for such a concept providing us with new ideas when considered in domains alien to it. In this section we address phenomena which live on the seam connecting quantum mechanics with other branches of science. We have chosen four characteristic fields illustrating in a striking way our point: (i) Classical electrodynamics, (ii) the classical-quantum transition, (iii) gravitation, and (iv) number theory. For each of these case studies we formulate in the form of a pregnant question a path out of the conundrum.

### 4.1. Can we create fractal radiation?[12]

The time evolution of a quantum state, dictated by the Schrödinger equation, can weave intricate structures in spacetime. Indeed, the probability density of an appropriately designed initial wave packet, when represented in a space spanned by the coordinate and time, displays infinitely many deep and narrow canals as well as sharp ridges that criss-cross each other in a regular way. These quantum carpets (*29*) emerge for example, in the quantum dynamics of a Gaussian caught in an infinitely deep potential well.

The canals exhibit an even finer structure (*30*) for a packet that is initially constant across the well. Indeed, due to the sharp corners of this mesa-shaped wave function, the bottoms of the canals, as well as the cuts through the probability density along the two coordinates of spacetime, are of fractal nature (*31*), with a Hausdorff dimension of 5/4 in the case of the canal along the main diagonal.

One possible application of such a fractal quantum carpet was the topic of a walk in Austin, Texas in the fall of 1997 with John Archibald Wheeler. He suggested to one the present authors (WPS) to consider the Smith-Purcell radiation (*32*) from an electron running along a valley with such a fractal metal floor.

The idea for his suggestion originated from the opening sentence of the Smith-Purcell (*32*) note:

> "If an electron passes close to the surface of a metal diffraction grating, moving at right angles to the rulings, the periodic motion of the charges induced on the surface of the grating should give rise to radiation. A simple Huygens construction shows the fundamental wave length to be… proportional to the distance between the rulings."

---

[11] Wolfgang P. Schleich
[12] Wolfgang P. Schleich

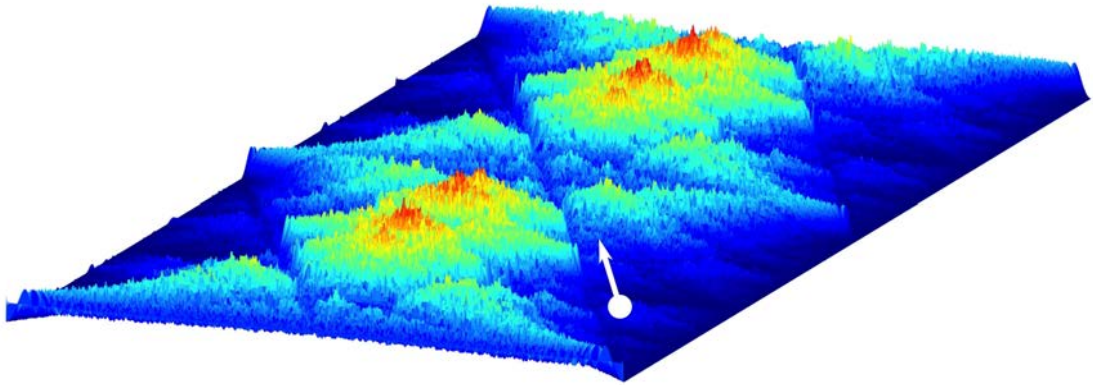

**Figure 17.** Fractal Smith-Purcell radiation emitted by an electron passing by a rough metal surface, provided for example by a fractal quantum carpet. Credit: Harald Losert.

Since the periodicity of the grating determines the wave length of the light, a grating with many periods, and in particular, of fractal nature, should lead to light with "fractal frequencies". Unfortunately, the task of calculating this radiation has never been completed and we need to ask the question summarized in Figure 17: Can we map the fractal behavior of a metal surface onto the frequency space of light, thereby achieving "fractal radiation"?

## 4.2. What is the elasticity of spins?[13]

Revival phenomena appear in a wide variety of physical systems ranging (*33*) from Rydberg atoms, via the Jaynes-Cummings model, to the Mott insulator phase transition. On first sight the periodic appearance (*34*) (*35*) of bursts of free-induction decay in a mixture of $^3$He-$^4$He seems to belong to this category of effects. However, we conjecture that they are a manifestation of Helmholtz oscillations (*36*).

This phenomenon is best illustrated by the example of an open bottle. The air inside the bottle-neck and the elasticity of the air inside the bottle represent a mass-spring system. An initial pressure difference between the inside and outside of the bottle leads to an excitation of this resonator resulting in air rushing back and forth through the bottle-neck. Were it not for dissipation, these oscillations would continue forever.

In the same spirit, the experiments of Ref. (*34*) (*35*), briefly summarized in Figure 18, rely on a container (bottom) being connected by a thin and short pipe to a reservoir (top) which is larger by a factor of ten. Completely filled with a dilute mixture of $^3$He-$^4$He both cavities are immersed in a strong magnetic field $H_0$ aligning the spins of the atoms. A coil around the small container at the bottom allows a complete inversion of the $^3$He-spins and establishes a nonequilibrium.

The surprising observation is the appearance of periodic bursts of free-induction decay measured by this coil. Three characteristic features stand out: (i) There exists a delay between the preparation of the spins and the first occurrence of the signal, (ii) the bursts are fragile when a gradient field $\nabla H$ is applied orthogonal to the strong magnetic field $H_0$ aligning the spins, but is not as sensitive when $\nabla H$ is parallel to $H_0$, and (iii) the time scales involved are unusually long, of the order of seconds and even hours.

We propose that the origin of this phenomenon is the formation (*37*), and the motion (*38*) of a domain wall separating the spins being up from the ones being down prepared by the initial pulse. Most likely the domain wall is formed in the pipe connecting the two cavities. The nonequilibrium state together with the disparity in the reservoir sizes forces the wall to move towards the smaller cavity, with the spins down flipping up as it progresses.

---

[13] Wolfgang P. Schleich, Harald Losert, Dean L. Hawthorne, Gavriil Shchedrin, Marlan O. Scully, and David M. Lee.

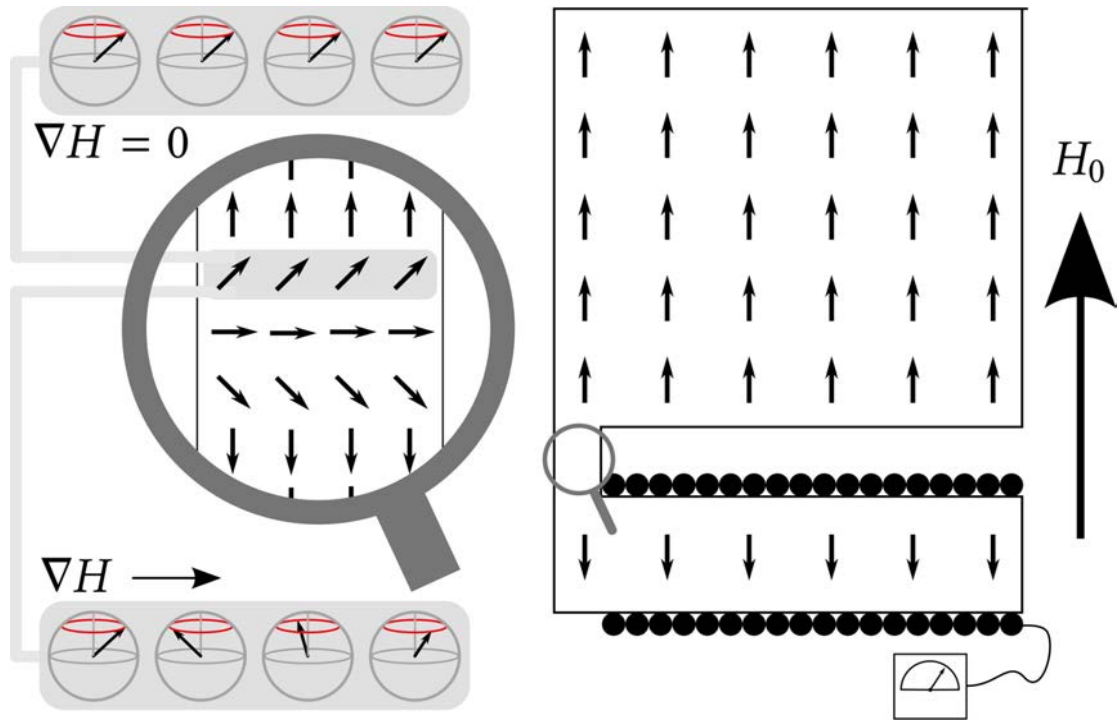

**Figure 18.** Helmholtz oscillations in a dilute mixture of liquid $^3$He-$^4$He explained by the formation and motion of a domain wall (left) in the narrow pipe connecting the top and the bottom containers. This wall separates the spins which are aligned with the strong homogeneous magnetic field $H_0$, from the ones which are oriented against it due to an initial π-pulse created by a coil around the bottom container. The difference in size of the spin systems, as well as their elasticity, leads to an oscillatory motion of the wall detected by the coil as a current. The wall is strongly influenced by a gradient magnetic field $\nabla H$ when aligned orthogonal to $H_0$. Credit: Harald Losert.

When the wall leaves the pipe and enters the lower cavity, it is faced with a majority of spins being down. It first flips them but then has to succumb to the overwhelming majority. This dissipation of the domain wall is followed by a diffusion process, resulting in a re-flipping of the spins in the tube, and the domain wall retreats to the tube.

The process repeats itself until the spin exchange approaches an equilibrium with a non-moving domain wall which significantly slows down the restoration of the initial state of the system, explaining the extreme lifetime.

Since in the pipe the domain wall is orthogonal to the aligning magnetic field $H_0$, a gradient field $\nabla H$ orthogonal to $H_0$ has a dramatic effect, forcing the spins to get out of step as indicated on the left of Figure 18. In contrast, a gradient parallel to $H_0$ has little influence since the domain wall is thin compared to the slope of the gradient, and the spins remain in phase.

The crucial step in obtaining a complete understanding of this phenomenon is to translate the mass-spring model of the Helmholtz oscillation into the movement and dissipation of the spin wave. The task is to extract it from the non-linear spin diffusion equation of Leggett (*39*). However, this problem corresponds to answering the question posed at the beginning of this section.

### *4.3. Is the event horizon of a black hole a beam splitter?[14]*

In 1973 Jacob Bekenstein conjectured (*40*) that the entropy of a black hole is proportional to the ratio of its surface area, determined by the event horizon, to the square of the Planck length. This fact is closely related (*41*) to the observation that the radiation emitted from an accelerated particle (*42*) or a black hole (*43*) is that of a black body.

Bekenstein also calculated (*44*) the quantum state of the electromagnetic field emitted by a black hole provided *n* photons are scattered off it. It is interesting that this grey-body radiation follows from a simple model (*45*) of the black hole with the event horizon serving as a beam splitter.

Indeed, two input modes and two output modes suffice to explain the essential features of grey-body radiation. The quantum states of the input modes are an *n* photon state and a thermal state. If one of the two output modes is absorbed, we obtain in the other mode the Bekenstein radiation.

The reflection and transmission coefficients of the beam splitter follow from the matching of the in- and out-going modes of the electromagnetic field at the event horizon with the familiar logarithmic phase singularity (*46*)

$$\psi(r) = (r - r_0)^{i\kappa} = e^{i\kappa \ln(r - r_0)} \tag{1}$$

following from the differential equation

$$(r - r_0)\frac{d\psi}{dr} = i\kappa \psi \tag{2}$$

Here $\psi$ is the approximate solution of the radial scalar wave equation in the presence of the Schwarzschild metric in the neighbourhood of the Schwarzschild radius $r_0$. Moreover, $\kappa$ contains the parameters defining the black hole.

It is interesting to note that the same singularity arises (*47*) in the context of tunnelling through a quadratic potential barrier when analyzed in rotated quadrature rather than position eigenstates. Indeed, the time-independent Schrödinger equation expressed in the rotated phase-space variables is identical to Eq. (2).

Moreover, the differential equation Eq. (2) on the real line rather than in the complex plane determines (*48*) the Stefan-Boltzmann law of black-body radiation, thereby connecting classical electrodynamics and classical thermodynamics. It is amazing that these two classical theories suffice to derive, apart from a numerical factor, this quantum law which involves Planck's constant explicitly.

Why not employ the beam-splitter model which was already successful in rederiving the celebrated Bekenstein formula to obtain the entropy of a black hole? In this case, we replace the *n*-photon state by the vacuum, and the grey-body radiation by the Hawking radiation corresponding to a thermal state of density operator $\rho'_{th}$. We again absorb the field in the other outgoing mode but search for the state $\rho_{th}$ of the input mode leading to $\rho'_{th}$ in the outgoing mode. This must be a thermal state of a larger photon number.

Hence, the entropy of this beam-splitter arrangement shown in Figure 19 should be that of a thermal state. The temperature is fixed by the mode matching at the horizon. One question remains: Does this idea lead to the desired formula?

---

[14] Wolfgang P. Schleich, David M. Lee, Marlan O. Scully, and Anatoly Svidzinsky

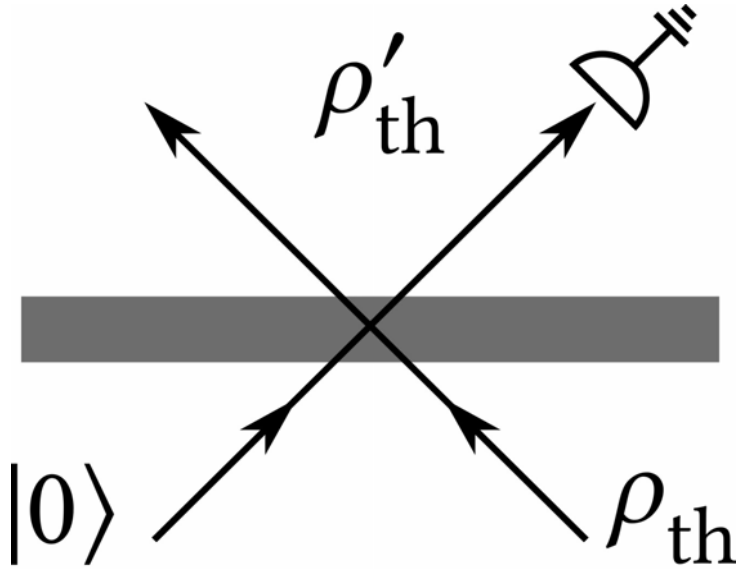

**Figure 19.** Model of a black hole as a beam splitter to obtain the Bekenstein entropy formula. Here the two input modes are in the vacuum state |0> and a thermal state $\rho_{th}$ of a given average number of photons, and the output modes are either absorbed or in a thermal state $\rho'_{th}$ corresponding to a number of photons that is smaller than of $\rho_{th}$. Credit: Harald Losert.

### 4.4. Do continental divides of the Newton flow offer a path towards the Riemann Hypothesis?[15]

On August 8, 1900, David Hilbert introduced ten of his famous 23 problems in mathematics at the International Congress of Mathematicians. At that time, none of them had been solved. Even today some are still unsolved, such as the problem number eight which deals with the Riemann Hypothesis: all non-trivial zeros of the Riemann zeta function $\zeta = \zeta(s)$ of the complex argument $s \equiv \sigma + i\tau$ lie on the critical line determined by the real part ½.

Closely related to the Riemann Hypothesis is the Hilbert-Pólya Conjecture, according to which the imaginary parts of these zeros correspond to the eigenvalues of a Hermitian, self-adjoint operator. Recently, motivated by PT symmetry such an operator has been proposed (*49*).

It is interesting that a point of view based on quantum dynamics rather than quantum *kinematics* yields insight into the distribution of the non-trivial zeros. Indeed, the probability amplitude for a time-evolved wave packet to contain its initial state represents (*50*) $\zeta$, provided we deal with a logarithmic energy spectrum, and an initial thermal phase state determined by this spectrum.

This realization of $\zeta$ by a single quantum system is only possible for $1 < \sigma$. In order to deal with the zeros in question, we need to consider the superposition of two such wave packets moving in phase space in opposite directions. However, in this case a physical realization requires two entangled oscillators rather than a single one. In this sense, the analytical continuation of $\zeta$ into the domain of the complex plane to the left of $\sigma = 1$ is analogous to entanglement.

More insight into the properties of $\zeta$, and its non-trivial zeros springs (*51*), (*52*) from the Newton flow of $\zeta$ which brings to light the lines of constant phase of a function by the inverse of the logarithmic derivative of the function. In this approach, the zeros are sinks of the flow, and the sources are either at the poles or at infinity.

Although the emerging patterns for $\zeta$ are aesthetically appealing they are rather convoluted. A

---

[15] Wolfgang P. Schleich, Iva Bezděková, Moochan B. Kim, Helmut Maier, and John W. Neuberger

much more symmetric structure emerges when we consider the product

$$\xi(s) \equiv \pi^{-s/2}(s-1)\Gamma(s/2+1)\zeta(s) \qquad (3)$$

where $\Gamma$ denotes the Gamma function.

In this case the flow lines are (i) symmetric with respect to the critical line, (ii) always approach it from $\sigma = \pm\infty$, and (iii) there the phase of $\xi$ increases without a bound.

Every flow line has to eventually terminate in a zero. However, when there is more than one zero flow lines exist that go first through the zero of the first derivative and only then approach the zero of the function. These curves shown in Figure 20 in green serve as continental divides for the flow and are called separatrices.

In Ref. (*53*) we have shown that if the phase difference between two consecutive separatrices is $\pi$ all zeros of $\xi$, and thus of $\zeta$, are simple zeros, and are located on the critical line. Hence, in this dynamical approach toward the Riemann Hypothesis the crucial question remains: How can we identify the separatrices in the sea of flow lines? Are the contour lines, that is, the lines of constant height of $\xi$ depicted in Figure 20 in magenta, a possible indicator?

### *4.5. Quantum tests of the equivalence principle with unprecedented sensitivity[16]*

Einstein's very successful theory of general relativity is built on the equivalence principle (*54*). It claims that the gravitational acceleration during free fall is universal, i.e. independent of the composition of attracting bodies. This feature makes gravity special amongst forces as it can be locally absorbed in space-time geometry. The simplicity of Einstein's theory is an appealing argument to trust in this postulate. This view might change, if one considers the challenge to reconcile gravity and quantum mechanics. So far, no theory masters this task in a satisfactory way. It might well be that a theory unifying all forces will lead to tiny violations (*55*), (*56*), (*57*), (*58*), (*59*), (*60*), (*61*). Other caveats relating to the validity of the postulate stem from cosmology and particle physics (*62*). Since its formulation, the postulate has been scrutinized by increasingly sophisticated experiments (*54*), and, so far, pendula (*63*) and laser ranging of the Earth and moon (*64*) set the most rigorous bounds. As we write, in 2017, the MICROSCOPE satellite is orbiting around the Earth to perform the most stringent test to date by comparing the free fall of test masses to parts in ten to the power of fifteen (*65*).

Matter wave interferometers can probe how gravity acts on the superposition of de Broglie waves and promise new insights and more stringent tests. These experiments have several interesting implications. They are performed with ensembles representing pure spin states and generally the species of choice can be employed both in atomic clocks and interferometers enabling tests of the universality of free fall as well as of the redshift. Both cases have the equivalence principle as a common root according to Schiff's conjecture (*62*). Starting from first proof-of-principle demonstrations laying the foundations for this research field at the interface between quantum mechanics and gravity (*66*), (*67*), (*68*), recent initiatives started to perform more stringent tests (*69*), (*70*), (*71*), (*72*), (*73*), (*74*).

Novel methods (*62*), (*20*), (*75*), (*76*), (*21*), (*77*), (*78*), (*79*), (*80*), (*81*), (*82*), (*83*), (*84*) (Figure 21), show great promise that quantum tests will not only reach the same level of performance as their classical counterparts, but that they may ultimately attain accuracies in the range 10 to the power of seventeen to eighteen (N. Gaaloul, in preparation).

---

[16] Ernst Rasel

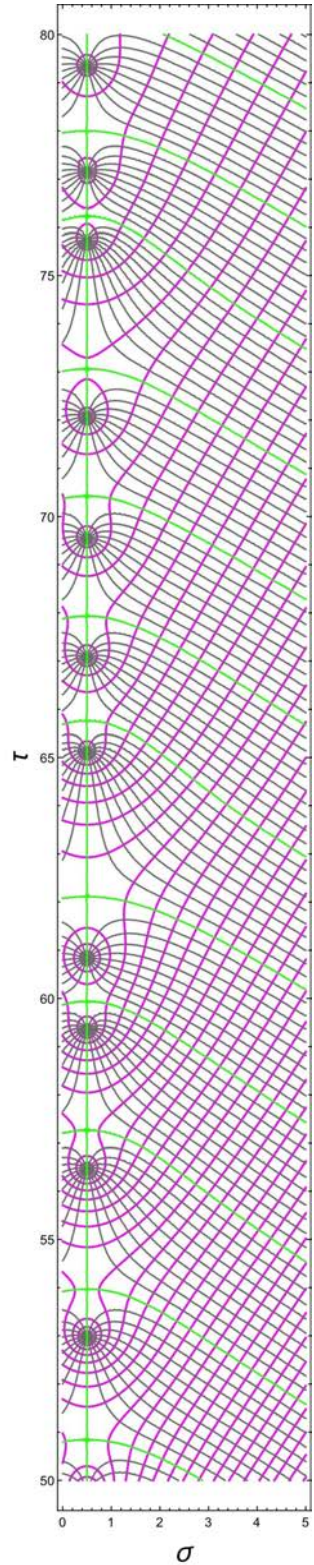

**Figure 20.** Geometrical approach towards the Riemann Hypothesis based on the lines of constant phase (grey curves) and of constant height (magenta curves) depicted in the complex plane $s \equiv \sigma + i\tau$ of $\xi = \xi(s)$ in the neighborhood of the critical line $\sigma = \frac{1}{2}$. The flows to the individual zeros (grey dots), in this domain of $\tau$-values all located on the critical line, are separated by continental divides (green lines) which merge with the critical line in points where the first derivative of $\xi$ vanishes. Credit: Iva Bezděková.

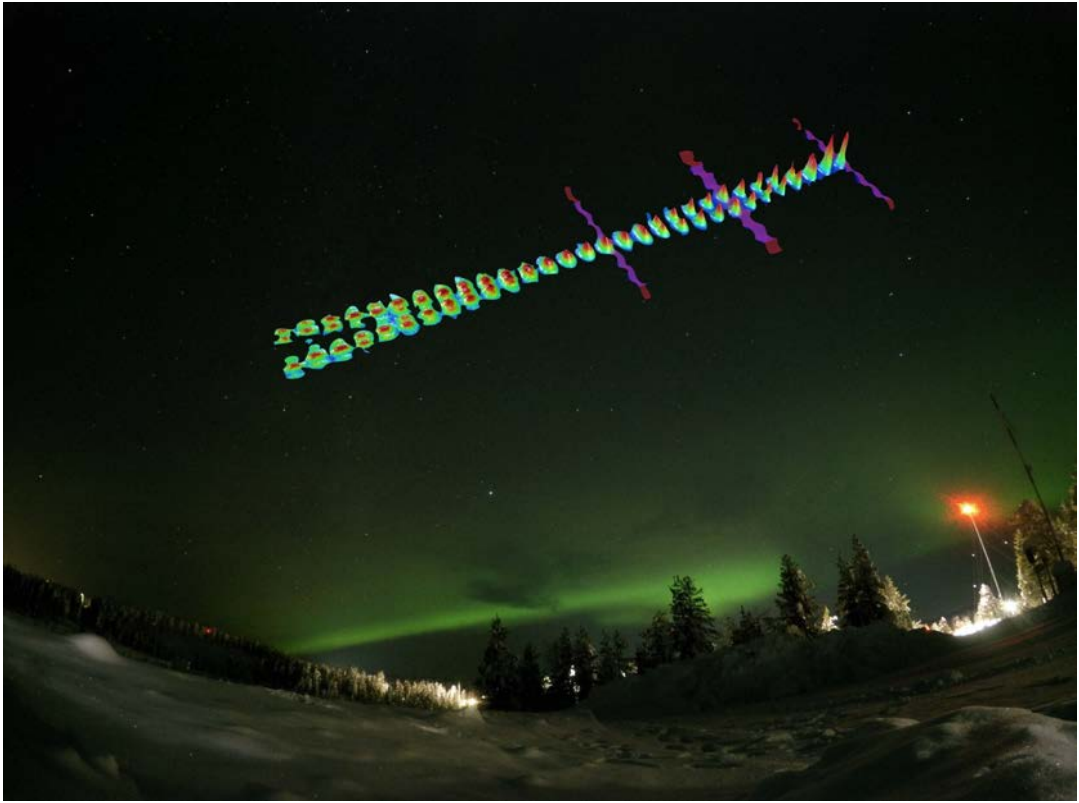

**Figure 21.** Artistic collage visualising a future space-born interferometer employing Bose-Einstein condensates (BEC) against a sky illuminated by the Northern lights in Sweden. The first Bose-Einstein condensate in space was created aboard the sounding rocket MAIUS-1, launched from Esrange, where this photograph was taken. BEC interferometers were proposed for satellite-based quantum tests of the equivalence principle like STE-QUEST. The duration of the free fall in space interferometers can be extended considerably in comparison to that in ground-based experiments, resulting in an enhanced sensitivity of the quantum tests. The BEC interferometer data, formed by light pulses, was obtained with the QUANTUS-1 apparatus. Credits: Ahlers and Schlippert (photograph) and Ernst Rasel.

## 5. (4.) What are the ultimate limits for laser photon energies?[17]

The wavelength of radiation fundamentally limits spatial resolution. Therefore, shortening of the wavelength leads to new records in all fields where high spatial resolution and precision are required, including microscopy, nanotechnology, tomography, imaging, lithography and so on. Shortening the wavelength means increasing the laser frequency and photon energy, resulting in much easier detection of photons owing to the reduced background noise and higher signal-to-noise ratio. However, there are some fundamental difficulties in reducing the laser wavelength to extreme ultraviolet (EUV or XUV) (50 nm – 15 nm), x-ray (15 nm – 0.1 Å) and gamma-ray (< 0.1Å) ranges and, accordingly, in increasing the photon energy to much higher values than 1 eV. Indeed, in their seminal paper "Infrared and Optical Masers" Charles Townes and Arthur Schawlow (*85*) noted:

> "Unless some radically new approach is found, they [maser systems] cannot be pushed to wavelengths much shorter than those in the ultraviolet region."

To appreciate the difficulties, it is necessary to recall that light amplification by stimulated emission of radiation (LASER) requires prevalence of the stimulated emission over the stimulated absorption, which implies a higher population of an upper energy state over a lower energy one: $n_2 > n_1$, i.e. a population inversion at the corresponding lasing transition. The latter condition requires an incoherent pump with a rate higher than a decay rate of the upper state. In the case of pumping by incoherent radiation, this condition requires a sufficiently large flux P of incoherent photons, $P > \frac{1}{\sigma_{res} T_1}$, where $\sigma_{res}$ is a resonant cross-section of the field-matter interaction and $T_1$ the lifetime of the upper state. The resonant cross-section and the lifetime decrease proportionally to the square and the cube of the wavelength λ, respectively, resulting in a rapid increase of the required pumping flux P ~ $\lambda^{-5}$. The existing sources of incoherent radiation in the high frequency ranges, such as x-ray tubes, LINACs and synchrotrons simply cannot provide the required flux. Besides, population inversion alone is not yet a sufficient condition for lasing. Indeed, the amplification requires a prevalence of a net resonant gain $G$ over the off-resonant losses caused by the photoionization and Compton scattering, $\beta$, that is $G > \beta$, where the net resonant gain is defined as

$$G = \frac{\lambda^2 (n_2 - n_1) \Delta\omega}{2\pi \Delta\omega_{tot}}.$$

(4)

---

[17] Olga Kocharovskaya

Here $\Delta\omega$ is a radiative linewidth of the resonant transition and $\Delta\omega_{tot}$ is a total linewidth which includes an inhomogeneous broadening. In the case of poor radiative broadening, gain scales as G ~ $\lambda^2$. Additional line broadening mechanisms reduce the gain by a factor $\Delta\omega / \Delta\omega_{tot}$, making it very difficult to overcome the off-resonant losses.

Finally the very fast (from pico to femtosecond) decay rate of the high-frequency atomic transition and radiation damage of mirrors by the flux of high-energy photons, as well as an absence of high finesse cavities in the high-frequency range, make a single-pass self-amplified spontaneous emission (SASE) regime a necessity, leading to an additional rather stringent requirement. Namely, the single-pass net gain should be reasonably high, $(G - \beta)L >> 1$, where $L$ is the length of an amplifying medium.

In spite of all these seemingly insurmountable problems, the persistent research efforts of the scientific community to shorten the laser wavelength during the last half century have led to a variety of lasers operating in the extreme ultraviolet and even x-ray ranges. The shortest laser wavelength achieved is 0.6 Å (SACLA, Japan). This corresponds to a photon energy 19.6 keV (which is about 20 000 times larger than the energy of an optical photon). What are the radically new approaches that led to such remarkable progress?

Currently, there are three types of lasers in the XUV (50 nm – 15 nm) and soft-x-ray (15 nm – 1 nm) ranges: (i) the table-top lasers based on population inversion between energy levels of ions in plasma, (ii) the large-scale soft-x-ray free-electron lasers based on collective emission of high quality electron-bunch trains accelerated to relativistic energies (~1 GeV) and wiggling in the periodic magnetic field of a long undulator, and (iii) the secondary lasers based on population inversion at an atomic inner-shell transition produced by the radiation of the x-ray free-electron laser.

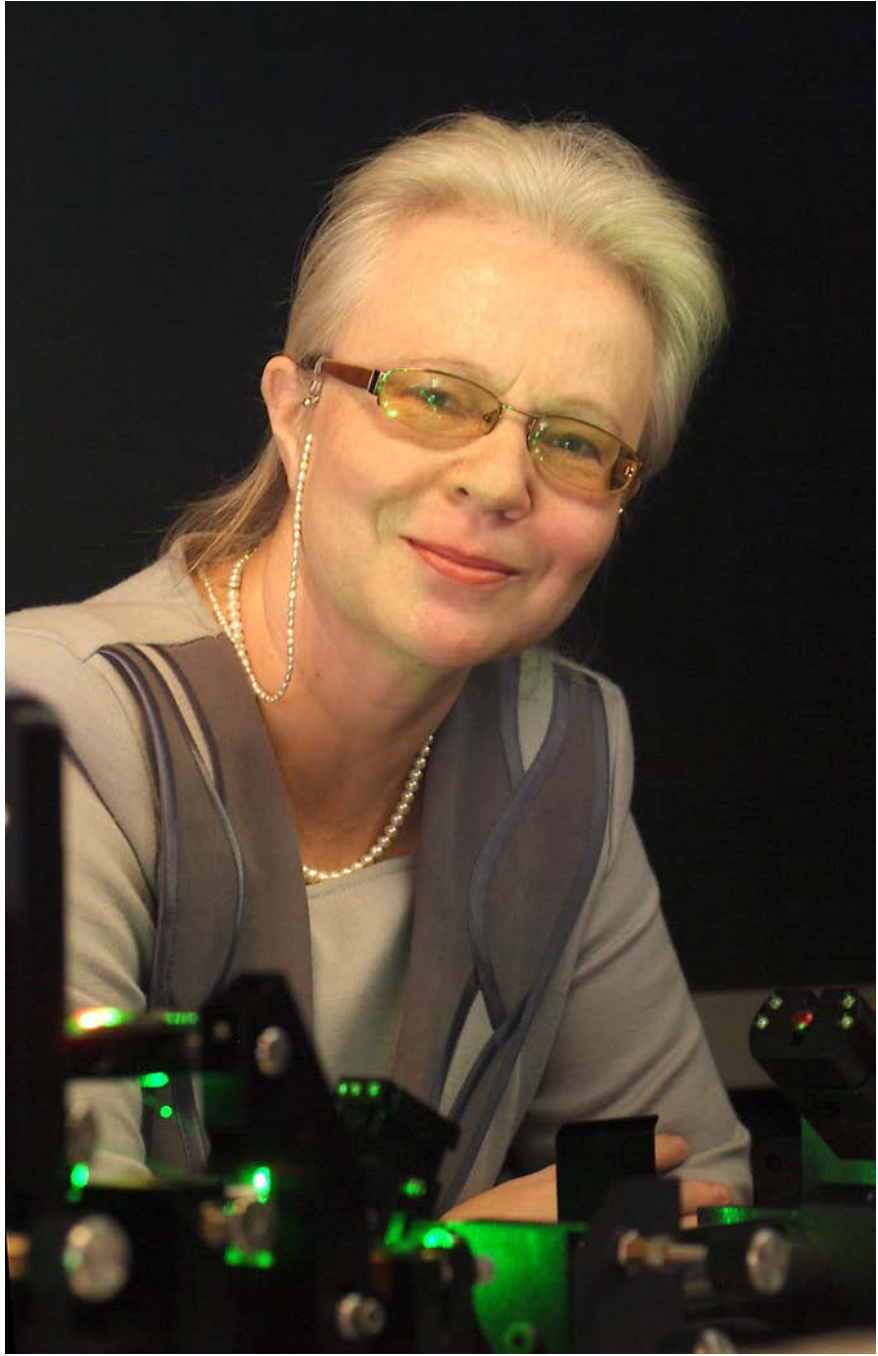

**Figure 22.** Olga Kocharovskaya in her laboratory. Credit: Institute for Quantum Science and Engineering.

There are two different types of table-top plasma-based x-ray lasers: recombination lasers with pumping via a three-body collisional recombination process and collisional lasers with pumping by electron collisional excitation (*86), (87)*. Interestingly, the building of a rather wide range of such plasma-based x-ray lasers is itself attributable to the development of the high power ultrashort IR and optical lasers widely used for production of plasma with the required parameters. In particular, the recombination x-ray lasers rely on fast optical laser ionization via tunneling, resulting in a complete stripping of all electrons without appreciable heating (*86)*. The shortest wavelength currently achieved in the recombination lasers is 4.03 nm (*88)*. The pulse duration in the table-top plasma lasers is longer than a picosecond. The pulse energy and the pulse repetition rate in the collisional plasma x-ray lasers are as high as several mJ and 100 Hz, respectively (*87*). Such plasma-based x-ray lasers have found numerous applications in chemistry, biology, medicine, nanoscience and materials science (*86*), (*87*).

Much shorter, up to several femtosecond pulses have been produced by soft x-ray free-electron lasers with a pulse energy of a few hundred *μJ* and a repetition rate of 10-100 Hz (*89*). The temporal coherent properties of such lasers are not very good owing to the shot noise in the electron bunch. The pulses are not Fourier-transform limited and differ from each other essentially in their intensity, temporal structure and spectral content. Seeding by a high harmonic source makes it possible to reduce the bandwidth and produce transform-limited pulses.

The secondary soft x-ray lasers are based on population inversion in the atomic innershell transitions and employ x-ray free-electron lasers for pumping. Such secondary sources have narrower radiation bandwidth than the primary ones. The first laser of this type had a wavelength of 1.46 nm (0.8 keV) at the K-shell transition in Ne+ ions pumped at 0.96 keV at the Linac Coherent Light Source (LCLS) in Stanford (*90)*. The only solid-state laser in the soft x-ray range was produced in silicon at 90 eV photon energy (wavelength ~8 nm) with pumping by a radiation at 115 eV at FLASH, DESY, Germany (*91)*.

Also worth mentioning is the recent development of a different type of coherent sources of radiation in the XUV and soft x-ray ranges, namely, the high harmonic generation sources (HHGS) (*92)*, (*93)*, (*94)*. These sources are based on the highly nonlinear process of phase matched high harmonic generation by incident IR, optical or VUV radiation in the noble gases or plasmas, caused by tunnel ionization followed by a recombination with the parent atoms or ions. In the XUV range, femtosecond trains of coherent pulses of ultrashort (~100 as) duration were produced (*92), (93), (94)*. Isolated transform-limited attosecond pulses of several mJ energy have also been produced using the temporal or polarization gating techniques (*95)*. Such pulses have found many applications for studies of ultrafast electron dynamics in atoms, molecules and solids with a record high temporal resolution. The spectrum of the high harmonics generated extends into the soft x-ray (*96)* and even hard x-ray ranges (to more than 1.6 keV photon energy (*97)*). However, the part of the pulse energy contained within these spectral ranges was dramatically small, of the order of nano or pico-J, and the sub-femtosecond pulses have not yet been demonstrated by a direct measurement in the time domain.

In the hard x-ray range, there are two types of lasers, namely, the X-ray Free-Electron Lasers (XFELs) and the atomic inner-shell lasers based on pumping by the XFELs. Currently, there are three XFELs operating in the hard x-ray range: LCLS in Stanford, USA (up to 11.2 keV photon energy), SACLA (the Spring-8 Ångström Compact free electron LAser) in Hyogo Prefecture, Japan (up to 19.6 keV) and PAL (Pohang Accelerator laboratory) in Seoul, Korea (up to 2.5 keV). Three more XFELs are expected to start operation by the end of this year: the European XFEL in Hamburg, Germany (up to 24.8 keV), the Swiss XFEL in Wurenlingen, Switzerland (up to 12.8 keV) and SINAP (Shanghai Institute of Applied Physics) in Shanghai, China (up to 12.4 keV).The shortest wavelength of 0.6 Å is achieved at SACLA, although it is expected that an even shorter wavelength, albeit only of 0.5 Å, will be achieved at the European XFEL after its official opening in September 2017. The European XFEL has the highest expected repetition rate of 27 000 pulses per second compared to 120 at LCLS and SACLA. The XFEL pulse duration is in the range 10 – 300 fs and the pulse energy is of the order of mJ. An intensity as high as $10^{17}$-$10^{20}$ W/cm$^2$ has been achieved, resulting in recent demonstrations of the second harmonic generation (*98), (99)*. The product of the duration and the spectral width of the XFEL's pulse is of the order of 100. Nearly transform-limited pulses of about 10 femtosecond duration were obtained at LCLS in a recently demonstrated self-seeded regime, resulting in a decrease of the bandwidth by a factor of about 50 (*100)*.

The shortest wavelength attained so far in lasers with population inversion at the inner-shell atomic transition is 1.54 Å (8 keV). This lasing was demonstrated by pumping K-shell electrons in Cu atoms with 1.4 Å radiation and seeding with 1.54 Å radiation (*101*). Both the seeding and the pumping radiation were produced by SACLA operated in a two-color mode.

There are certainly many prospects for further development of existing coherent sources in the x-ray range. For example, LSLC II (an upgrade of LCLS) plans an increase in the repetition rate of up to one million pulses per second. Shortening of the table-top x-ray pulses or amplification of the high harmonics in a resonant passive or active plasma modulated by an IR laser field could lead to intense attosecond table-top x-ray sources (*102*), (*103*).

Building the secondary x-ray lasers with population inversion at the atomic innershell transitions pumped with XFELs may stimulate a revisiting of the gamma-ray laser problem, at least, for the photon energies below 24 keV (*104*). Note that, historically, when mentioning the gamma-ray laser, the reference is to a laser operating at the nuclear (as opposed to the electronic) transition and not necessarily in the usual gamma-ray range. In spite of considerable effort, the problem has remained unsolved for more than fifty years (see the reviews (*105*), (*106*), (*107*)). It was included in the famous Ginzburg list of 30 of the most important problems of modern physics (*108*).

### *5.1. What are the prospects for production of coherent sources in the gamma-ray range with a wavelength λ < 0.1 Å and photon energy $\hbar\omega >$ 100 keV ?*[18]

None of the existing coherent x-ray sources (including the table-top plasma lasers, high-harmonic generation and XFELs) looks very promising for upgrading to photon energies much higher than 25 keV. Indeed, modern XFELs are based on RF linear accelerators (LINAC) accelerating electron beams to several GeV energies at a rate of MV/m and, hence, require lengths of the order of 1 km. Besides, these are very expensive state of the art facilities.

On the other hand, there is hope that the recently developed laser plasma wakefield accelerators may open a path towards compact XFELs (*109*), (*110*). With this approach, an intense laser pulse drives the plasma density wakes to produce (by charge separation) a strong longitudinal electric field that can reach a magnitude of a few 100 GV/m, dramatically reducing the length of the system. It still remains to improve the quality of the accelerated electron beams produced (to reduce the emittance, the divergence and the relative energy spread) sufficiently to demonstrate lasing based on such compact accelerators in the VUV, XUV and, eventually, x-ray ranges. However, once this goal has been achieved, such lasers could potentially be scaled to the gamma-ray range.

Another possible way to attain lasers in the gamma-ray range is by producing a population inversion at high energy nuclear transitions. The possible candidates for incoherent pumping are the sources of intense gamma-ray radiation in the 100 keV – 10 MeV photon energy range. They are based on the back Compton scattering of intense IR/optical radiation on the electron beams accelerated to relativistic GeV energies. Two types of sources of this kind have been developed, respectively using traditional long RF-cavity accelerators (*111*) and compact laser wakefield acceleration (*112*), (*113*). The latter, and most recent approach already provides similar or even superior results compared to the traditional technique.

Finally, it is worth mentioning an interesting theoretical proposal, a positronium BEC annihilation γ-laser with photon energy of 0.5 MeV (*114*), (*115*), (*116*). However, all major stages of this proposal, including preparation of the dense gas of triplet positronium atoms, cooling it to the Bose-Einstein condensed state, transfer of the positronium atoms from a triplet into a singlet state by chirped terahertz radiation from a powerful gyrotron source, and the subsequent stimulated electron-positron annihilation, need experimental verification and further study.

---

[18] Olga Kocharovskaya

Following the above-cited careful conclusion made over fifty years ago by the pioneers of masers and lasers (*85)*, one may state that lasers can hardly be pushed into the gamma-ray range "unless some radically new approach is found". At the same time, the impressive development of X-ray lasers provides hope that radically new approaches can be found.

## 6. (5.) What are the ultimate limits to temporal, spatial and optical resolution?

### 6.1. What determines the temporal and spatial resolution?[19]

There has recently been great interest in increasing the information capacity of quantum communications systems by encoding more than bit of information per photon. Recent work (*117*) has shown that it is possible to encode large amounts of data onto a single photon by exploiting the spatial degree of freedom. This form of encoding creates new opportunities for both classical (*118*) and quantum (*119*) technologies. One specific implementation that has been demonstrated by a number of groups is the encoding of information in the orbital angular momentum (OAM) states of light, such as the Laguerre-Gaussian modes (*118*), (*119*), (*120*), (*121*), (*122*).

In recent work, some of us exploited OAM to demonstrate a free-space quantum key distribution (QKD) system that carries more than one bit of information per detected photon (*119*). More generally, by using a high-dimensional state to encode multiple bits on each photon, fewer photons may be needed to achieve a specified information transfer, with an overall potential reduction in energy consumption and/or increase in levels of security.

These spatial encoding ideas raise the more general question: exactly how much information can reliably be carried by a single photon? The answer to this question remains open; nonetheless, here we make a few comments based on our current understanding.

One can encode information in the various degrees of freedom of the photon: polarization, wavelength (or carrier frequency), time bins (which is not entirely independent of wavelength), and, pertinent to our study, in the transverse spatial structure of the photon. By the transverse spatial structure, we mean in the transverse momentum (i.e. direction of propagation), or transverse position of the photon.

Whereas polarization gives two degrees of freedom (that is, one can encode one bit of information using the polarization of light), carrier frequency (or equivalently wavelength) provides an extremely large information content, but it is not unlimited. This use of carrier frequency is known as wavelength-division multiplexing (WDM). For example, one can divide the optical spectrum into a large number of frequency channels of width $\delta\omega$. We let $\Delta\omega$ be the full range of frequencies that is utilized. The number of degrees of freedom of such an encoding scheme is thus $N = \Delta\omega / \delta\omega$. This number can be very large, but we need to recall that the channel width $\delta\omega$ cannot be smaller than $1/T$, where $T$ is the time duration of the measurement. We thus find that $N_{max} = \Delta\omega T$, the conventional time-bandwidth product. Note that if we were to use only a single channel, but of width equal to $\Delta\omega$, we would need a time duration no less than $T'=1/\Delta\omega$ to measure the photon. Therefore, if we use the same integration time $T$, we would obtain $N_{max} = T/T'$ which leads to the same result as before: $N_{max} = \Delta\omega T$ (*123)*. We thus conclude that, from a conceptual point of view, there is no advantage to the use of WDM. However, WDM is widely and successfully used in practical communications, even though it does not allow one to overcome a time-bandwidth product (i.e., the Heisenberg limit). The practical advantage of WDM is that one utilizes the entire time-bandwidth product without the need for ultrafast modulators and similar components. Similar benefits accrue to other encoding methods as well.

As we mentioned above, one can also encode in the transverse spatial structure of the photon.

---

[19] Robert Boyd, Miles Padgett, and Alan E. Willner

The use of this degree of freedom to increase information content is known as spatial division multiplexing. This degree of freedom is independent of polarization or wavelength and hence offers a method to increase further the information content of the link. There has recently been considerable interest in exploiting this spatial capability, and there has been considerable debate regarding the best strategy to adopt (*124*), (*125*), (*126*), (*127*). Various proposals have been made regarding encoding in plane waves, in Laguerre-Gaussian modes (as illustrated in Figure 23), in Hermite-Gaussian (HG) modes, more complex patterns (known as photonic lanterns) or in an entirely different strategy in which the available telescope aperture is simply divided into multiple sub-apertures. Straightforward, rough order-of-magnitude calculations indicate that all of these methods generally provide the same information capacity. Specifically, the number *N* of low-loss channels that can be implemented is given by $N \approx (D_1 D_2/\lambda L)^2$ where $D_1$ ($D_2$) is the diameter of the transmitting (receiving) aperture, $\lambda$ is the wavelength of the light, and $L$ is the distance between the sender and receiver. We note that the combination is the Fresnel number of the transmission link. Which of the various encoding schemes is optimum will depend on issues of available technology and operating parameters, and involves detailed consideration of the circumstances under which the communication link is to be employed. One crucial issue is the size and shape of the transmitting and receiving apertures; for example, one may want to avoid the use of square-shaped (e.g., HG) beams when using circular apertures. Other crucial considerations include the availability of required components (switches, modulators, sorters), the nature of the aberrations present in the transmission channel (how these map onto the chosen modal set), and possible requirements of rotational symmetry of reference frames or alignment errors. The optimum choice of encoding method to maximize real-world performance remains an issue of ongoing investigation.

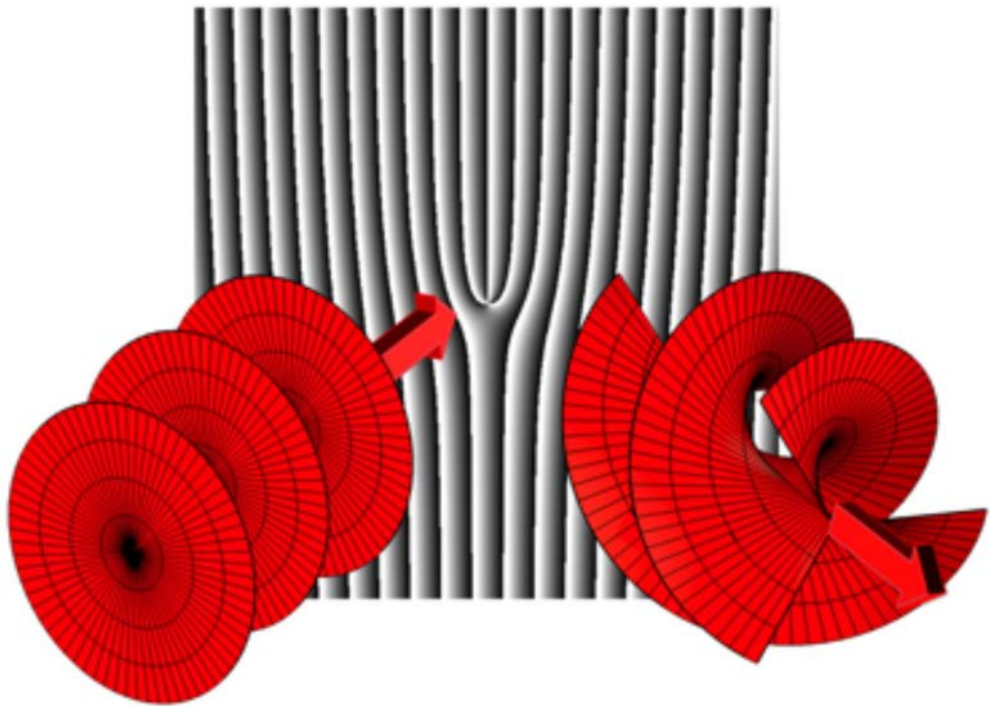

**Figure 23.** A laser pulse containing one or several photons (left) falls onto a computer-generated hologram. The diffracted wave (right) acquires twisted wavefronts and carries orbital angular momentum (OAM). Each possible value of OAM creates an independent communication channel onto which information can be encoded. Credit: Miles Padgett.

## *6.2. What can we say about optical resolution?[20]*

There is no physical limit to optical resolution — at least in principle. Precisely tailored interference of light diffracted from a mask can create an arbitrarily small optical hotspot (focus) which is not limited by the Abbe diffraction limit. This phenomenon is known as superoscillation and was first described for optical systems by Berry and Popescu in 2006 (*128*) when they showed that optical waves could form arbitrarily small spatial energy localizations that propagate far from a source, and without the need for evanescent waves. They argued that existence of superoscillations is underpinned by the fact that a band-limited signal can oscillate arbitrarily fast for arbitrarily long finite intervals, as in the Wigner representations of the local Fourier transform in the 'phase space'. The Wigner function can have both positive and negative values, which causes subtle cancellations in the Fourier integration over all of the function. In fact, this work on optical superoscillations was inspired by earlier analysis by Aharonov et al. who found that the weak measurement of a quantum system can have expectation values much higher than the spectrum of the operator (*129*).

Optical superoscillations were observed experimentally by the Southampton group in 2007 (*130*). In a superoscillatory focus, only a small fraction of energy goes into the hotspot, while the majority of the light's energy is distributed into a broad "halo" surrounding the hotspot. The area between the hotspot and the surrounding halo is often referred to as the "field of view". Computer modeling shows that objects smaller than the field of view can be imaged in a scanning mode with resolution corresponding to the size of the hot spot. For instance a pair of 9 nm x 9 nm holes in an opaque screen, spaced by 28 nm, can be perfectly resolved by a superoscillatory imaging apparatus operating at the wavelength of 400 nm, thus giving resolution exceeding one tenth of the wavelength (*131*). In the last few years, a practical label-free super-resolution imaging technology based on superoscillations combined with confocal detection has been developed and resolution up to one six of the wavelength has already been demonstrated (*132*). This technique is now deployed in biological imaging (*133*).

So, wave optics as such does not impose any fundamental limit on optical resolution: the main challenges in using superoscillations for real-life imaging are technological. Indeed, although instructive algorithms to design a superoscillatory mask which generates arbitrarily small spots have been developed (*134*), such masks require complex grey-scale density and retardation profiles that cannot be fabricated with sufficient finesse by current technology, and less precise binary masks and pixelated spatial light modulators are used instead. Future developments in nanofabrication will hopefully overcome these problems. Another, and perhaps more serious, problem is to separate the desired weak signal coming from the subwavelength hotspot from the intense scattered light of the halo. In reality, one may have to deal with many orders of magnitude difference in intensity between the halo and the superoscillatory hotspot. For weakly scattering samples such as a group of holes in the screen that is smaller than the field of view this is not a big problem; for larger samples, scattering from the halo can become a real limit for producing accurate images of the sample. Therefore, although superoscillatory technology does not appear to have a fundamental limit on resolution, it has a practical limit of the size of the sample that can be adequately imaged. The colossal advantage of superoscillatory imaging is that it is a far-field to far-field technique and it works without the need to label samples with dyes or luminescent quantum dots used in the STORM and STED super-resolution techniques. Also, no prior knowledge of the sample or image deconvolution procedures are required. Moreover, being a linear technique it can be used with optical sources of very low intensity that are not harmful to living matter. Indeed, single photon super-oscillations have been demonstrated (*135*).

---

[20] N.I. Zheludev

## 7. (6.) What novel roles will atoms play in technology?[21]

The past few decades have seen remarkable advances in the ability to control and cool atoms in the gas phase. These advances have been enabled by the method of laser cooling, which produces ultra-cold atomic gases (*136*). These atoms are then further cooled by evaporation, creating Bose-Einstein condensates and the so-called atom laser (*137*). Current applications of atoms in technology include atomic clocks for global positioning systems, atomic interferometry for inertial navigation and remote sensing, atomic magnetometers for medicine, and ion beams for microscopy and nanofabrication (*138*), (*139*), (*140*), (*141*). Other applications, such as neutral atom microscopy and atom lithography have yet to cross over from basic scientific research to real-life applications. In our view, the future impact of atoms on technology hinges on two features: atomic flux and atom optics. The flux of laser-cooled atoms has been limited to around 109 atoms per second, by one factor: the optical density of resonant photons. As the optical density increases above unity, multiple scattering becomes dominant and the method of laser cooling fails as a result of collective instabilities. The flux of laser-cooled atoms, together with the large loss of atoms during evaporative cooling, has limited the atom laser to a flux of around $10^5$ atoms per second. These fluxes are very low, limiting the real-life applications of ultra-cold atoms and the atom laser. New methods are being developed as an alternative to laser cooling and evaporative cooling that could provide a breakthrough in flux: magnetic stopping of supersonic beams, and a new method of cooling that only requires a laser for optical pumping of atoms, combined with state-dependent magnetic kicks (*142*). While optical pumping also uses resonant photons, a magnetic field gradient can be applied to spectrally broaden the atomic resonance, dividing the sample into optically thin slices. The above methods are also applicable to most elements in the periodic table, unlike laser cooling, which is limited to a small subset of elements. There is a second limitation on atom optics, which may seem surprising given the great advances in this area and its application towards atom interferometry. However, one key element has been missing: an aberration-corrected lens for atoms. This is a crucial component in optical imaging and in the electron microscope, providing diffraction-limited resolution. Recent work on a pulsed magnetic hexapole offers a solution to aberration-corrected atom imaging, and may enable the development of a neutral atom microscope with atomic resolution (*143*). Imaging of atoms with this new approach is illustrated in Figure 24. The combination of nanoscale imaging with increased flux may also make it worth revisiting atom lithography.

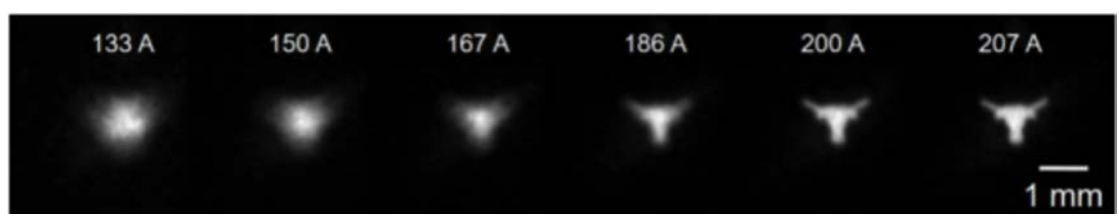

**Figure 24.** Imaging of metastable neon atoms using a pulsed hexapole magnetic lens. The complex pattern formed is easily recognized in the last frame (Image courtesy of Erik Anciaux, Raizen Laboratory).

There is another feature of atoms which has no parallel with electrons or photons: isotopes of the elements. Here, it is the nuclear properties of the atoms that are important, determined by a different number of neutrons in the nucleus. Most elements in the periodic table have multiple stable isotopes, while radioisotopes are created by nuclear transmutation or fission. Isotopes are a great natural resource, with life-saving applications in healthcare, such as imaging of disease, targeted cancer therapy, and diagnosis of malnutrition with stable tracers. Isotopes are also used in industry, for instance, for oil and gas exploration, and for national security in detecting dangerous materials. While there are already great applications in technology, isotopes are still a mostly untapped resource owing to the difficulty and cost of separation. The main method used today, the Calutron, was invented in the 1930s, and is very inefficient and expensive. This method relies on ionization of neutral atoms by electron bombardment, and separation by charge-to-mass ratio (*144*). The only large-scale Calutrons in operation today are in Russia, and even these machines were built over 60 years ago. A new method was recently developed that is much more efficient than the Calutron, and will make isotopes readily available for technology. This method, Magnetically Activated and Guided Isotope Separation (MAGIS) relies on optical pumping of atomic beams and separation by magnetic-moment-to-mass ratio in a novel guide of permanent magnets (*145*). This is more than just a long-term dream: the method will soon be implemented at a non-profit entity, the Pointsman Foundation, which will produce isotopes for medicine (*146*). Within the next five years production lines should be completed, assuring a worldwide supply of key isotopes. One example is Ytterbium-176, which is the stable precursor of the radioisotope Lutetium-177, a most promising agent in targeted cancer therapy. Beyond existing uses of isotopes, new applications are under development. These include imaging and treatment of heart disease with targeted radioisotopes, and inhibition of biofilms with pure beta emitters to reduce infection. We are on the cusp of an exciting era in which atomic isotopes will drastically improve our lives.

---

[21] Mark Raizen

## 8. (7.) What applications lie ahead for nitrogen-vacancy centers in diamond?

### *8.1. Computing and sensing*[22]

Why has the nitrogen-vacancy (NV) center in diamond shown so much promise for diverse applications like sensing, imaging, and quantum computing? Is it the fact that it can be spin-polarized with incoherent green light? Is it the fact that a single NV can be easily seen? Is it the long coherence time? Is it the fact that all of these work at room temperature? Certainly these are all essential for the most interesting applications but the NV is not unique in having these features.

Where the NV has been unique is having a large optical readout contrast that allows the spin state to be quickly determined, even for single NVs at room temperature. Ideally the different ground state spins should produce as big a difference in fluorescence brightness as possible. Single NVs typically have a spin readout contrast in the range of 30-40 %, where a spin contrast of 100 % corresponds to one spin state fluorescing brightly and the other not at all. Ensembles usually are in the range of 5-15 % at zero magnetic field (*147*). Basically the NV is on the border. A higher spin contrast makes it easier to achieve single shot readout, which is so important for many quantum applications. Conversely a lower spin contrast leads to excessively long readout times and severely limits the most interesting applications.

Unfortunately there are still questions about exactly why the NV has such a large spin contrast, and so improving it or finding a superior replacement is problematic. Nonetheless there are existence proofs. Specifically there have been recent reports of new color centers in diamond with spin contrast of 50 % (*148*) and even 100 % (*149*). Although these new centers do not have as long a spin lifetime as the NV, the benefits of single-shot readout would more than compensate in many applications. Once these new centers are properly identified and their spin contrast physics understood, the search for a NV replacement can begin.

Originally the NV was explored in the context of quantum information applications, as a possible replacement for trapped ions/atoms. Here it was postulated that NV must have the proper level structure for polarization selective spin-photon entanglement, as well as a cycling transition for single-shot readout. At cryogenic temperature (below ~10 K) it was discovered that the NV did indeed have the required properties (*150)*. In 2011, single-shot readout of NV centers was demonstrated at 4 K (*151)* opening the door to cryogenic quantum information applications. However, like other solid state optical emitters, the NV absorption line suffers from undesirable inhomogeneous broadening, and even less desirable spectral diffusion (*152)* . The prospects for overcoming these limitations were not good, because after many decades of research on solid state optical emitters, only a handful were found that had small enough inhomogeneous broadening to be interesting (*153)*, (*154)*. But all of these were based on weak optical transitions, and therefore detection of single emitters would be difficult. Fortunately, since spectral diffusion of NVs was found to be very slow, resonant excitation schemes allowed for the effective elimination of inhomogeneous broadening (*155)*, enabling experiments that had been out of reach for trapped atoms/ions, such as deterministic teleportation (*155)* and a loophole-free Bell test (*156)*.

---

[22] Philip Hemmer

Recently another color center in diamond, silicon-vacancy (SiV), has shown a remarkably narrow optical line with almost no spectral diffusion, yet with a strong optical transition (*157)*, (*158)*. Although in hindsight the requirements for such an optical emitter seem obvious (specifically inversion symmetry in a crystal host with minimal spin impurities) it was the discovery of the SiV that identified these criteria and opened the door to the search for alternatives to the NV. For example, the recently discovered germanium-vacancy (GeV) center in diamond (*159)* has shown an even stronger coupling to optical photons than the SiV. This coupling is so strong that high cooperativity in atom-cavity coupling will be possible, and if a version with a longer spin lifetime can be identified, then it might actually replace trapped ions/atoms for many quantum communication/computing applications.

## *8.2. Imaging[23]*

In recent years NV color centers in diamond have attracted intense interest as precision quantum sensors with wide-ranging applications in both the physical and life sciences. Most prominently, NV-diamond has been shown to provide a combination of magnetic field sensitivity and nanoscale spatial resolution that is unmatched by any existing technology — including SQUIDs, atomic magnetometers, and magnetic resonance force microscopy — while operating over a wide range of temperatures from cryogenic to well above room temperature in a robust, solid-state system (*160)*, (*161)* , (*162)*. Importantly, since V centers are atomic-sized defects and can be localized very close to the diamond surface, they can be brought to within a few nanometers of the sample of interest, greatly enhancing the sample's magnetic field at the position of the NV sensor (e.g., dipole fields fall off as $1/r^3$) and enabling nanometer scale resolution. For magnetic field sensing, one optically measures the effect of the Zeeman shift on the NV groundstate spin levels. Similarly, NV-diamond can provide nanoscale electric field sensing via a linear Stark shift in the NV ground-state spin levels induced by interactions with the crystalline lattice (*163)*, as well as nanoscale temperature sensing via a change in the zero-magnetic-field splitting between the NV spin levels (*164)*. In addition, NV-diamond has other enabling properties for both physical and life science applications, including: fluorescence that typically does not bleach or blink; ability to be fabricated into a wide range of forms such as nanocrystals, atomic force microscope (AFM) tips, and bulk chips with NVs a few nanometers from the surface or uniformly distributed at high density; compatibility with most materials (metals, semiconductors, liquids, polymers, et al.); benign chemical properties; and good endocytosis and no cytotoxicity for diamond nanocrystals and other structures used in sensing and imaging of living biological cells and tissues.

---

[23] Ronald Walsworth

Applications of NV-diamond quantum sensors are rapidly advancing and diversifying, with translation of the technology into other fields accelerating, and commercialization beginning. Some recent highlights and their future potential are outlined here. NMR detection of a single protein (*165*) and MRI of a single proton with Ångström resolution (*166*) may lead to structure determination of individual proteins and other materials of interest with atomic-scale resolution (*167*). Noninvasive sensing and imaging of biomagnetism in living cells and whole animals with submicron resolution, e.g., see Figure 25, provides a powerful new platform for studies in cell biology (*168*), genetics (*169*), and brain function and disease (*170*), (*171*), (*172*), as well as lab-on-a-chip bioassays (*173*). In vivo nanodiamonds, which have been used to map temperature and chemical changes in living human cells (*174*), may be applied in the near term to guide thermoablative therapy for tumors and other lesions, and in the longer term to monitor and even repair in vivo damage at the cellular and molecular level. Mapping of heterogeneous magnetic materials within primitive meteorites (*175*) and early Earth rocks (>4 billion years old) (*176*) with submicron resolution is already providing breakthrough advances in the understanding of the formation of the solar system and Earth. Similarly, imaging patterns of nanoscale magnetic fields is being successfully applied to a wide variety of advanced materials such as magnetic insulators undergoing spin injection (*177*), skyrmions (*178*), graphene (*179*), spin-torque oscillators, and multiferroics. This method may fill a critical technical need in the exploration and targeted development of smart materials for challenges in energy, the environment, beyond-Moore's-law information processing, and more. In addition, robust bulk diamond and nanodiamond sensors are being developed for extreme environments, which could provide unique tools for sensing buried ordinance and natural resources, as well as navigation, underground, deep underwater, in conditions of extreme heat, radiation, pressure, etc.

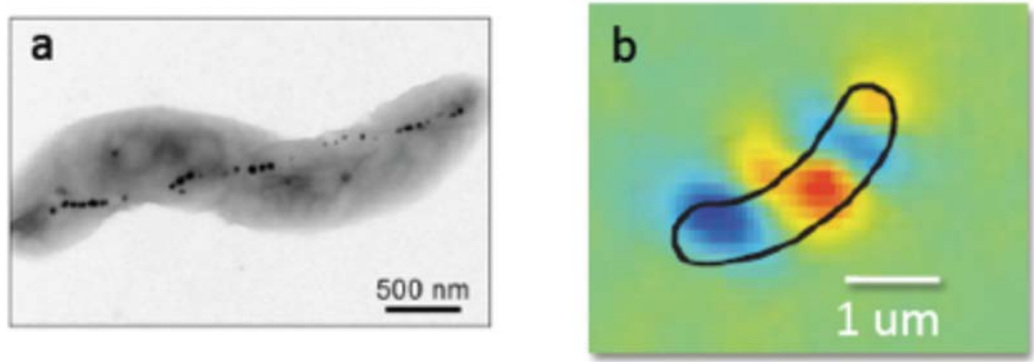

**Figure 25.** Application of NV-diamond magnetic imaging to cell biology. (a) Transmission electron microscope (TEM) image of a magnetotactic bacterium (MTB). Magnetite nanoparticles in the magnetosome chain appear as spots of high electron density. (b) NV-diamond magnetic image of one MTB on the surface of the diamond chip, showing magnetic field patterns produced by the magnetosome with sub-cellular (400 nm) resolution. Cell outline (in black) is from a bright-field optical image of the MTB. Wide-field magnetic images of many MTB in a population provide new biological information, such as the distribution of magnetic moments from individual bacteria in a particular MTB species. (Adapted from (*180*).)

## 8.3. Dark matter searches[24]

Weakly interacting massive particles (WIMPs) are a prominent class of dark matter candidates. These particles naturally arise in a number of theories of physics beyond the standard model and if they exist, it can be shown that production processes in the early universe will generate a cosmic abundance for the WIMP that is comparable to the observed dark matter density. There are a number of experiments presently underway to detect the rare scattering of the WIMP. In these experiments, the WIMP (a particle with mass ~10 - 100 GeV) scatters off a nucleus, depositing a small amount of energy (~10 keV). This energy is detected using a variety of sensitive techniques. The next generation of WIMP detection experiments will be sensitive to the coherent scattering of solar neutrinos (*181*), creating an irreducible background for WIMP detection since the scattering topology of WIMPs and neutrinos is identical.

---

[24] Surjeet Rajendran

This background can be overcome if the direction of the nuclear recoil from a scattering event is identified. The direction of the solar neutrino flux is known while the dark matter flux should be isotropic. By vetoing events that correspond to incident particles emerging from the known location of the Sun, the isotropic dark matter flux can be identified. It is important to be able to measure the recoil direction at solid state densities since large target masses are necessary to detect the small WIMP cross-sections. Crystal defects such as nitrogen vacancy centers in diamond, paramagnetic F-centers in metal halides and defects in silicon carbide could potentially be used to identify the direction of WIMP induced nuclear recoil (*182*). The detection idea is the following: when the WIMP scatters, the induced nuclear recoil creates a tell-tale damage cluster, localized to within 50 nm, that correlates well with the direction of the recoil. This damage cluster induces strain in the crystal, changing the energy levels of crystal defects such as nitrogen vacancy centers. This level change can be measured optically (or through paramagnetic resonance) making it possible detect the strain environment around the defect in a solid sample. It is experimentally possible to create a high density of these defects, and nanoscale resolution of defect properties has also been demonstrated (*183*), (*184*), (*185*), (*185*) (*186*), (*187*), (*174*). To identify the direction of a nuclear recoil, we can first use conventional WIMP detection ideas such as the collection of ionization/scintillation to identify potential dark matter events. Once an event is identified, the defects in the vicinity of the event will be interrogated (for example, optically) to determine the strain environment, thus identifying the direction of the recoil (see Figure 26). If successful, this concept would open a new path to continue to probe the theoretically well motivated WIMP. It would also be a novel application of a quantum sensor such as the spin of a nitrogen vacancy center for a particle physics application.

## 9. (8.) What is the future of quantum coherence, squeezing and entanglement for enhanced superresolution and sensing?[25]

Superresolution imaging and sensing is a frontier of optical science and technology, offering untapped potential for unprecedented structural sensitivity and capacity to understand the fundamental physics of biological systems on molecular and atomic scales. At these scales, quantum-mechanical uncertainty, superposition, coherence, and possibly entanglement, will come into play. Quantum optical methods have been already applied to microscopic systems; however, they can now be extended to nanoscopic imaging and sensing, and pave the way towards quantum biophysics.

The developments in quantum optics can lead to much higher superresolution. This is considered in three parts: (a) Classical structured illumination and quantum correlations; (b) quantum light for illumination with a variety of correlation and homodyne measurements; and (c) quantum coherence to obtain nanoscale localization.

---

[25] Girish Agarwal

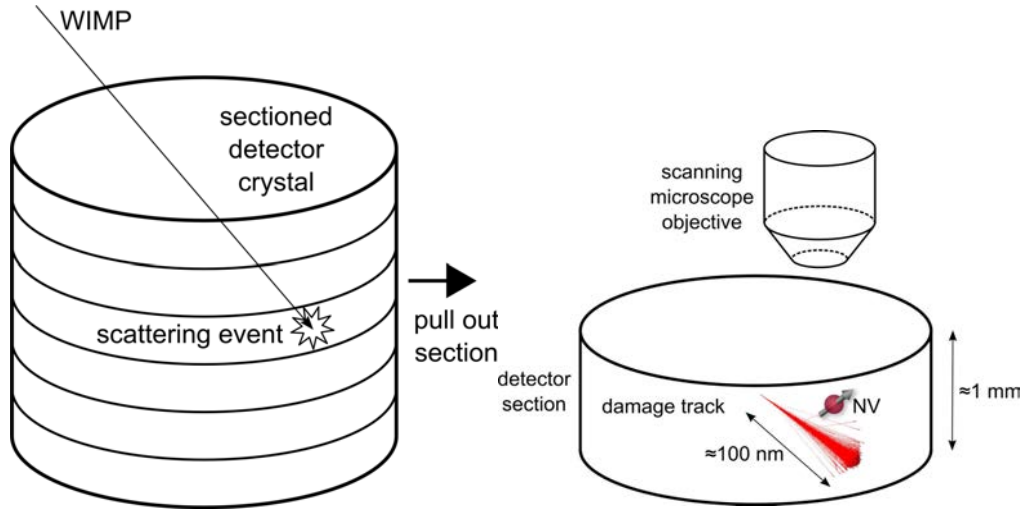

**Figure 26.** (a) Sectioned detector: Event identified by conventional methods. The section of interest is studied separately in (b). (b) Super-resolution techniques applied to sense the localized crystal damage optically. Credit: Surjeet Rajendran.

## 9.1. How can intensity-intensity correlations and structured illumination be employed far beyond the diffraction limit?[26]

High resolution is especially important in biology where, for example, one would like to understand cells at a nanoscale. To achieve this, structured illumination (*188*) has been very widely used for resolution enhancement in bioimaging. Studies of this kind measure the mean intensity of light and depend on the use of the signal, which could depend in a nonlinear way on the illumination intensity. Now, considering a two-dimensional structured illumination $I_{\text{str}}(\boldsymbol{r},t) = I_0[\frac{1}{2} + \frac{1}{2}\cos(\boldsymbol{k}_0\boldsymbol{r} + \varphi)]$ with spatial frequency $\boldsymbol{k}_0$ and adjustable phase $\varphi$, a linear response of the fluorophores, and ordinary intensity measurements, one obtains a doubled resolution by the principle of Moiré fringes. The illumination pattern and the investigated sample produce beat patterns in the object and the image plane such that initially unobservable spatial frequencies in reciprocal space are shifted by the amount $k_0 = |\boldsymbol{k}_0| = (k_x^2 + k_y^2)^{1/2}$ moving them into the observable region, thereby making them accessible. In principle, this technique provides an unlimited resolution, albeit at much higher intensities. This is based on the nonlinear response of the fluorophore emission. However, one would prefer not to use intense beams as one would like to probe, in vivo or in vitro, bio samples without damaging them. The benefits to be obtained by using structured illumination in nonlinear spectroscopy (*189*), like CARS (coherent anti-Stokes Raman spectroscopy), are under study.

There are developments in quantum optics whereby quantum emitters make it possible to obtain higher resolution. The key idea is that quantum emitters exhibit anti-bunching. Thus, by using the measurements of the intensity $G^{(1)}(r)$ and the intensity-intensity autocorrelation $G^{(2)}(r,r)$ in the image plane, one can achieve superresolution, in the sense that one effectively replaces the point spread function (PSF) by its square in the signal $A^{(2)}(r) = (G^{(1)}(r))^2 - G^{(2)}(r,r)$. This reduces the FWHM of the effective PSF by a factor of $\sqrt{2}$ improving the resolution by the same factor. Making use of higher-order correlations leads to further improvement in the resolution, scaling as $\sqrt{N}$. For *N*=2, this technique was implemented in wide-field and confocal microscopy setups imaging distributions of quantum dots (*190*),

---

[26] Girish Agarwal

(*191*). Clearly one will do even better by combining all of these different techniques, i.e. structured illumination, nonlinearities and the measurements involving, not just intensity, but all order correlations for quantum systems like membrane proteins and others. These newer techniques fall under the category of "Superresolution via Structured Illumination Quantum Correlation Microscopy (SIQCM)". We have preliminary results on this new technique (*192*), which is expected even to work for the three-dimensional imaging.

A key result is that for linear low-intensity excitation and linear optical detection, the simultaneous use of structured illumination and correlations leads to a theoretically unlimited resolution power, with the improvement scaling favourably as $m + \sqrt{m}$, depending on the correlation order, $m$. Fig. 27 illustrates how correlations of different order enable one to probe increasingly large Fourier components, leading to superresolution.

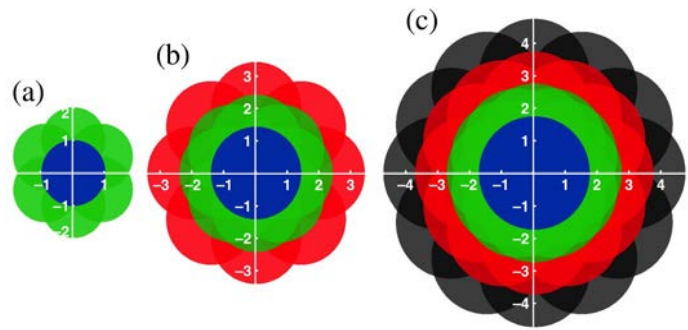

**Figure 27.** (a)-(c) Comparison of the observable regions in reciprocal space provided by ordinary structured illumination microscopy [SIM], structured illumination quantum correlation microscopy [SIQCM] of order two and three (left to right). The axes are normalized by the modulus of the highest spatial frequency transmitted through the imaging system. The green disks stem from the densities, while the red and black disks originate from the higher harmonics arising in the SIQCM-different order signals. Note that the size of the individual disks is enlarged from (a) to (c) because of correlations of different order (after (*192*))

## *9.2. Quantum metrology: excitation with quantum light[27]*

We considered illumination with a classical light in the previous section; however, studies over the last decade have shown how the resolution capabilities of an interferometer can be improved through excitation with a quantum light (*193*), (*194*), (*195*), (*196*), (*197*). These studies have also demonstrated superresolution as well as super-sensitivity. The super-sensitivity results from the use of entangled photon pairs in a novel way.

The entangled photons are produced by a down-converter, which can even be seeded to take advantage of both coherent and entangled light (*194*). Several experiments have demonstrated the super-resolution and super-sensitivity with entangled photons (*196*), (*197*), (*198*). In an interesting bio-application, entangled photons were used to measure the concentration of blood protein (*199*). One can also use quantum correlations of higher order to produce results similar to those obtained with NOON states without actually using NOON states (*200*). Newer possibilities to get the phase information would be to combine the intensity correlation spectroscopy with homodyne measurements or utilize orbital angular momentum of the electromagnetic fields (*201*). Further one can use strongly squeezed light with 8-10 dB squeezing. This has been used for example in the ultrasensitive measurement of the displacement of a cantilever, and for sensitive tracking of a biological particle (*202*), (*203*). This can be done in four wave mixing processes which are especially efficient in atomic vapors. Another intriguing possibility is to arrange down-converters in an interferometric configuration (*204*), (*205*) to make an SU[1,1] interferometer which can give super-resolution and super-sensitivity. Guided by the success of structured illumination in classical microscopy, one would think that making a quantum light, with a mode function with spatial structure, can surpass what has been achieved so far. Clearly quantum light is going to contribute to a whole lot of ultra precision measurements leading to both fundamental science and applications in diverse fields.

---

[27] Girish Agarwal

### 9.3. Quantum coherence for deep sub-wavelength localization and tracking[28]

The discovery of coherent population trapping (*206*) is highly significant as it opens up the possibility of producing coherence between levels that belong to the ground manifold, and thus have very little decoherence, leading to long-lived quantum coherence. Quantum coherence is used extensively in many spectroscopic applications. It has provided unprecedented sensitivity in nonlinear spectroscopy methods like CARS (*207*), (*208*). This coherence is a nonlinear function of the intensities of the fields applied to create it. Thus, a structured light pump would produce localization capabilities (*209*), (*210*), (*211*) far beyond the diffraction limit — specifically this produces localization at a scale of the order of $\Delta x = \lambda/\pi\sqrt{1+I_p/I_s}$, with $I_s$ and $I_p$ being the intensities of the standing wave pump field (see Fig. 28) and the weak probe field, respectively (*211*). The physics of this technique is quite distinct from that of the Nobel prize winning work of Hell on STED (*212*), so it is rather surprising that the localization length has an intensity dependence which is identical to that which occurs in STED. Yavuz's group at Wisconsin has already performed extensive studies of this coherence to produce localization at the scale $\lambda/13$ for an ultracold sample of atoms (*213*). The resolution achieved is quite significant despite the extremely complex nature of the ultracold system. Additional improvements can be obtained by using different types of structured illumination or by using sideband modulation of the exciting fields (*214*). Instead of atomic levels, one can use vibrational states and electronic states of a protein molecule (see Figure 28).

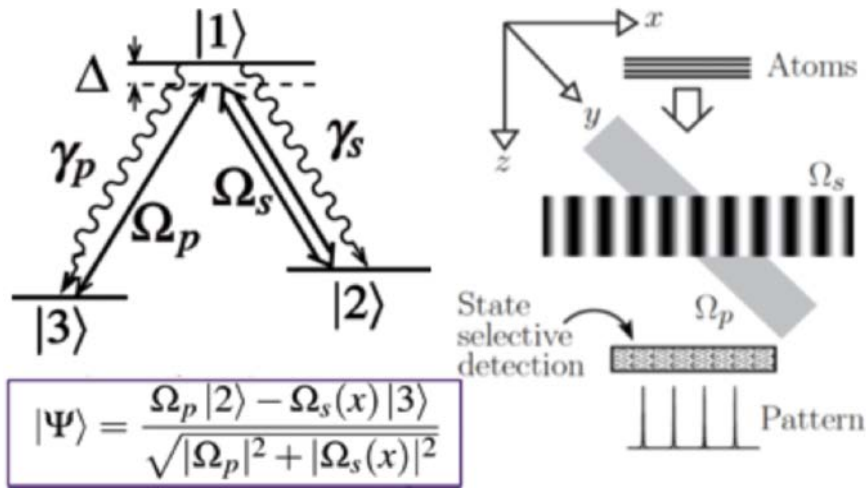

**Figure 28.** (Left) Energy diagram for coherent population trapping using vibrational, $|2\rangle$ and $|3\rangle$, and electronic, $|1\rangle$, states of a molecule. (Right) Modulated standing wave excitation for enhanced spatial resolution imaging [redrawn after (*211*) and reproduced with permission].

---

[28] Girish Agarwal

## 10. (9.) How can we solve (some of) humanity's biggest problems through new quantum technologies?

### 10.1. Light, Energy, Civilization: What novel approaches can provide feasible clean generation of electricity?[29]

Civilization is presently in the hunter/gatherer mode of energy production. Nonetheless, the continual drop in cost of solar panels will lead to an agrarian model in which energy that is harvested from the sun optically will satisfy all of society's needs.

Solar panels are optical. By recognizing the optical physics in solar cells, scientists are, for the first time, approaching the theoretical limit of ~33.5% efficiency from a single bandgap.

At the same time, solar panels have dropped in price by a factor of about three per decade, for the last four decades, cumulatively amounting to an approximately one hundred-fold reduction in real price. Since solar panels are manufactured in factories under controlled conditions, where continuous improvement is possible, these panels will continue to drop in price until solar electricity becomes the cheapest form of primary energy (likely to occur around 2030). At that point, solar electricity will become cheap enough to be converted into fuels, which can be stored summer-to-winter. The creation of fuel requires panels that are three to four times cheaper than today's already depressed solar panel cost, while maintaining the highest efficiency.

The highly successful petroleum industry is over 150 years old. It has taken advantage of technology, but it appears resistant to disruptive technical changes that could sweep it away, as so many industries have been irrevocably changed or entirely eliminated by the advance of technology. Nonetheless, the application of solar electricity to create fuel could sweep away the petroleum exploration industry, which I call the "hunter/gatherer" mode.

Future solar cells will all have direct bandgaps, allowing them to be very thin. The cost of the material elements composing the cell will be small, since a film as thin as 100 nm can fully absorb sunlight using light trapping. Even if the chemical elements were to be expensive, there would be so little material used in such thin photovoltaic films that the cost would be low. Indeed there are methods to produce free-standing single crystal thin films, economically.

The key to high performance from a solar cell is external luminescence efficiency, an insight which has produced record open-circuit voltage and power efficiency. This has everything to do with light extraction, in agreement with the mantra, "a great solar cell needs to also be a great light emitting diode"; the scientific principle behind this counter-intuitive requirement for luminescent emission is illustrated in Figure 29.

With high efficiency, and low cost, in hand, by following the scientific principle in Figure 29, solar electricity will be brought from the open field to nearby locations where it will be used for electrolysis of $CO_2$ solutions, and the recycling of atmospheric carbon. There have been great strides in electrolysis which can produce various proportions of $H_2$, $CH_4$, and higher hydrocarbons as products. The carbon-carbon bond is particularly prized, since such compounds can be readily converted into diesel fuel and jet fuel. The study of such selective electro-catalytic surfaces is still in its infancy. Even if $H_2$ were the only electrolysis product, there are industrial methods of using $H_2$ to reduce $CO_2$, and make useful liquid fuels, among many other materials.

The ability to create fuels would increase the size of the photovoltaic panel industry at least 10-fold, allowing the adoption of new cell technology that is better than the current outdated 1950s crystalline silicon solar cell technology.

---

[29] Eli Yablonovich

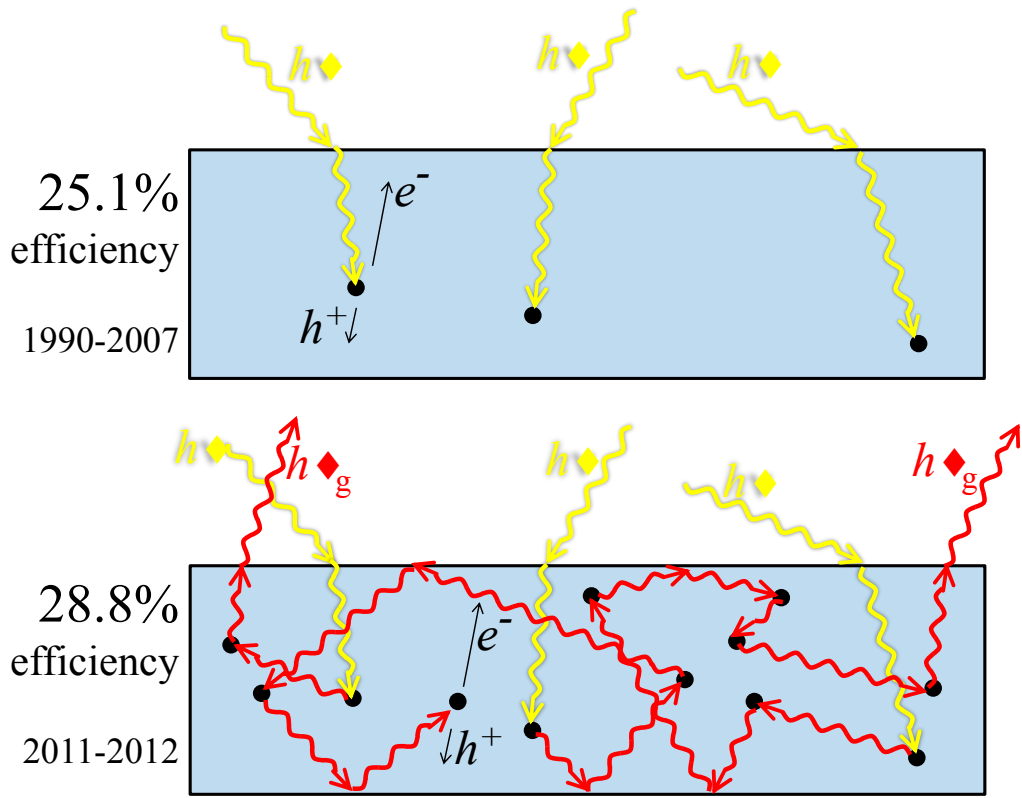

**Figure 29.** The physical picture of high efficiency solar cells, compared to conventional cells. In high efficiency solar cells, good luminescent extraction is a requirement for the highest open circuit voltages. One-sun illumination is accompanied by up to 40 suns of trapped infrared luminescence, leading to the maximum external fluorescence efficiency. Credit: Eli Yablonovitch.

Thus we see that the application of optical science in making solar cells more efficient, and lower in cost, will produce a revolution in mankind's energy source, playing a role analogous to the agricultural revolution of 10,000 years ago.

### 10.2. What are the applications of nonlinear science and what potential do they have?[30]

Linear science is based on the premise that the input-output relation is proportional. This way of thinking is intuitive, rational and powerful. However, it works mostly only locally and, much of the time, only within a small range and/or scale. The theory of linear systems has gone through several centuries of development already and has now reached a saturation point. A great majority of systems in natural and social sciences are actually nonlinear. A modern scientist, engineer and mathematician must be able to "*think nonlinearly*" in problem solving.

In interplanetary exploration, for example, when attempting to reach a far distant planet, "linear thinking" would engender the need to build ever larger and more powerful rocket boosters in order for the spacecraft to be propelled to an increasingly remote place. However, "nonlinear thinking" provides a different approach, enabling us to use interplanetary motion configurations and gravity assists to fling the spacecraft to the farther destination(s) without a much larger rocket booster containing proportionally more fuel. This accounts for how NASA has succeeded in the Cassini-Huygens (Figure 31) (*216)* and other space missions.

---

[30] Goong Chen

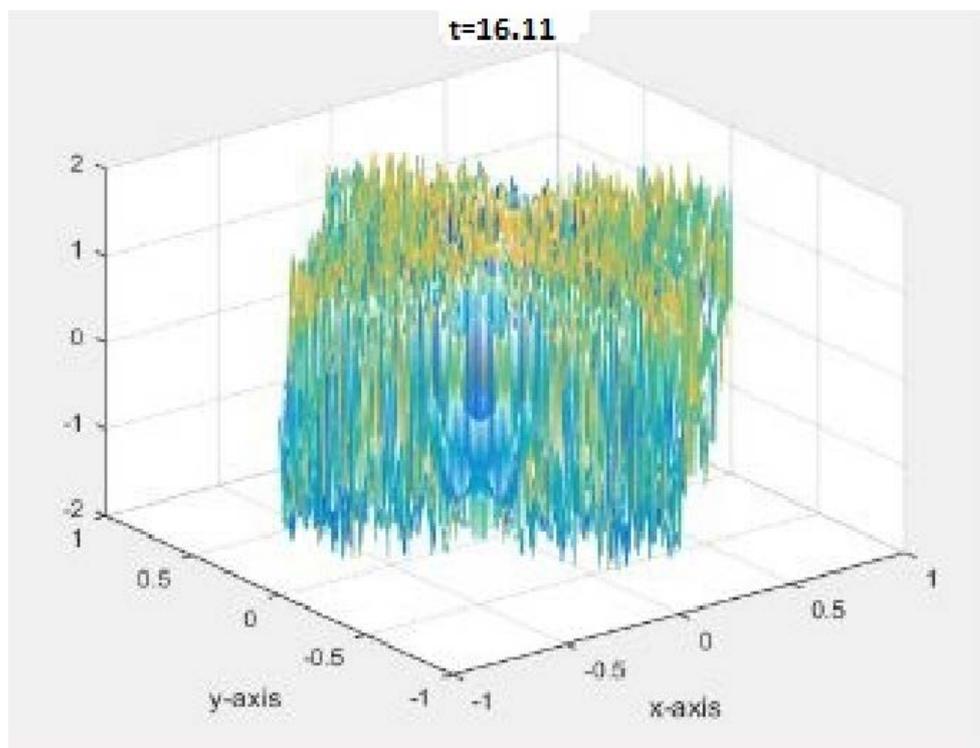

**Figure 30.** A snapshot of chaotic vibrations on a 2D domain excerpted from (*215*). The main objective of the work is to show the complexity of spatiotemporal chaos and to perform mathematical analyses of chaos in multidimensional settings; at present, it has no specific applications.

There will be more use and application of the fundamental nonlinear properties of *bifurcations, periodicity, pattern formations, ergodicity, chaos, and turbulence* (*217*) to advance physics, engineering and technology. For example, recently, turbulent mixing of fuel drop spray in the combustion engine has been found to greatly enhance combustion efficiency. Along similar lines of thought, there must also be more ways to understand, utilize and even control turbulent vortex shedding over airfoils to further dramatically increase aircraft stability, maneuverability, lift over drag ratio, energy efficiency, and overall performance. Such control could potentially be brought about through nonlinear feedback of smart sensors/actuators, shape designs, and deformable materials, leading to new generations of flying machines that can fly as gracefully, powerfully, and energy effectively as an eagle.

### 10.2.1. *What is the most urgent undertaking in science and technology?*[31]

*Climate change* and *global warming* are the greatest threat to the quality of life and long-term survival of human civilization. If Earth's average temperature continues to rise as trends have indicated during the past decade, we will soon see massive die-offs of species. There is a pressing need for a nonlinear shortcut of chemical processes to effectively capture carbon in the atmosphere. For example, can we mimic, speed-up or even improve photosynthesis in a new generation of nonlinear chemical reactors through novel designs of diffusion, convection, mixing, and reaction? Another research front would be quantum control of laser-induced chemical bond breaking (*218*), but this technology appears to be decades away from being realizable.

---

[31] Goong Chen

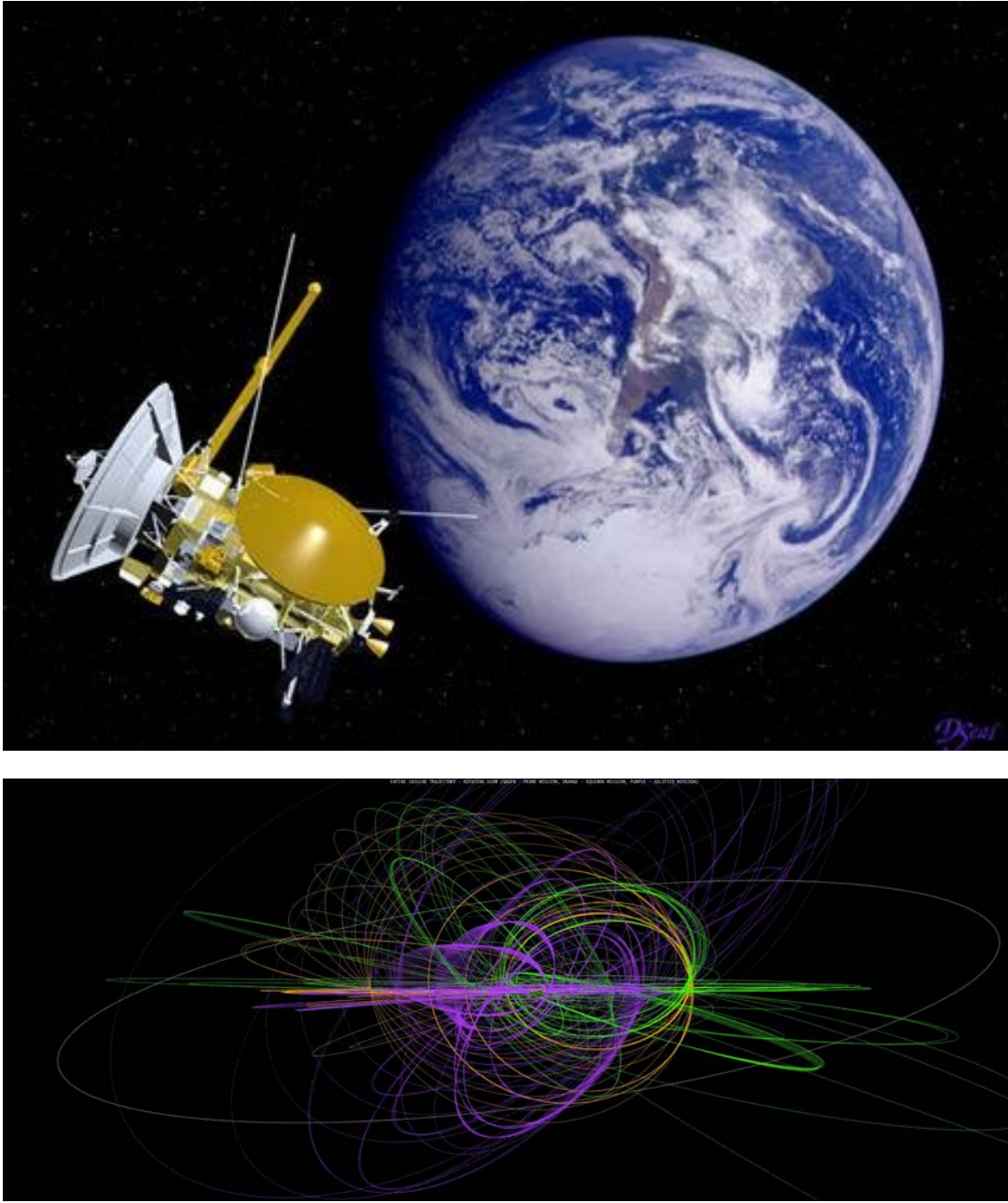

**Figure 31.** Nonlinear trajectories in the Cassini-Huygen's space mission: (a) The earth-moon fly-by of the Cassini spacecraft gave a 5.5 km/second speed boost, so although the craft was once again only 1,100 km from Earth after almost 2 years of travel, its speed was considerably greater . (b) Mission planners referred to the trajectories of Cassini as "the ball of yarn". All orbits between July 1, 2004 and September 15, 2017 are shown in this image. Cassini's mission terminated in September, 2017. Credit: NASA.

### *10.2.2. How else can nonlinear science help us?*[32]

*Nonlinear smart feedback control* could be the field within nonlinear science with the greatest development potential. Here we briefly explain the important concepts:

*Self-regulation:* This is exemplified by a van der Pol oscillator (*219)*, where the electric current or voltage will rise if it is too low, and fall if it is too high. Such a self-regulating servomechanism is usually modeled by *van der Pol-type nonlinearities*. This enables a given system to operate within a prescribed advantageous range.

---

[32] Goong Chen

*Self-organization:* If an initially disordered system can spontaneously convert itself to some overall form of order, then it has self-organizing capability. For example, our human digestive system can process a wanton ensemble of foodstuff in the stomach into various nutrients suitable for absorption. This self-organization is based on multiple-interactions including enzymes-catalysts-neurons, balances of exploitation and exploration, and strong nonlinear interactions of positive or negative feedbacks.

*Self-assembly:* This may be viewed as the outcome of self-organization, it results in the formation some kind of ordered structures, patterns, or objects.

*Self-replication:* This is a capability or phenomenon where an organism or structure, usually small, such as a virus, cell, or nanostructure, replicates, producing an almost identical copy of itself.

*Self-manufacturing:* This concept, that of a self-replicating machine, was proposed, advanced and examined by Homer Jacobsen, Edward F. Moore, Freeman Dyson, John von Neumann, K. Eric Drexler, et al. The machine is an autonomous robot that is capable of reproducing itself by using raw materials found in the environment, presumably on an extensive scale. This concept is increasingly close to becoming a reality after the recent advances in 3D lithography. One may think about "the mining of moons and asteroid belts for ore and other materials, the creation of lunar factories, and even the construction of solar power satellites in space" (*220*).

## 11. (10.) What new understanding of materials and biological molecules will result from their dynamical characterization with free electron lasers?[33]

The studies of materials and molecules with fast intense laser pulses is a rapidly developing frontier of science, with the latest capabilities involving both large-scale and compact facilities. Lasing can now be achieved with microwave, terahertz, infrared, visible, ultraviolet, and X-ray radiation. All of these ranges of photon energies have important potential applications, even though X-rays are emphasized below because wavelengths comparable to an atomic spacing are needed to obtain understanding of structures and mechanisms at the atomic level. Three prominent X-ray free electron laser facilities and experimental results are represented by Figures 32, 33, 34, 35, and 36. It is valid to use the term "laser" for newer systems that employ different principles from the original lasers – with a broader interpretation of the terms in this acronym – since the result is still (nearly) coherent, monochromatic, and unidirectional electromagnetic radiation.

In a free-electron laser, relativistic electrons are made to wiggle back and forth as they pass through an undulator consisting of magnets with alternating polarity. These accelerated charges emit radiation, which becomes increasingly intense and acts back on the electrons. A detailed treatment shows that the electrons are forced into microbunches with a spacing equal to one wavelength of the radiation. When the electrons are stopped after an appropriate distance, the radiation that emerges is then nearly monochromatic and coherent. If the electrons are extremely relativistic, the radiation can also be nearly unidirectional (~100 μm across) with a very short wavelength (down to ~ 1 Å at the most energetic current facilities).

The duration of an X-ray pulse can be femtosecond-scale, and when the duration is ~20 femtoseconds it is possible to have diffraction before destruction. This means that the structure of a single molecule or nanoparticle has now become experimentally accessible. In previous studies of the structure of biological molecules, it was necessary that they form crystals, since even the brightest x-ray sources required many molecules in a crystalline array to yield enough diffracted intensity for a structural determination. But X-ray free electron lasers, with coherent beams, have a billion times the intensity of previous X-ray sources. They can then, in principle, obtain enough diffracted intensity to obtain the structure of a single nanoparticle or molecule.

---

[33] Sebastian Lidström and Roland E. Allen

The dream, already partially realised, is that "movies" can be made by observing the behavior of thousands of individual molecules that are successively observed (with diffraction before destruction). The difficulty is in assembling these "snapshots" to form a "movie" – i.e., a sequence which actually mimics the natural dynamics of the molecule. But simpler time-dependent processes can already be studied (*221*).

Enormous sophistication has been employed in obtaining the structures of a vast number of proteins (see Figure 37) and other large molecules, by finessing the crystallography phase problem with various tricks. But still more sophistication is required to obtain the structure and dynamics of single molecules that are subjected to femtosecond-scale pulses with unprecedented intensity, when the orientation and other initial conditions are different for each molecule before it is blasted apart in less than 100 femtoseconds. Figure 38 illustrates results obtained with the LCLS for individual mimivirus particles (*223*) using a new algorithm (*224*). As the authors say about this remarkable new capability, "Since each particle is randomly orientated when exposed to the x-ray pulse, the relative orientations of the particles have to be retrieved from the diffraction data alone."

The principal inventor of the free-electron laser was John Madey in the early 1970s (*225*), (*226*), following the construction of the first undulator by Hans Motz and coworkers in the early 1950s (*227*), (*228*).

A fascinating history of the X-ray free-electron laser (XFEL) has been provided by Pellegrini (*229*). This article also summarizes the up-to-date status of this area in early 2012. The current spectacular capabilities of the XFELs at Stanford, DESY, RIKEN, and elsewhere were made possible by the brilliant experimental innovations of many people, as summarized in this article. Among the theoretical contributions, beyond those of Motz and Madey, were the recognition that the XFEL can be described classically to a good approximation (*230*), (*231*) and that "very high X-ray dose rates and ultrashort exposures may provide useful structural information before radiation damage destroys the sample" (*232*).

As another example of the potential for FELs in a biological context, we quote from a recent paper which discusses XFEL based structural studies of e.g. bacteriorhodopsin and a bacterial photosynthetic reaction center (*233*):

> "The breath taking development of XFEL sources that generate femtosecond X-ray pulses with peak brilliance one billion times that available from synchrotron radiation has generated tremendous excitement across many scientific disciplines. Membrane protein structural biologists have exploited these new opportunities to study protein structure and dynamics in completely novel ways. We can now routinely collect 'damage free' structural data from micron or submicron scale crystals, perform time-resolved studies of protein dynamics on an ultrafast time scale, and a promising technique of XFEL based 2D membrane protein diffraction is emerging."

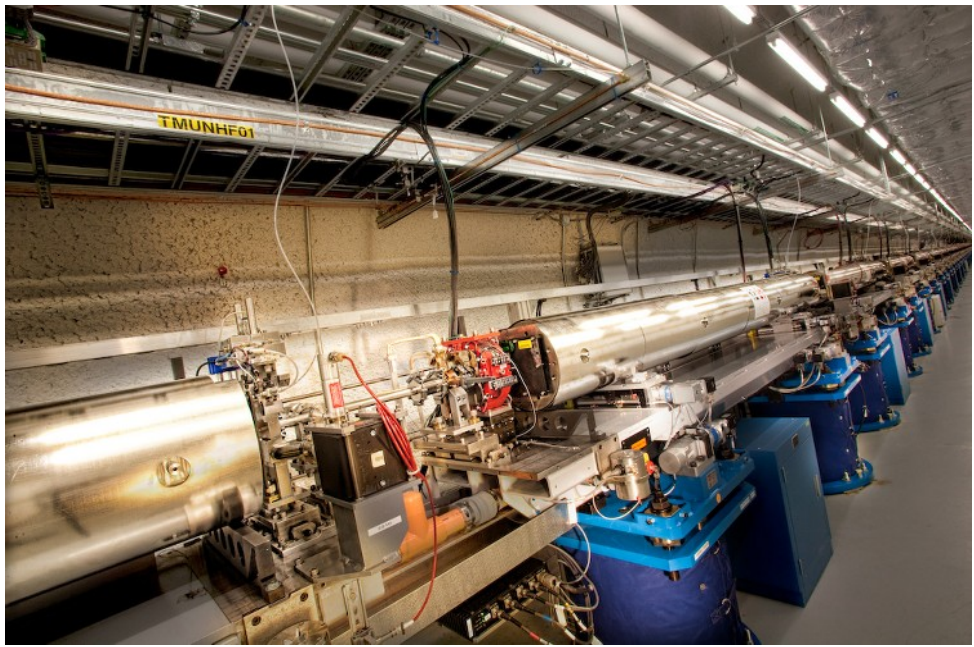

**Figure 32.** The Linac Coherent Light Source (LCLS) at Stanford University for high-energy X-ray laser light. The LCLS has provided the brightest, shortest pulses of laser X-rays for studies of materials and biological molecules. With these ultrafast pulses it is possible to make molecular movies that capture dynamics at the atomic scale. Credit: SLAC National Accelerator Laboratory.

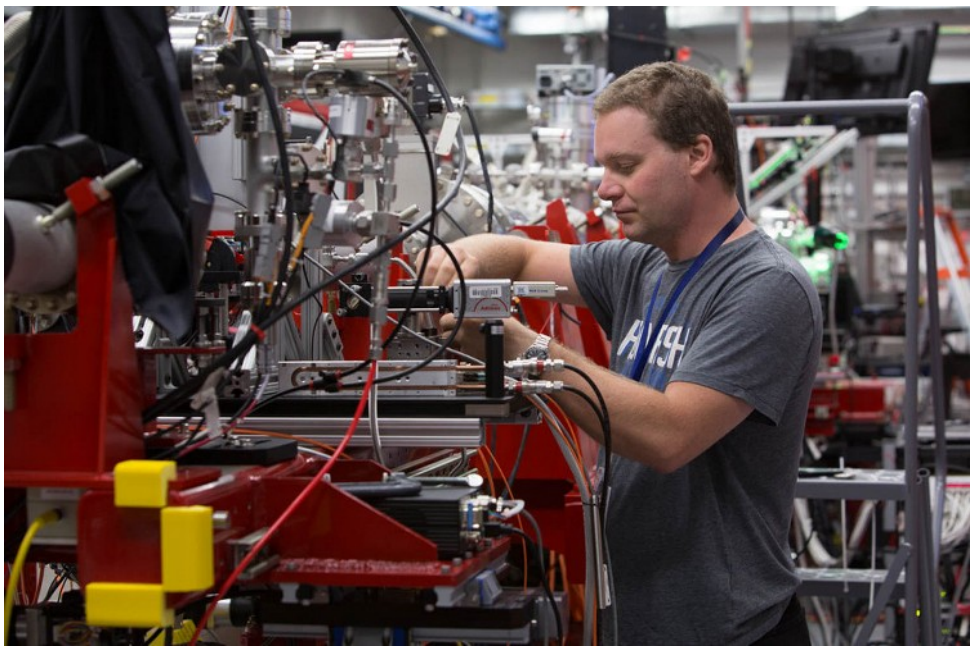

**Figure 33.** Protein Crystal Screening Program at the LCLS X-ray free-electron laser at Stanford. Credit: Matt Beardsley.

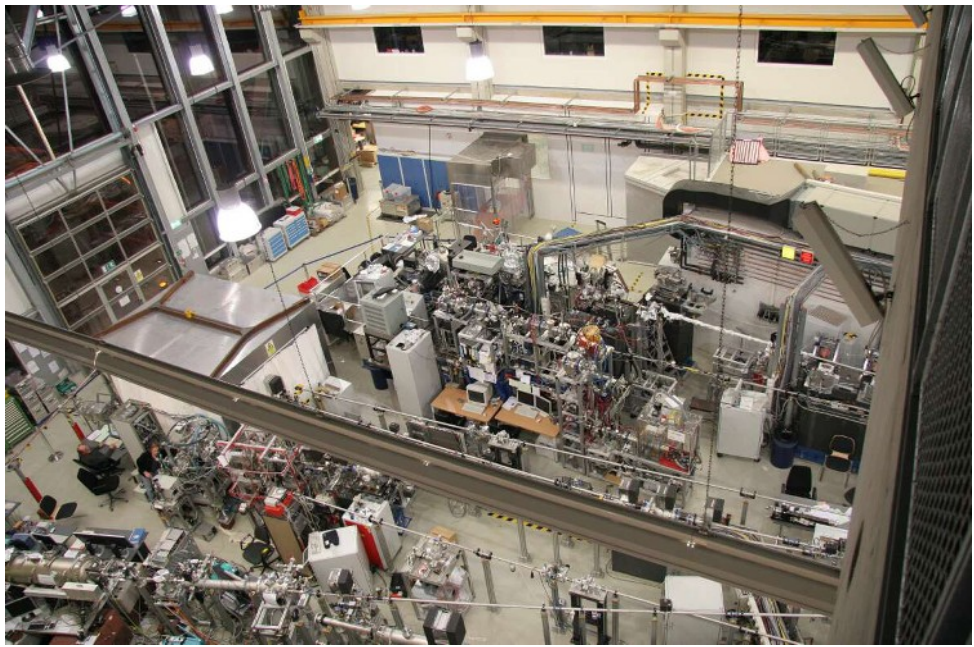

**Figure 34.** Experimental hall at the free-electron laser FLASH in Hamburg, Germany. Credit: DESY.

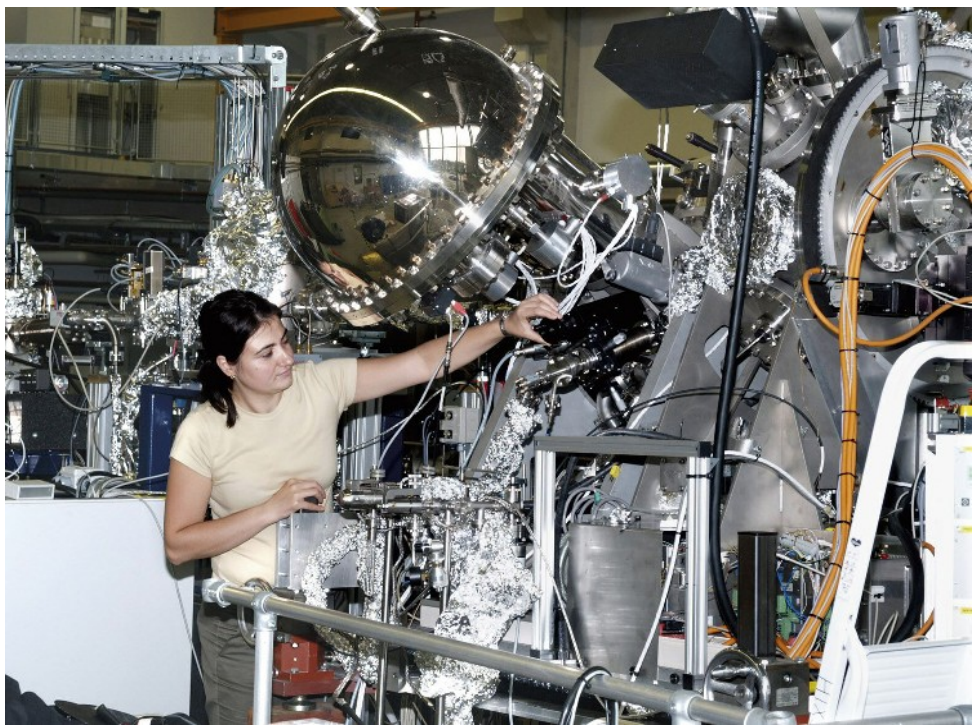

**Figure 35.** An experiment at the FLASH X-ray free-electron laser in Hamburg. Credit: DESY.

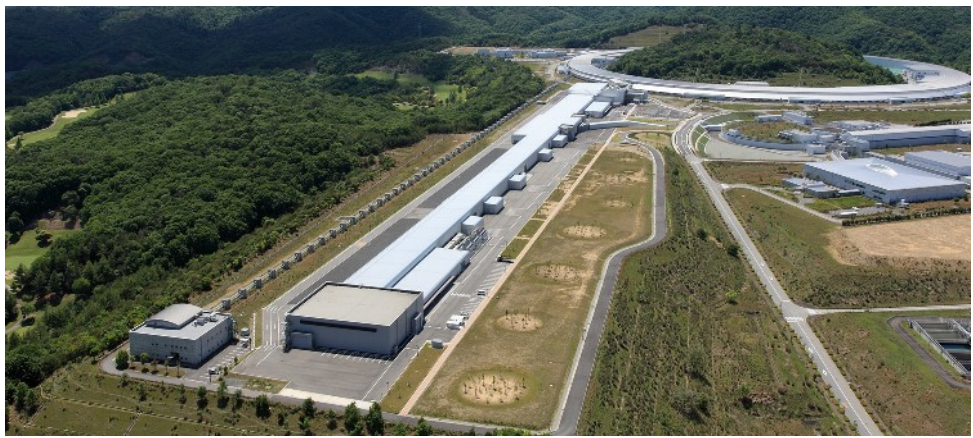

**Figure 36.** The X-ray free-electron laser (XFEL) facility in Japan called SACLA is operated by RIKEN and located in Harima Science Garden City. Credit: RIKEN.

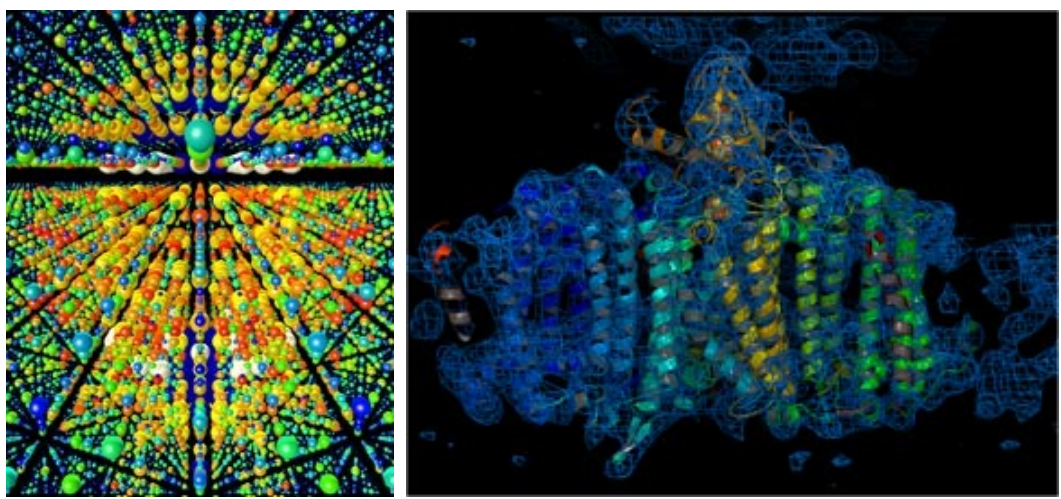

**Figure 37.** (a) Three-dimensional rendering of the X-ray diffraction pattern for the Photosystem I protein, reconstructed from more than 15,000 single nanocrystal snapshots taken at the LCLS. Credit: Thomas White, DESY. (b) A reconstructed image of the Photosystem I complex. Credit: Raimund Fromme, Arizona State University. Based on (*222)*.

## 12. (11.) What new technologies and fundamental discoveries might quantum optics achieve by the end of this century?

### *12.1. Will we have a quantum internet?[34]*

Quantum information science has transformed the way we think about information. Whereas classical bits can only be 0 or 1, quantum bits can also take up any superposition state in between (colloquially speaking "being 0 and 1 at the same time"). Furthermore, quantum entanglement between several qubits makes the individual states fully undetermined, but enforces strong correlations between measurement outcomes on them, even when the qubits are vastly separated from each other. The presence of entanglement can be detected using Bell's inequality – originally conceived to test the notion of local realism *(234)*. Decades of experiments testing Bell's inequality *(235)*, *(236)*, *(237)*, *(238)* – culminating in the first Bell tests free of any experimental loopholes in 2015 (*156*), (*239*), (*240*), (*241*) – have pushed control over remote entangled qubits to a level where large-scale applications may come into view. One highly ambitious application would be a quantum internet.

---

[34] Ronald Hanson

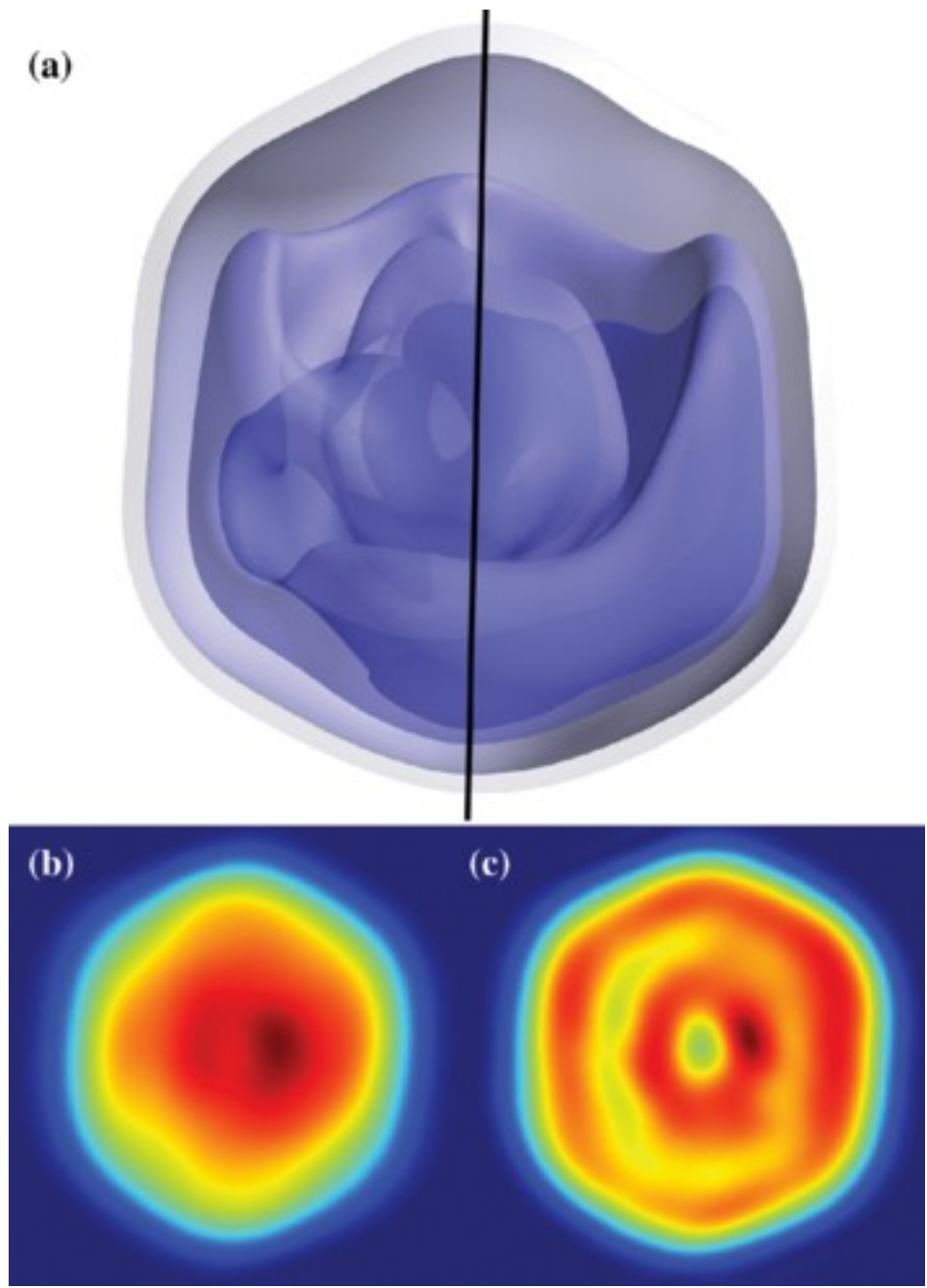

**Figure 38.** Reconstructed electron density. (a) The electron density of the mimivirus is recovered to a full-period resolution of 125 nm. The figure shows a series of isosurfaces, where blue represents denser regions and white represents lower density. The reconstruction shows a nonuniform internal structure, and the line indicates the pseudo-fivefold axis. (b) A projection image of the recovered electron density. (c) A slice through the center of the recovered electron density.

A quantum internet is a network that enables the generation of entanglement between qubits at any two places on Earth. It would function in parallel and in close conjunction with the Internet that is in place today – not replacing any of the current Internet's functions but rather providing fundamentally new opportunities. If realized, a quantum internet would allow for private communication secured by the laws of physics (*242*), for synchronization protocols that could lead to enhanced timekeeping (*243*) or better telescopes (*244*), for communication protocols that would be substantially more efficient than classical counterparts and for a means to link future quantum computers (*245*). All of these promises hinge on entanglement being generated, stored and processed in small qubit processors around the globe.

Although this is at the moment still a vision, the current pace of progress is encouraging. Researchers have been able to establish long-lived entanglement between qubit systems over macroscopic distances (*246*) up to a kilometer (*156*). The same systems show promise for the processing of quantum states for correcting errors (*247*), distilling stronger entanglement from several pairs of weaker entangled qubits (*248*) and for running local qubit programs to enhance quantum internet functionality (*249*)). At the same time, photonic entangled states have been distributed over long distances (recently over more than 1000 km using a satellite system (*250*), showing a pathway for establishing entanglement on a truly global scale. While all these experiments are at the proof-of-concept level, it is clear that exciting times are ahead that might – ultimately – lead to a global quantum internet.

### *12.2. Maxwell's demon for light – an approach to calculating the entropy of light[35]*

The concept of entropy has been a useful tool in multiple fields of science and engineering. Attempts to apply this concept to studies of light date back to the early 20th century when Plank, and then Einstein, through their analysis of disordered, thermal light were led to the notion of a photon – the quantum of the electromagnetic field. In more recent times, quantum heat engines (*251*) and the entropy of laser light (*252*) have been under study. In particular, laser entropy is deemed to be related to the finite amount of disorder introduced – by virtually unavoidable, incoherent, spontaneous emission – into the coherent stream of stimulated-emission photons. The interplay between spontaneous and stimulated emission leads to an imperfect predictability of the phase of an electromagnetic wave produced by a laser, and results in the Schawlow-Townes limit for laser linewidth (*85*).

Instead of trying to quantify the finite amount of disorder intrinsic to the quantum nature of laser light, we envision an alternative approach: Start with fully-disordered thermal light, and calculate (or measure) the amount of work it takes to "de-modulate" the fluctuations out of the field pattern, producing a laser-like coherent wave. The demodulation process is akin to the workings of a Maxwell's demon with a capable being or device now exerting its action on a stream of light, instead of a gas of molecules (*253*).

Let us analyze what it would take to construct such a de-modulation device, restricting ourselves to consideration of the temporal and spectral characteristics of light only, not spatial ones. We can get away from spatial randomness or incoherence by first filtering our source of thermal light through a spatial filter, i.e. a pinhole, much like the one Thomas Young used in the early 19th Century to filter sunlight for his famous double-slit experiment. We start by noting that spatially filtered thermal light, such as sunlight, represents a random amplitude and phase modulation of an electromagnetic wave, characterized by a large total bandwidth. For the remainder of this section, we will attempt to design a proof-of-principle experiment aiming to take incoherent thermal light (i.e., sunlight) as an input, and reduce it to a perfectly coherent wave, through filtering and de-modulation.

---
[35] Alexei Sokolov

This appears to be doable, even if cumbersome on a practical level. Active amplitude and phase modulators will of course have a bandwidth limit – a conservative present day number is ~100 GHz, although one can envision near-future devices attaining ~10 THz bandwidth. The visible band of the electromagnetic spectrum is approximately 300 THz wide. The first step in our task will be to separate this band into the proper number of narrow channels – be it 30 or 3000 (x2 for the two orthogonal states of polarization) – such that each individual channel can be comfortably handled by high-speed modulators. In the most general picture, we will have a spectral disperser (based on a diffraction grating, or on multilayer dielectric thin-film technology) that will separate our multiple channels, and a spectral combiner which will be the disperser run in reverse. In between the disperser and the combiner, we will have individual access to each of the wave channels, and we will implement a device that will actively compensate for light fluctuations.

We note a connection between the "Maxwell's demon for light" discussed here, and the so-called arbitrary waveform generation through programmable pulse shaping. If one is able to fully measure the amplitude and phase of a light wave, then this can be run through a pulse shaper in reverse to generate a coherent pulse of light.

Two questions arise: Given the quantum nature of light: (1) Can we acquire sufficient information about the fluctuating field through a measurement that does not result in complete absorption of the field? And (2) can we, in any meaningful way, compensate absorption by gain? One is tempted to answer both questions in the negative, since the common wisdom of quantum mechanics says one cannot fully characterize a quantum state without destroying it, and one cannot amplify a state without inevitably introducing noise. Researchers continue to question and explore both issues, in the realm of theories of weak measurements on one hand and noiseless amplification on the other. We, however, will not rely on these bootstraps and proceed to describe a classical device capable of active filtering.

Our proposed device (Figure 39) has a resonant optical cavity at its core. Pre-filtered thermal light (one channel of the spectrally dispersed sunlight discussed above) is coupled into the cavity, through a partially transmitting input mirror. The electromagnetic energy comes out of the cavity through a similar partially transmitting output mirror. In the absence of any fluctuations (when the input light is perfectly monochromatic) the cavity field will build up to a certain (high) steady-state value, and the output will be equivalent to the input. Our conjecture is that, in the presence of input phase and amplitude fluctuations, one can implement active feedback to minimize and virtually eliminate the reflection off the input mirror, thereby "forcing" the total electromagnetic energy through the resonant cavity. In the (conceptually) simplest configuration, the resonant cavity will consist of two mirrors with low transmissivity and high reflectivity (and negligible absorption), with the output mirror fixed and the input mirror's position and transmissivity dynamically adjusted, in an automated way, so as to minimize reflected power and thereby compensate for the fluctuations of the input field's amplitude and phase.

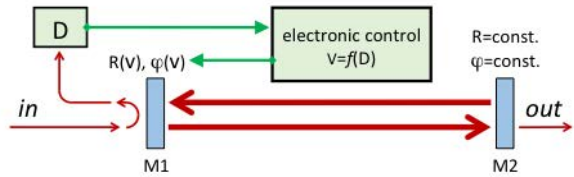

**Figure 39.** Maxwell's demon for light: Simplified schematics for the device envisioned and discussed in the text. Cavity mirror M1 is controlled electronically so as to minimize the total reflected power, while mirror M2 is fixed, ensuring stable monochromatic output. Credit: Alexei Sokolov.

### 12.3. Amplification by (cooperative) spontaneous emission?[36]

The laser uses stimulated emission to get amplification of radiation. But there is another way to get amplification of radiation via cooperative Dicke superradiance (*254*), (*255*). Thus, we ask: Can we get a new kind of high frequency light source amplification using superradiance? And if so, is it related to negative mass?

### 12.4. Will tomorrow's particle accelerators be based on lasers?

Particle accelerators based on microwave fields are the rule of the day. But we can, in principle, use laser fields which are much stronger than microwave fields (*256*), (*257*). Can we make real particle accelerators from lasers? Is it possible that the accelerators of tomorrow will be laser particle accelerators?

### 12.5. Will new optical techniques enable us to go beyond the Rayleigh limit?

Can we use new optical techniques which allow us to use entanglement and quantum coherence to exceed the Rayleigh limit? We think we can, and one step in this direction was the Nobel Prize in 2014. Zubairy, et al., have gone further (*209*). How far can we go? (*258*)

### 12.6. Quantum coherence in lasers conceptually related to the Higgs?

The laser operates via the Einstein stimulated emission coefficient and Bose condensation operates in a similar way. It is striking that the Bose condensation has so many features which make it look like an atom laser. This is the basis for Dan Kleppner's calling the BEC an atom laser. Can this laser-BEC analogy give insight into BCS and the Higgs particle? (*259*)

### 12.7. Why is the many particle Lamb shift divergence free?

The Lamb shift was the beginning of quantum field theory and renormalization physics. It is interesting therefore that a many atom Dicke superradiant Lamb shift does not involve infinite integrals. Maybe there is a connection here between superradiance and a more general picture of the vacuum. Is it possible that renormalization in a many particle superfluid type system would yield a different kind of quantum field theory not involving renormalization? (*260)*, (*261)*.

---

[36] Marlan Scully, 12.3 – 12.9

### *12.8. How does Bayesian logic impact quantum thinking?*

The quantum eraser is a paradigm shift in the measurement process. Typically people have thought that the process of measurement (e.g., which slit a photon goes through in a Young's apparatus) involves scrambling phases and substantially perturbing the photon. It has been shown that this paradigm is useful but not inevitable and that the simple acquisition of information and entanglement is sufficient to produce the "measurement rules out interference paradigm". Is it possible that these before and after considerations can be rewritten in terms of conditional Bayesian logic and provide a different window on the very concept of time? (*262*)

### *12.9. Noise enhanced engine and bio-efficiency?*[37]

Quantum noise allows us to improve quantum heat engine efficiency. Can we also use quantum noise to improve on biological efficiency and information processing? We think we can and the book by McFadden, *Life on the Edge: The Coming Age of Quantum Biology* (*263*), supports this supposition.

On this work and that of researchers in the Netherlands, Sweden and Russia, who detected quantum beating in plant photosystem reaction center II, McFadden and Al-Khalili go on to say:

> "And remember that photosynthetic reaction centers evolved between two and three billion years ago. So for nearly the entire history of our planet, plants and microbes seem to have been utilizing quantum-boosted heat engines – a process so complex and clever that we have yet to work out how to reproduce it artificially – to pump energy into carbon and thereby make all the biomass that formed microbes, plants, dinosaurs and, of course, us. Indeed, we are still harvesting ancient quantum energy in the form of fossil fuels that warm our homes and our cars and drive most of today's industry. The potential benefits of modern human technology learning from ancient natural quantum technology are huge."

## 13. (12.) What novel topological structures can be created and employed?

### *13.1. Orbital angular momentum, flying doughnuts, …*[38]

Transverse plane waves are not the only exact propagating solutions of Maxwell equations. Topologically more complex spatial constructs such as twistors, localized and propagating field configurations derived from the Hopf fibration, and related knotted vector fields are now attracting considerable attention (*264).*

*Flying doughnuts*

Another interesting class of topological structures is electromagnetic doughnut pulses (*265*). They are single-cycle, time-space non-separable, propagating solutions of Maxwell's equations of toroidal topology. They were first identified by Hellwarth and Nouchi in 1996 (*266*) as a subset of the more general family of finite energy electromagnetic directed energy pulses introduced by Ziolkowski in 1989 (*267*). Although electromagnetic doughnut pulses have not yet been observed experimentally, it is clear now that they can be generated from short transverse oscillations in a singular metamaterial converter (*268).* The electrodynamics of doughnut pulses is under construction (*269)*, with issues such as their waveguiding and propagation in dispersive media and through inhomogeneous media and interfaces now being investigated. Doughnut pulses have a strong component of electric (or magnetic) field along the direction of propagation and thus interact with matter in a peculiar way. They can be filtered from transverse pulses and then detected by a conventional radiation detector, which opens new opportunities for telecommunications. Moreover, the non-separable space-time dependence, which distinguishes doughnut pulses from the vast majority of electromagnetic waveforms, allows for novel schemes of information transfer and spectroscopy.

---

[37] Marlan Scully, 12.3 – 12.9
[38] N.I. Zheludev

Electromagnetic doughnuts are good broadband probes for localized dynamic toroidal excitations (classically represented by an oscillating poloidal current on a torus), the often overlooked component of the multipole expansion. Indeed, account of toroidal transitions in matter, e.g. in large molecules with substantially non-local response (*270*) and in artificial nanostructures (*271*), should have far-reaching implications for the interpretation of spectroscopic data that should also help in detecting molecules with elements of toroidal symmetry. Intriguingly, doughnut pulses can be used to pump energy into electromagnetic anapoles (*272*), non-trivial non-radiating charge-current configurations of destructively interfering electric and toroidal dipoles (see Figure 40). The latter are of particular interest as the excitation underpinning narrow EIT-like resonances and as well-isolated quantum qubits.

### *13.2. Topological photonics[39]*

Topology is a branch of mathematics that deals with conserved quantities that do not change when physical objects are continuously deformed, no matter how much. For example, the number of holes (or handles) of a complex connected surface can be characterized by its genus. This topological index does not change as the surface is deformed without introducing any cuts. Therefore, it can be thought of as being topologically robust. It is not surprising that topology finds a welcoming home in physics, which has a rich tradition (going back to Emma Noether's theorem) of appealing to conserved quantities including energy, momentum, angular momentum, and many others. It is easy to see, however, how topological indices are more robust than, for example, angular momentum: the class of deformations that preserve the former is much broader than those preserving the latter. For example, the stringent condition for the conservation of the angular momentum is that the system remains rotationally invariant. The sea change occurred in 1980s, when it was realized (*273*), (*274*) that topological invariants can be introduced for single-particle electron states of a two-dimensional electron gas (2DEG) in a combined periodic potential and out-of-plane magnetic field. These integer topological indices (known as Chern numbers) originate from the quantum nature of the electrons, specifically from their wave-like behavior in the periodic potential of the lattice. The electron's propagation through the lattice is described by its wave function $\psi(\vec{x}; \vec{p})$, where $\vec{x}$ is the real-space coordinate and $\vec{p}$ is the Bloch quasi-momentum in a periodic lattice. The Chern number characterizes the phase accumulation of the wave function as the Bloch wavenumber encircles the Brillouin zone of the crystalline lattice. Thus, the key feature of all continuous topological phases is that their topological indices are related to the behavior of the wave functions in the *momentum* space. While such accumulation of what is commonly known as the Berry phase (*275*) can occur even in one dimension (*276*), the most common and widely studied physical systems are two-dimensional. Although an external magnetic field is sufficient for producing non-trivial topological phases in a 2DEG, it is not necessary. Other types of interactions, such as for example spin-orbit coupling, can also give rise to insulating topological phases. The conservation of another discrete degree of freedom, the valley, produces yet another type of a topological phase, which is responsible for the emerging field of valleytronics. Regardless of the nature and dimensionality of topological phases, they share a remarkable property: the emergence of robust edge states that exist at the domain wall between two distinct topological phases that are characterized by different Chern numbers. In condensed matter physics, the existence of such robust edge states is responsible for a wide range of exotic phenomena, such as the quantum Hall effect, spin-polarized and valley-polarized currents, and many others. The resulting topological phases have the unique property of being insulating in the bulk and conducting at their interfaces owing to the presence of topologically protected edge states (in the case of 2D topological phases) or surface states (in the case of 3D topological phases). All the above-mentioned topological effects are the consequence of the wave nature of the electrons. Because light is a wave, many topological phenomena that exist in condensed matter physics have been emulated in photonics. A new class of photonic structures, photonic topological insulators (PTIs) (*277*), (*278*), (*279*), (*280*), (*281*), (*282*), (*283*), (*284*), (*285*), (*286*), (*287*), (*288*), (*289*) have been recently realized. PTIs enable reflections-free propagation of topologically protected edge waves (TPEWs) (*284*), (*287*), (*290*), (*291*) along almost arbitrarily shaped domain walls separating the PTIs with different topological indices. The three basic condensed matter systems supporting topological insulating phases – quantum Hall (QH), quantum spin-Hall (QSH), and quantum valley-Hall (QVH) TIs – have all been emulated in photonics. In some ways, PTIs enabled some of the features that have not yet been obtainable with condensed matter TIs. For example, interfaces between heterogeneous PTIs have been recently proposed (*292*). Such photonics structures have important applications, such as the development of broadband non-reciprocal devices (*292*) as well as ultra-compact resonant cavities (*289*). With the emergence of 3D PTIs and other exotic photonic concepts that rely on topological protection, we are undoubtedly seeing a renaissance in photonics that is likely to translate into a variety of novel devices.

---

[39] Gennady Shvets

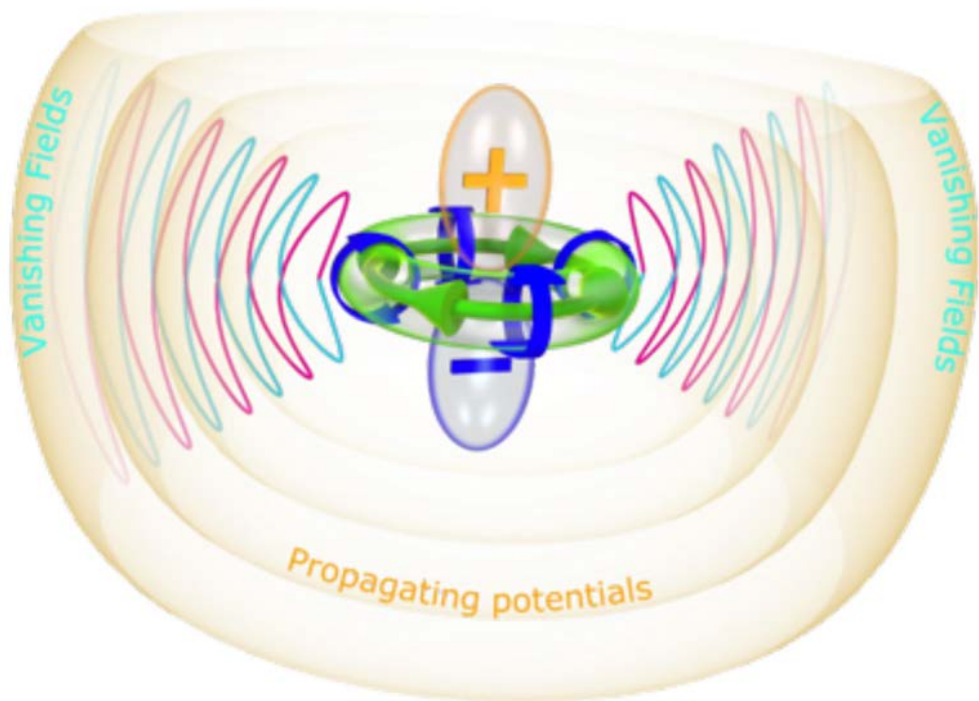

**Figure 40.** Artistic impression of electromagnetic anapole which is a combination of oscillating electric and toroidal dipoles that does not emit electromagnetic fields, but nevertheless creates an oscillating vector potential outside of the structure.

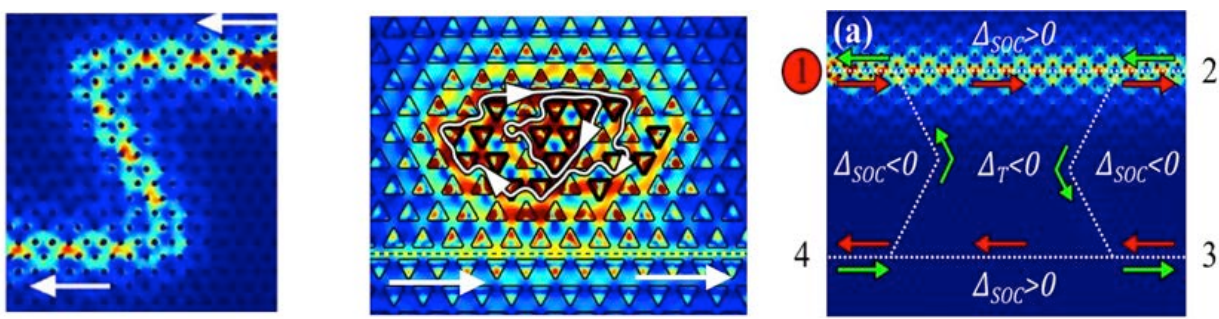

**Figure 41.** Applications enabled by topological photonics. (Left) Reflectionless propagation of topologically protected edge waves along the domain wall between two topological phases with opposite Chern indices. Note the lack of reflection from the sharp 120 degree turns of the domain wall *(287)*. (Middle) Topological cavity: edge waves propagating along the domain wall between two valley-Hall photonic topological phases couple into a cavity, where the light propagates along highly-compact zigzag path, thereby making the cavity highly compact *(289)*. (Right) A compact four-port circulator that enables perfect electromagnetic wave propagation from Port 1 to Port 2. Waves propagating in the opposite direction from Port 2 are deflected into Port 3 *(292)*.

## 14. Conclusion: The magic of quantum and classical light[40]

As we read through the amazing contributions throughout this paper, beginning with the new window on the universe introduced by Rai Weiss and his collaborators, and ending with the novel quantum devices suggested by Gennady Shvets, it is impossible not to feel a sense of awe at the hundreds of new directions for quantum and classical light, radiating from the minds of the authors of this paper and their colleagues. One anticipates that the youngest participants at the 2017 Snowbird conference will achieve successes that will continue the legacy of the previous decades, which have changed the course of fundamental science, technology, and human society.

The use of questions to focus thinking *(293)*, *(294)*, *(295)*, *(296)* is an honorable tradition that extends back at least to Newton's *Opticks*, where he posed 31 queries including the following:

1. Do not Bodies act upon Light at a distance, and by their action bend its Rays; and is not this action (caeteris paribus) strongest at the least distance?
12. Do not the Rays of Light in falling upon the bottom of the Eye excite Vibrations in the Tunica Retina? Which Vibrations, being propagated along the solid Fibres of the optick Nerves into the Brain, cause the Sense of seeing.
30. Are not gross Bodies and Light convertible into one another, and may not Bodies receive much of their Activity from the Particles of Light which enter their Composition?

We can only hope that the questions raised here remain equally meaningful after the passing of centuries, or at least have some significance during the course of the present century. But it is clear that quantum (and classical) optics has already undergone a historic blossoming into realms of understanding and application that are almost magical, and that the research programs of the authors in this article (and their colleagues) promise equally exciting developments in the future. We interpret the toast of Figure 42 as a celebratory salute to both the achievements of the past and those which are yet to come.

---

[40] Suzy Lidström and Roland E. Allen

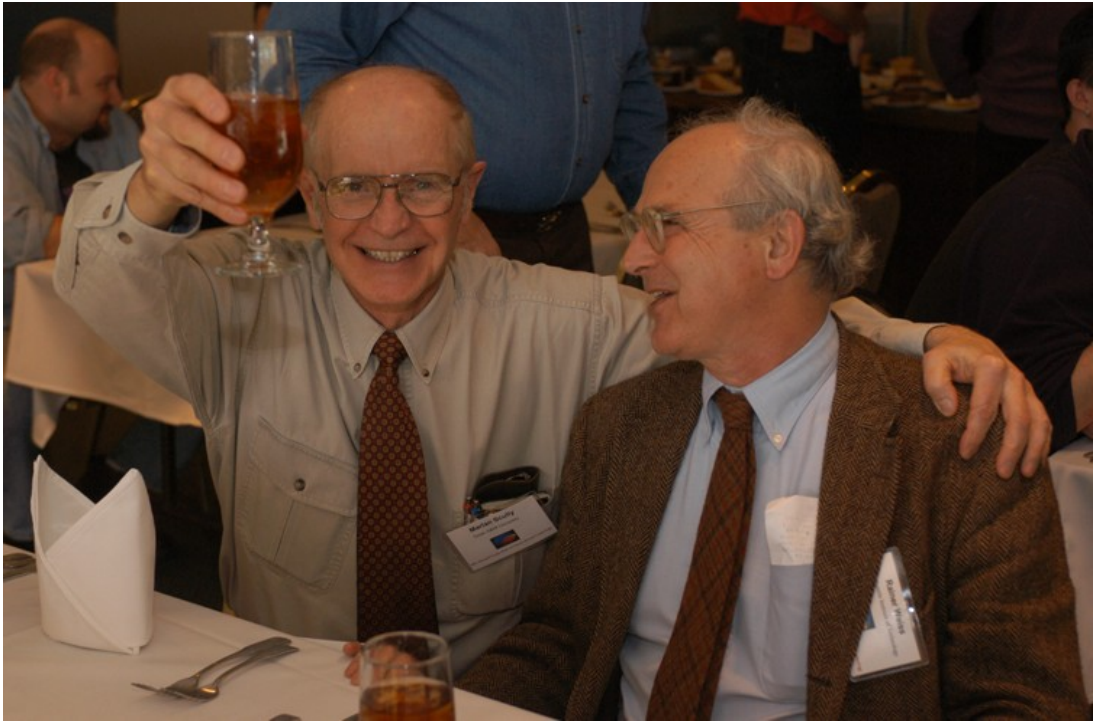

**Figure 42.** A prescient toast to the achievements of Rai Weiss and his colleagues, who have now introduced the era of gravitational wave astronomy. Photograph: Roland Allen.

## Acknowledgements

SL would like to thank Texas A&M for the hospitality shown during the completion of this work. SKL is grateful to REA for excellent supervision of his undergraduate project. WS and colleagues acknowledge many fruitful and stimulating discussions with our colleagues in particular, P. C. Abbott, J. Ankerhold, G. Agarwal, P. M. Alsing, J. S. Ben-Benjamin, C. Bokas, C. Feiler, D. M. Greenberger, M. Knobl, D. Lebiedz, H. Montgomery, G. Nunes Jr., H. Paul, S. T. Stenholm, S. Varro, M. S. Zubairy and J. Zuber. IB is grateful for the financial support from the Grant Agency of the Czech Technical University in Prague, Grant No. SGS16/241/OHK4/3T/14. HL is grateful for the financial support by the German Science Foundation (DFG) within SFB/TRR21. DML and DLH thank the US National Science Foundation for support over the years including most recently NSF grant DMR 1707565. WPS is grateful to the Hagler Institute for Advanced Study at Texas A&M University for a Faculty Fellowship and to Texas A&M University AgriLife Research for its support. G. Chen is grateful to Qatar National Research Fund Grant # NPRP 8-028-1-001 for partial financial support. OK thanks the NFS for financial support under NSF grant numbers PHY-1307346 and PHY-1506467. AS is grateful to the Welch Foundation and the NSF for support. ER acknowledges the German Space Agency (DLR) for funds provided by the Federal Ministry of Economic Affairs and Energy (BMWi) from the German Bundestag under Grant No. DLR 50WM1131–1137 as well as support from DFG through the CRCs geo-Q and dq-mat, and FPM and QUANOMET. The Office of Naval Research (Award No. N00014-16-1-3054) and the Robert A. Welch Foundation (Award A-1261) are kindly acknowledged by MS and AS.

## References

(1) Dimopoulos, S.; Graham, P.; Hogan, J.; Kasevich, M. General relativistic effects in atom interferometry. *Phys. Rev. D* **2008**, *78* (4), 042003. DOI:10.1103/PhysRevD.78.042003.
(2) Graham, P.; Hogan, J.; Kasevich, M.; Rajendran, S. A New Method for Gravitational Wave Detection with Atomic Sensors. *Phys. Rev. Lett.* **2013**, *110*, 171102. arXiv:1206.0818 DOI:10.1103/110.171102
(3) Kolkowitz, S.; Pikovski, I.; Langellier, N.; Lukin, M.; Walsworth, R.; Ye, J. Gravitational wave detection with optical lattice atomic clocks. *Phys. Rev. D* **2016**, *94* (12), 124043.
(4) Abbott, B.P.; et al; LIGO-consortium; VIRGO-consortium. Observation of Gravitational Waves from a Binary Black Hole Merger. *Phys. Rev. Lett.* **2016**, *116*, 061102.
(5) Abbott, B.; et al; LIGO-consortium; VIRGO-consortium. Tests of general relativity with GW150914. *Phys. Rev. Lett.* **2016**, *116*, 221101.
(6) Acernese, F.; Agathos, M.; Agatsuma, K.; et al. *Classical and Quantum Gravity* **2015**, *32*, 024001.
(7) Grote, H.; et al. *Classical and Quantum Gravity* **2010**, *27*, 084003.
(8) Sesana, A.; et al. Multi-band gravitational wave astronomy: science with joint space and ground based observations of black hole binaries. *Journal of Physics: Conf. Series* **2017**, *840*, 012018. DOI: 10.1088/1742-6596/840/1/012018.
(9) Cutler, C.; Thorne, K. An overview of gravitational-wave sources. arXiv: gr-qc/0204090 **2002**.
(10) eLISA Consortium; Amaro-Seoane, P.; Aoudia, S.; et al. The Gravitational Universe. arXiv:1305.5720 **2013**.
(11) Amaro-Seoane, P.; Aoudia, S.; Babak, S.; Binétruy, P.; Berti, B.; Bohé, A.; Caprini, C.; Colpi, M.; Cornish, N.J.; Danzmann, K.; et al. *Classical and Quantum Gravity* **2012**, *29*, 124016.
(12) Armano, M.; et al. *Phys. Rev. Lett.* **2016**, *116*, 231101.
(13) Canuel, A., C.; Bertoldi, L.A.; Bertoldi, L.A.; et al. Exploring gravity with the MIGA large scale atom interferometer. arXiv:1703.02490 **2017**.
(14) Hogan, J.M.; Kasevich, M.A. *Phys. Rev. A* **2016**, *94*, 033632.
(15) Yu, N.; Tinto, M. *Gen. Relativ. Gravit.* **2011**, *43*, 1943.
(16) Hogan, J.M.; Johnson, D.M.; Dickerson, S.; et al. *Gen. Relativ. Gravit.* **2011**, *43*, 1953.
(17) Hohensee, M.; Lan, S.Y.; Houtz, R.; et al. *Gen. Relativ. Gravit.* **2011**, *43*, 1905.
(18) Dimopoulos, S.; Graham, P.W.; Kasevich, M.A.; Rajendran, S. Atomic gravitational wave

interferometric sensor. *Phys. Rev. D* **2008**, *78*, 122002.

(19) Kovachy, T.; Asenbaum, P.; Dickerson, S.; Sugarbaker, A.; Hogan, J.; Kasevich, M.A. Quantum superposition at the half-metre scale. *Nature* **2015**, *528*, 530. doi:10.1038/nature16155.

(20) Dickerson, S.M.; Hogan, J.M.; Sugarbaker, A.; Johnson, D.M.S.; Kasevich, M.A. *Phys. Rev. Lett.* **2013**, *111*, 083001.

(21) Müntinga, H.; et al. *Phys. Rev. Lett.* **2013**, *110*, 093602.

(22) Wilczek, F. *Phys. Rev. Lett.* **2012**, *109*, 160401.

(23) Shapere, A.; Wilczek, F. *Phys. Rev. Lett.* **2012**, *109*, 160402.

(24) Watanabe, H.; Oshikawa, M. *Phys. Rev. Lett.* **2015**, *114*, 251603.

(25) Wilczek, F. *Phys. Rev. Lett.* **2013**, *111*, 250402.

(26) Zhang, J.; Hess, P.W.; Kyprianidis, A.; et al. Observation of a discrete time crystal. *Nature* **2017**, *543*, 217–220.

(27) Choi, S.; et al. *Nature* **2017**, *543*, 221–225.

(28) Else, D.V.; Bauer, B.; Nayak, C. Prethermal Phases of Matter Protected by Time Translation Symmetry. *Phys. Rev. X* **2017**, *7*, 011026.

(29) Stifter, P.; Leichtle, C.; Schleich, W.; Marklof, J. Das Teilchen im Kasten: Strukturen in der Wahrscheinlichkeitsdichte. *Zeitschrift für Naturforschung* **1997**, *52* (5), 377–385.

(30) Berry, M.; Marzoli, I.; Schleich, W. Quantum Carpets, Carpets of Light. *Physics World* **2001**, *14* (6), 39–44.

(31) Berry, M.; Klein, S. Integer, Fractional and Fractal Talbot Effects. *J. Mod. Optics* **1996**, *43*, 2139–2164.

(32) Smith, S.J.; Purcell, E.M. Visible Light from Localized Surface Charges Moving Across the Grating. *Phys. Rev.* **1953**, *92*, 1069.

(33) Schleich, W. *Quantum Optics in Phase Space*; Wiley-VCH, 2001.

(34) Nunes, Jr., G.; Jin, C.; Putnam, A.M.; Lee, D.M. Longitudinal spin diffusion and nonlinear spin dynamics in a dilute $^3$He-$^4$He solution. *Phys. Rev. Lett.* **1990**, *65*, 2149–2152.

(35) Nunes, Jr., G.; Jin, C.; Hawthorne, L.; Putnam, A.M.; Lee, D.M. Spin-polarized $^3$He-$^4$He solutions: Longitudinal spin diffusion and nonlinear spin dynamics. *Phys. Rev. B* **1992**, *46*, 9082–9103.

(36) Sato, Y.; Packard, R.E. Superfluid helium quantum interference devices: physics and applications. *Rep. Prog. Phys.* **2012**, *75*, 016401.

(37) Ragan, R.J.; Schwarz, D.M. Castaing Instabilities in Longitudinal Spin-Diffusion Experiments. *J. Low Temp. Phys.* **1997**, *109*, 775–799. .

(38) Hawthorne, D.L.; Wilde, S.; Lee, D.M. Simulations of Non-linear Spin Dynamics in Spin Polarized Dilute $^3$He-$^4$He Mixtures. *J. Low Temp. Phys.* **2013**, *171*, 165–170.

(39) Leggett, A.J. Spin diffusion and spin echoes in liquid $^3$He at low temperature. *J. Phys. C* **1970**, *3*, 448.

(40) Bekenstein, J. Black Holes and Entropy. *Phys. Rev. D* **1973**, *7*, 2333–2346.

(41) Scully, M.O.; Lee, D.; Schleich, W.P.; Svidzinsky, A. Black hole acceleration radiation: a quantum optical perspective. **2017,** http://www.pqeconference.com/pqe2018/abstractd/639.pdf.

(42) Scully, M.O.; Kocharovsky, V.; Belyanin, A.; Fry, E.; Capasso, F. Enhancing Acceleration Radiation from Ground-State Atoms via Cavity Quantum Electrodynamics. *Phys. Rev. Lett.* **2003**, *91*, 243004.

(43) Hawking, S.W. Particle Creation by Black Holes, *Commun. Math. Phys.* **1975**, *43*, 199–220.

(44) Bekenstein, J.; Meisels, A. Einstein A and B Coefficients for a Black Hole. *Phys. Rev. D* **1977**, *15*, 2775–2781.

(45) Buzek, V.; Krähmer, D.S.; Fontenelle, M.T.; Schleich, W.P. Quantum Statistics of Grey-Body Radiation. *Phys. Lett. A* **1998**, *239*, 1–5.

(46) Leonhardt, U. A laboratory analogue of the event horizon using slow light in an atomic medium. *Nature* **2002**, *415*, 406–409.

(47) Heim, D.M.; Schleich, W.P.; Alsing, P.M.; Dahl, J.P.; Varro, S. Tunnelling of an Energy Eigenstate Through a Parabolic Barrier Viewed from Wigner Phase Space. *Phys. Lett. A* **2013**, *377*, 1822–1825.

(48) Paul, H.; Greenberger, D.M.; Stenholm, S.T.; W. P. Schleich, W.P. The Stefan-Boltzmann Law: Two Classical Laws Give a Quantum One. *Phys. Scrip.* **2015**, *T165*, 014027.

(49) Bender, C.M.; Brody, D.C.; Müller, M.P. Hamiltonian for the Zeros of the Riemann-Zeta function. *Phys. Rev. Lett.* **2017**, *118*, 130201.
(50) Feiler, C.; Schleich, W.P. Entanglement and Analytical Continuation: An Intimate Relation told by the Riemann Zeta Function. *New J. Phys.* **2013**, *15*, 063009.
(51) Neuberger, J.W.; Feiler, C.; Maier, H.; Schleich, W.P. Newton Flow of the Riemann Zeta function: Separatrices Control the Appearance of Zeros. *New J. Phys.* **2014**, *16*, 103023.
(52) Neuberger, J.W.; Feiler, C.; Maier, H.; Schleich, W.P. The Riemann Hypothesis illuminated by the Newton Flow of Zeta. *Phys. Scrip.* **2015**, *90* (10), 108015.
(53) Schleich, W.P.; Bezdekova, I.; Kim, M.B.; Abbott, P.C.; Maier, H.; Montgomery, H.; and Neuberger, J.W. Equivalent Formulations of the Riemann Hypothesis Based on Lines of Constant Phase. To be published, submitted 2017.
(54) Will, C. The Confrontation between General Relativity and Experiment, *Living Rev. Relativity* **2006**, *9*, 3.
(55) Will, C.M. *Theory and Experiment in Gravitational Physics*; Cambridge University Press, **1993**.
(56) Taylor, T.; Veneziano, G. *Phys. Lett. B* **1988**, *213*, 450.
(57) Damour, T.; Polyakov, A.M. *Nucl. Phys. B* **1994**, *423*, 532.
(58) Dimopoulos, S. Macroscopic Forces from Supersymmetry, *Phys. Lett. B* **1996**, *379*, 105.
(59) Antoniadis, I.; et al. *Nucl. Phys. B* **1998**, *516*, 70.
(60) Rubakov, V. *Phys. Usp* **2001**, *44*, 871.
(61) Maartens, R.; Koyama, K. *Living Rev. Relativity* **2010**, *13*, 5.
(62) STE-QUEST Assessment Study Report (Yellow Book), ESA/SRE, **2013**.
(63) Schlamminger, S.; et al. *Phys. Rev. Lett.* **2008**, *100*, 041101.
(64) Williams, J.; et al. *Phys. Rev. Lett.* **2004**, *93*, 261101.
(65) Touboul, P.; Rodrigues, M. *Class. Quantum Grav.* **2001**, *18*, 2487.
(66) Colella, R.; Overhauser, A.; Werner, S. *Phys. Rev. Lett.* **1975**, *34*, 1472.
(67) Peters, A.; Chung, K.Y.; Chu, S. *Nature* **1999**, *400*, 849.
(68) Fray, S.; et al. *Phys. Rev. Lett.* **2004**, *93*, 240404.
(69) Varoquaux, G.; Nyman, R.; Geiger, R.; Cheinet, P.; Landragin, A.; Bouyer, P. *New J. Phys.* **2009**, *11*, 113010.
(70) Merlet, S.; et al. *Metrologia* **2010**, *47*, L9.
(71) Bonnin, A.; Zahzam, N.; Bidel, Y.; Bresson, A. Simultaneous dual-species matter-wave accelerometer. *Phys. Rev. A* **2013**, *88*, 043615.
(72) Schlippert, D.; et al. *Phys. Rev. Lett.* **2014**, *112*, 203002.
(73) Tarallo, M.; Mazzoni, T.; Poli, T.; Sutyrin, D.; Zhang, X.; Tino, G. *Phys. Rev. Lett.* **2014**, *113* (2), 023005.
(74) Zhou, L.; et al. Test of Equivalence Principle at $10^{-8}$ Level by a Dual-Species Double Diffraction Raman Atom Interferometer. *Phys. Rev. Lett.* **2015**, *115*, 013004.
(75) Sugarbaker, A.; Dickerson, S.; Hogan, J.; Johnson, D.; Kasevich, M. *Phys. Rev. Lett.* **2013**, *111*, 113002.
(76) van Zoest, T.; et al. *Science* **2010**, *328*, 1540.
(77) Aguilera, D.; et al. *Class. Quantum Grav.* **2014**, *31*, 159052.
(78) Schubert, C.; et al. arXiv:1312.5963 **2013**.
(79) Lévèque, T.; Gauguet, A.; Michaud, F.; Pereira Dos Santos, F.; Landragin, A. *Phys. Rev. Lett.* **2009**, *103*, 080405.
(80) Rosi, G.; et al. *Nat. Comms.* **2017**, *8*, 15529.
(81) Hartwig, J.; et al. *NJP* **2015**, *17*, 035011.
(82) Ahlers, H.; et al. *Phys. Rev. Lett.* **2016**, *116*, 173601.
(83) Kovachy, T.; Hogan, J.M.; Sugarbaker, A.; Dickerson, S.M.; Donnelly, C.A.; Overstreet, C.; Kasevich, M.A. Matter Wave Lensing to Picokelvin Temperatures. *Phys. Rev. Lett.* **2015**, *114*, 143004.
(84) Roura, A. *Phys. Rev. Lett.* **2017**, *118*, 160401.
(85) Schawlow, A.L.; Townes, C.H. Infrared and optical masers. *Phys. Rev.* **1958**, *112*, 1940.
(86) Suckewer, S.; Jaeglé, P. *Laser Phys. Lett.* **2009**, *1*, 26.
(87) Reagan, B.A.; Berrill, M.; Wernsing, K.A.; Baumgarten, C.; Rocca, J.J. *Phys. Rev. A* **2014**, *89*, 053820.
(88) Suckewer, S. X-ray lasers 2016. In *Proceedings in Physics*; Kawachi, T., Bulanov, S., Daido,

H., Kato, Y., Eds.; Springer, **2018**; Vol. 202.
(89) Schmusser, P.; Dohlus, M.; Rossbach, J. The UV and soft-X-ray FEL in Hamburg: Introduction to physical principles, experimental results, technological challenges. *STMP* **2008**, *229*, 121.
(90) Röhringer, N.; et al. *Nature* **2012**, *481*, 488.
(91) Beye, M.; et al. Stimulated X-ray emission for materials science. *Nature* **2013**, *501* (7466), 193.
(92) Agostini, P.; DiMauro, F.F. *Rep. Prog. Phys.* **2004**, *67*, 813.
(93) Corkum, P.B.; Krausz, F. *Nature Physics* **2007**, *3*, 381.
(94) Krausz, F.; Ivanov, M. *Rev. Mod. Phys.* **2009**, *81*, 163.
(95) Chini, M.; Zhao, K.; Chang, Z. *Nat. Photonics* **2014**, *8*, 178.
(96) Li, J.; et al. *Appl. Phys. Lett.* **2016**, *108*, 231102.
(97) Popmintchev, T.; et al. *Science* **2012**, *336*, 1287.
(98) Shwartz, S.; et al. *Phys. Rev. Lett.* **2014**, *112*, 163901.
(99) Fuchs, M.; et al. *Nature Phys.* **2015**, *11*, 964.
(100) Amann, J.; et al. *Nature Photonics* **2012**, *6*, 693.
(101) Yoneda, H.; et al. *Nature* **2015**, *524*, 446.
(102) Akhmedzhanov, T.R.; Antonov, V.; Kocharovskaya, O. *Phys. Rev. A* **2016**, *94*, 023821.
(103) Akhmedzhanov, T.R.; et al. *Phys. Rev. A* **2017**, *96*, 033825.
(104) Gunst, J.; Litvinov, Y.A.; Keitel, C.H.; Pálffy, A. *Phys. Rev. Lett.* **2014**, *12*, 082501.
(105) Baldwin, G.; Solem, J. *Rev. Mod. Phys.* **1997**, *69*, 1085.
(106) Kocharovskaya, O.; Kolesov, R.; Rostovtsev, Y. *Laser Physics* **1999**, *9*, 745.
(107) Rivlin, L.A. *Quantum Electronics* **2007**, *37*, 723.
(108) Ginzburg, V.L. *Rev. Mod. Phys.* **2004**, *76* (3), 981.
(109) Esarey, E.; Schroeder, C. B.; Leemans, W. P. *Rev. Modern Phys.* **2009**, *81,* 1229.
(110) Couprie, M.E.; Loulergue, A.; Labat, M.; Lehe, R.; Malka, V. *J. Phys. B* **2014**, *47*, 234001.
(111) Habs, D.; Tajima, T.; Schreiber, J.; Barty, C.P.; Fujiwara, M.; Thirolf, P.G. *EPJ D* **2009**, *55*, 279.
(112) Phuoc, K.T.; et al. *Nat. Photonics* **2012**, *6*, 308.
(113) Liu, J.C.; et al. *Opt. Lett.* **2014**, *39*, 4132.
(114) Platzman, P.M.; Mills, A.P.J. *Phys. Rev. B* **1994**, *49*, 454.
(115) Avetissian, H.K.; Avetissian, A.K.; Mkrtchian, G.F. *Phys. Rev. Lett.* **2014**, *113*, 023904.
(116) Wang, Y.H.; Anderson, B.M.; Clark, C.W. *Phys. Rev. A* **2014**, *89*, 043624.
(117) Malik, M.; Shin, H.; O'Sullivan, M.; Zerom, P.; Boyd, R.W. *Phys. Rev. Lett.* **2010**, *104*, 163602.
(118) Wang, J.; Yang, J.Y.; Fazal, I.M.; Ahmed, N.; Yan, Y.; Huang, H.; et al. *Nature Photonics* **2012**, *6*, 488.
(119) Mirhosseini, M.; Magana-Loaiza, O.S.; O'Sullivan, M.N.; Rodenburg, B.; Malik, M.; Lavery, M.P.; et al. *New J. Physics* **2015**, *17*, 033033.
(120) D'Ambrosio, V.; et al. *Nature Communications* **2012**, *3*, 961.
(121) Trichili, A.; Salem, A.B.; Dudley, A.; Zghal, M.; Forbes, A. *Optics Letters* **2016**, *41*, 3086.
(122) Krenn, M.; et al. *Pub. Nat. Acad. Sciences (USA)* **2016**, *113*, 13648.
(123) Proakis, J.; Salehi, M. *Digital Communications*, 5th ed.; New York: McGraw-Hill Education, **2007**.
(124) Boyd, R.W.; Rodenburg, B.; Mirhosseini, M.; Barnett, S.M. *Optics Express* **2011**, *19*, 18310.
(125) Andersson, M.; Berglind, E.; Björk, G. *New J. Phys.* **2015**, *17*, 043040.
(126) Zhao, N.; Li, X.; Li, G.; Kahn, J. Capacity limits of spatially multiplexed free-space communication. *Nature Photonics* **2015**, *9*, 822–826.
(127) Bash, B.A.; Chandrasekaran, N.; Shapiro, J.H.; Guha, S. arXiv:1604.08582 **2016**.
(128) Berry, M.V.; Popescu, S. Evolution of quantum superoscillations and optical superresolution without evanescent waves. *J. Phys A* **2006**, *39* (22), 6965–6977.
(129) Aharonov, Y.; Popescu, S.; Röhrlich, D.; Vaidman, L. Measurements, Errors, and Negative Kinetic-Energy. *Phys Rev A* **1993**, *48* (6), 4084–4090.
(130) Huang, F.M.; Zheludev, N.; Chen, Y.F.; de Abajo, F.J.G. Focusing of light by a nanohole array. *Applied Physics Letters* **2007**, *90* (9), 9.
(131) Rogers, E.T.F.; Zheludev, N.I. Optical super-oscillations: sub-wavelength light focusing and

super-resolution imaging. *Journal of Optics* **2013**, *15* (9), 1–23.
(132) Rogers, E.T.F.; et al. *Nature Materials* **2012**, *11* (5), 432–435.
(133) Rogers, E.T.F.; et al. Super-Oscillatory Imaging of Nanoparticle Interactions with Neurons. *Biophys. J.* **2015**, *108* (2), 479a.
(134) Huang, F.M.; Zheludev, N.I. Super-Resolution without Evanescent Waves. *Nano Letters* **2009**, *9* (3), 1249.
(135) Yuan, G.; et al. Quantum super-oscillation of a single photon, *Light-Sci. Appl.* **2016**, *5*, 1–6.
(136) Metcalf, H.; van der Straten, P. *Laser Cooling*; Springer-Verlag, **1999**.
(137) Ketterle, W. Nobel lecture: When atoms behave as waves: Bose-Einstein condensation and the atom laser. *Rev. Mod. Phys.* **2002**, *74*, 1131.
(138) Ludlow, A.D.; Boyd, M.M.; Ye, J.; Peik, E.; Schmidt, P.O. Optical atomic clocks. *Rev. Mod. Phys.* **2015**, *87*, 637.
(139) Cronin, A.D.; J., S.; Pritchard, D.E. Optics and interferometry with atoms and molecules. *Rev. Mod. Phys.* **2009**, *81*, 1051.
(140) Shah, V.; Knappe, S.; Schwindt, P.D.D.; Kitching, J. Subpicotesla atomic magnetometry with a microfabricated vapour cell. *Nature Photonics* **2007**, *1*, 649.
(141) Hanssen, J.L.; McClelland, J.J.; Dakin, E.A.; Jacka, M. Laser-cooled atoms as a focused ion beam source. *Phys. Rev. A* **2006**, *74*, 063416.
(142) Raizen, M.G. Comprehensive control of atomic motion. *Science* **2009**, *324, 1403-1406*.
(143) Gardner, J.R.; Anciaux, E.M.; Raizen, M.G. Neutral atom imaging using a pulsed electromagnetic lens. *J. Chem. Phys.* **2017**, *146*, 081102.
(144) Love, L.O. Electromagnetic separation of isotopes at Oak Ridge: An informal account of history, techniques, and accomplishments. *Science* **1973**, *182, 343-352*. DOI: 10.1126/science.182.4110.343.
(145) Mazur, T.R.; Klappauf, B.; Raizen, M.G. Demonstration of magnetically activated and guided isotope separation. *Nature Physics* **2014**, *10* (8), 601. doi:10.1038/nphys3013.
(146) The Pointsman Foundation, www.pointsman.org.
(147) Acosta, V.M.; Bauch, E.; Ledbetter, M.P.; Santori, C.; Fu, K.M.C.; Barclay, P.E.; Beausoleil, R.G.; et al. Diamonds with a high density of nitrogen-vacancy centers for magnetometry applications. *Phys. Rev. B* **2009**, *80* (11), 115202.
(148) Lee, S.Y.; et al. Readout and control of a single nuclear spin with a metastable electron spin ancilla. *Nat. Nano.* **2013**, *8* (7), 487–492.
(149) Jelezko, F.; et al. Private communication.
(150) Ph, T.; et al. Spin-flip and spin-conserving optical transitions of the nitrogen-vacancy centre in diamond. *New Journal of Physics* **2008**, *10* (4), 045004.
(151) Robledo, L.; et al. *Nature* **2011**, *477*, 574.
(152) Robledo, L.; et al. Control and Coherence of the Optical Transition of Single Nitrogen Vacancy Centers in Diamond. *Phys. Rev. Lett.* **2010**, *105* (17), 177403.
(153) Thiel, C.W.; Böttger, T.; Cone, R.L. Rare-earth-doped materials for applications in quantum information storage and signal processing. *Journal of Luminescence* **2011**, *131* (3), 353.
(154) Ahlefeldt, R.L.; Manson, N.B.; Sellars, M.J. Optical lifetime and linewidth studies of the transition in: A potential material for quantum memory applications. *Journal of Luminescence* **2013**, *133*, 152.
(155) Pfaff W.; Hensen B.J.; Bernien H.; van Dam S.B.; Blok M.S.; Taminiau T.H.; Tiggelman M.J.; Schouten R.N.; Markham M.; Twitchen D.J.; et al. Quantum information. Unconditional quantum teleportation between distant solid-state quantum bits. *Science* **2014**, *345*, 532–535. 10.1126/science.1253512.
(156) Hensen, B.; Bernien, H.; Dréau, A.E.; et al. Loophole-free Bell inequality violation using electron spins separated by 1.3 kilometres. *Nature* **2015**, *526*, 682.
(157) Rogers, L.J. All-Optical Initialization, Readout, and Coherent Preparation of Single Silicon-Vacancy Spins in Diamond. *Phys. Rev. Lett.* **2014**, *113* (26), 263602.
(158) Neu, E.; Agio, M.; Becher, C. Photophysics of single silicon vacancy centers in diamond: implications for single photon emission. *Optics Express* **2012**, *20* (18), 19956.
(159) Siyushev, P.; Metsch, M.H.; et al. Optical and microwave control of germanium-vacancy center spins in diamond. *Phys. Rev. B* **2017**, *96*, 081201.
(160) Taylor, J.M.; et al. *Nature Physics* **2008**, *4*, 810.

(161) Maze, J.R.; et al. *Nature* **2008**, *455*, 644.
(162) Childress, L.; Walsworth, R.L.; Lukin, M.D. Atom-like crystal defects: From quantum computers to biological sensors. *Physics Today* **2014**, *67,* 38. 10.1063/PT.3.2549.
(163) Dolde, F.; et al. Electric-field sensing using single diamond spins. *Nature Physics* **2011**, *7*, 459.
(164) Acosta, V.; Bauch, E.; Ledbetter, M.P.; Waxman, A.; Bouchard, L.S.; Budker, D. Temperature Dependence of the Nitrogen-Vacancy Magnetic Resonance in Diamond. *Phys. Rev. Lett.* **2010**, *104*, 070801.
(165) Lovchinsky, I. *Science* **2016**, *351*, 836.
(166) Sushkov, A.O.; Lovchinsky, I.; Chisholm, N.; Walsworth, R.L.; Park, H.; Lukin, M.D. *Phys. Rev. Lett.* **2014**, *113,* 197601.
(167) Ajoy, A.; Bissbort, U.; Lukin, M.D.; Walsworth, R.L.; Cappellaro, P. *Phys Rev X* **2015**, *5*, 011001.
(168) Sage, D.L.; Arai, K.; Glenn, D.R.; et al. Optical magnetic imaging of living cells. *Nature* **2013**, *496*, 486–489. 10.1038/nature12072.
(169) Rahn-Lee, L. *PLOS Genetics* **2015,** *11,* 1004811.
(170) Barry, J.F.; et al. *PNAS* **2016**, *113*, 14133.
(171) Davis, H.C.; Ramesh, P.; Bhatnagar, A.; Lee-Gosselin, A.; et al. *arXiv:1610.01924* **2016**.
(172) Waddington, D.E.J. et al, *Nature Photonics* **2012**, *6*, 488–496.
(173) Glenn, D.R.; et al. Single-cell magnetic imaging using a quantum diamond microscope. *Nature Methods* **2015**, *12*, 736.
(174) Kucsko, G.; Maurer, P.C.; Yao, N.Y. Nanometre-scale thermometry in a living cell. *Nature* **2013**, *500*, 54–58.
(175) Fu, R.R.; et al. *Science* **2014**, *346*, 6213.
(176) Fu, R.R.; et al. *Earth and Planetary Science Letters* **2017**, *458*, 1.
(177) Du, C.; Van der Sar, T.; Zhou, T.X.; Upadhyaya, P.; Casola, F.; Zhang, H.; Onbasli, M.C.; Ross, C.A.; Walsworth, R.L.; Tserkovnyak, Y.; et al. *Science* **2017**, *357*, 195–198.
(178) Dovzhenko, Y.; Casola, F.; Schlotter, S.; Zhou, T.X.; Buttner, F.; Walsworth, R.L.; Beach, G.S.D.; Yacoby, A. *arXiv:1611.00673* **2016**.
(179) Brenneis, A. Ultrafast electronic readout of diamond nitrogen–vacancy centres coupled to graphene. *Nature Nanotechnology* **2015**, *10*, 135.
(180) Le Sage, D.; et al. *Nature* **2013**, *496*, 486.
(181) Billiard, J.; Strigari, L.; Figueroa-Feliciano, E. Implication of neutrino backgrounds on the reach of next generation dark matter direct detection experiments. *Phys. Rev. D* **2014**, *89*, 023524.
(182) Rajendran, S.; Zobrist, N.; Sushkov, A.; Walsworth, R.; Lukin, M. A method for directional detection of dark matter using spectroscopy of crystal defects. *Phys. Rev. D* **2017**, 96, 035009.
(183) Togan, E.; Chu, Y.; Imamoglu, A.; Laser, M.K.L. Cooling and real-time measurement of the nuclear spin environment of a solid-state qubit. *Nature* **2011**, *478*,497–501.
(184) Bassett, L.C.; Heremans, F.J.; Yale, C.G.; Buckely, B.B.; Awschalom, D.D. Electrical Tuning of Single Nitrogen Vacancy Center Optical Transitions Enhanced by Photoinduced Fields. *Phys. Rev. Lett.* **2011**, *107, 266403*.
(185) Jarmola, A.; et al. Longitudinal spin-relaxation in nitrogen-vacancy centers in electron irradiated diamond. *Appl. Phys. Lett.* **2015**, *107*, 242403.
(186) Rittweger, E.; Han, K.Y.; Irvine, S.E.; Eggeling, C.; Hell, S.W. STED microscopy reveals crystal colour centers with nanometric resolution. *Nature Photonics* **2009**, *3*, 144.
(187) Arai, K.; et al. Fourier Magnetic imaging with nanoscale resolution and compressed sensing speed-up using electronic spins in diamond. *Nature Nanotechnology* **2015**, *10*, 859–864.
(188) Gustafsson, M.G.L. Surpassing the lateral resolution limit by a factor of two using structured illumination microscopy. *J. Microsc* **2000**, *198*, 82–87.
(189) Hayek, K.M.; Littleton, B.; Turk, D.; McIntyre, T.J.; Rubinsztein-Dunlop, H. A method for achieving super-resolved widefield CARS microscopy. *Opt. Express* **2010**, *18*, 19263– 19272.
(190) Schwartz, O.; Levitt, J.M.; Tenne, R.; Itzhakov, S.; Deutsch, Z.; Oron, D. Superresolution Microscopy with Quantum Emitters. *Nano Lett.* **2013**, *13*, 5832–5836.
(191) Monticone, D.G.; Katamadze, K.; Traina, P.; et al. Beating the Abbe Diffraction Limit

in Confocal Microscopy via Nonclassical Photon Statistics. *Phys. Rev. Lett.* **2014**, *113*, 143602.

(192) Classen, A.; Zanthier, J.; Scully, M.O.; Agarwal, G.S. Superresolution via structured illumination quantum correlation microscopy. *Optica* **2017**, *4*, 580–587.

(193) Boto, A.N.; Kok, P.; Abrams, D.S.; Braunstein, S.L.; Williams, C.P.; Dowling, J.P. Quantum interferometric optical lithography: Exploiting entanglement to beat the diffraction limit. *Phys. Rev. Lett* **2000**, *85*, 2733.

(194) Kolkiran, A.; Agarwal, G.S. Heisenberg limited Sagnac interferometry. *Opt. Express* **2007**, *15*, 6798–6808.

(195) Kolkiran, A.; Agarwal, G.S. Quantum interferometry using coherent beam stimulated parametric down-conversion. *Opt. Express* **2008**, *16*, 6479–6485.

(196) Nagata, T.; Okamoto, R.; O'Brien, J.L.; Sasaki, K.; Takeuchi, S. Beating the Standard Quantum Limit with Four-Entangled Photons. *Science* **2007**, *316*, 726–729.

(197) Rosen, S.; Afek, I.; Israel, Y.; Ambar, O.; Silberberg, C. Sub-Rayleigh Lithography Using High Flux Loss-Resistant Entangled States of Light. *Phys. Rev. Lett.* **2012**, *109*, 103602.

(198) Rozema, L.A.; et al. Scalable Spatial Superresolution Using Entangled Photons, *Phys. Rev. Lett* **2014**, *112*, 223602.

(199) Crespi, A.; et al. Measuring protein concentration with entangled photons. *Appl. Phys. Lett.* **2012**, *100*, 233704.

(200) Thiel, C.; Bastin, T.; Martin, J.; Solano, E.; Zanthier, J.; Agarwal, G.S. Quantum imaging with incoherent photons. *Phys. Rev. Lett.* **2007**, *99*, 133603.

(201) Jha, A.K.; Agarwal, G.S.; Boyd, R.W. Supersensitive measurement of angular displacements using entangled photons. *Phys. Rev. A* **2011**, *83, 053829*.

(202) Pooser, R.C.; Lawrie, B. Ultrasensitive measurement of microcantilever displacement below the shot-noise limit. *Optica* **2015**, *2*, 393–399.

(203) Taylor, M.A. Biological measurement beyond the quantum limit. *Nat. Photon.* **2013**, *7*, 229–233.

(204) Plick, W.N.; Dowling, J.P.; Agarwal, G.S. Coherent-light-boosted, sub-shot noise, quantum interferometry. *New J. Phys* **2010**, *12*, 083014.

(205) Hudelist, F.; Kong, J.; Cunjin, L.; et al. Quantum metrology with parametric amplifier-based photon correlation interferometers. *Nat. Commun.* **2014**, *5*, 3049.

(206) Arimondo, E. Coherent population trapping in laser spectroscopy. *Progress in Optics* **1996**, *35*, 257−354.

(207) Scully, M.O.; et al. FAST CARS: Engineering a laser spectroscopic technique for rapid identification of bacterial spores. *Proc. Natl. Acad. Sci.* **2002**, *99*, 10994.

(208) Pestov, D.; et al. Optimizing the laser-pulse configuration for coherent Raman spectroscopy. *Science* **2007**, *316*, 265–268.

(209) Kiffner, M.; Evers, J.; Zubairy, M.S. Resonant interferometric lithography beyond the diffraction limit. *Phys. Rev. Lett.* **2008**, *100*, 073602.

(210) Liao, Z.; Al-Amri, M.; Zubairy, M.S. Quantum lithography beyond the diffraction limit via Rabi oscillations. *Phys. Rev. Lett.* **2010**, *105*, 183601.

(211) Agarwal, G.S.; Kapale, K.T. Subwavelength atom localization via coherent population trapping. *J. Phys. B* **2006**, *39*, 3437–3446.

(212) Hell, S.W. Nanoscopy with focused light. *Angew. Chem. Int. Ed.* **2015**, *19*, 8054. (Nobel lecture).

(213) Miles, J.A.; Simmons, Z.J.; Yavuz, D.D. Subwavelength Localization of Atomic Excitation Using Electromagnetically Induced Transparency. *Phys. Rev. X* **2013**, *3, 031014*.

(214) Kapale, K.T.; Agarwal, G.S. Subnanoscale resolution for microscopy via coherent population trapping. *Opt. Lett* **2010**, *35*, 2792–2794.

(215) Li, L.; Tian, J.; Chen, G. Chaotic vibration of a two-dimensional non-strictly hyperbolic equation, preprint.

(216) NASA's Cassini-Huygens Mission: https://en.wikipedia.org/wiki/Timeline-of-Cassini\%E2\%80\%93Huygens/.

(217) Wiggins, S. *Introduction to Applied Nonlinear Dynamical Systems and Chaos*; Texts in Applied Mathematics.; Springer, **2003**.

(218) Bandrauk, A.D.; Delfour, M. *Quantum Control: Mathematical and Numerical Challenges*,

*CRM Proceedings and Lecture Notes* **2003**.
(219) Stoker, J.J. *Nonlinear Vibrations in Mechanical and Electrical Systems*; Wiley: New York, **1992.**
(220) Self-replicating machine: https://en.wikipedia.org/wiki/Self-replicating_machine.
(221) Feldhaus, J.; et al. AMO science at the FLASH and European XFEL free-electron laser facilities. *J. Phys. B* **2013**, *46*, 164002.
(222) Chapman, H.N.; Fromme, P. Femtosecond X-ray protein nanocrystallography. *Nature* **2011**, *470*, *73*.
(223) Ekeberg, T. Three-Dimensional Reconstruction of the Giant Mimivirus Particle with an X-Ray Free-Electron Laser. *Phys. Rev. Lett.* **2015**, *114*, 098102.
(224) Loh, D.N.T.; Elser, V. Reconstruction algorithm for single-particle diffraction imaging experiments. *Phys. Rev. E* **2009**, 80, 026705.
(225) Madey, J. Stimulated emission of bremsstrahlung in a periodic magnetic field. *J. Appl. Phys* **1971**, *42, 1906.*
(226) Deacon, D.; Elias, L.; Madey, J.; Ramian, G.; Schwettman, H.; Smith, T. First Operation of a Free-Electron Laser. *Phys. Rev. Lett.* **1977**, *38*, 892.
(227) Motz, H. Applications of the Radiation from Fast Electron Beams. *Journal of Applied Physics* **1951**, *22, 527.*
(228) Motz, H.; Thon, W.; Whitehurst, R.N. Experiments on Radiation by Fast Electron Beams. *Journal of Applied Physics* **1953**, *24, 7.*
(229) Pellegrini, C. The history of X-ray free electron lasers. *European Physical Journal* **2012**, *37*, 659.
(230) Hopf, F.A.; Meystre, P.; Scully, M.O.; Louisell, W.H. Classical theory of a free-electron laser. *Opt. Commun* **1976**, *18, 413.*
(231) Hopf, F.; Moore, G.; Scully, M.; Meystre, P. Nonlinear theory of free-electron devices. *Phys. Quantum Electron.* **1978**, *5*, 41.
(232) Neutze, R.; Wouts, R.; van der Spoel, D.; Wecker, E.; Hajdu, J. Potential for biomolecular imaging with femtosecond X-ray pulses. *Nature* **2000**, *406*, 752.
(233) Neutze, R.; Brändén, G.; Schertl, G.F. Membrane protein structural biology using X-ray free electron lasers. *Current Opinion in Structural Biology* **2015**, *33*, 115.
(234) Bell, J.S. *Physics* **1964**, *1*, 195.
(235) Aspect, A.; Dalibard, J.; Roger, G. *Phys. Rev. Lett.* **1982**, *49*, 1804.
(236) Weihs, G.; Jennewein, T.; Simon, C.; Weinfurter, H.; Zeilinger, A. *Phys. Rev. Lett.* **1998**, *81*, 5039.
(237) Zeilinger, A. Light for the quantum. Entangled photons and their applications: a very personal perspective. *Phys. Scr.* **2017**, *92*, 072501.
(238) Tittel, W.; Brendel, J.; Zbinden, H.; Gisin, N. *Phys. Rev. Lett.* **1998**, *81*, 3563–3566.
(239) Shalm, L.; et al. *Physical Review Letters* **2015**, *115*, 250402.
(240) Giustina, M.; et al. *Physical Review Letters* **2015**, *115*, 250401.
(241) Rosenfeld, W., et al., **2016**, arXiv:1611.04604.
(242) Ekert, A.; Renner, R. *Nature* **2014**, *507*, 443.
(243) Komar, P. *Nat. Phys.* **2014**, *10*, 582.
(244) Gottesman, D.; Jennewein, T.; Croke, S.; Croke, S. *Phys. Rev. Lett.* **2012**, *109*, 070503.
(245) Kimble, H. *Nature* **2008**, *453*, 1023; Cirac, J., Ekert, A., Huelga, S., & Macchiavello, C. *Phys. Rev. A* **1999**, 59, 4249.
(246) Moehring, D.; et al. *Nature* **2007**, *449, 68.*
(247) Cramer, J. *Nature Communications* **2016**, *7*, 11526.
(248) Kalb, N.; Reiserer, A.A.; Humphreys, P.C. Quantum Spin Hall Effect in Graphene. *Science* **2017**, *356,* 226801.
(249) Debnath, S.; Linke, N.; Figgatt, C.; et al. *Nature* **2016**, *536*, 63–66.
(250) Jin, Y.; et al. *Science* **2017**, *356*, 1140.
(251) Scovil, E.D.; Schulz-DuBois, E.O. Three-Level Masers as Heat Engines. *Phys. Rev. Lett.* **1959,** 2, 62.
(252) Scully, M.O.; Lamb, W.E. Quantum Theory of an Optical Maser. *Phys. Rev. Lett.* **1966,** 16, 853.
(253) Maxwell, J.C. Theory of Heat. 1872. *Nature* **2008**, 455, 644.

(254) Scully, M.O.; Svidzinsky, A.A. The Super of Superradiance. *Science* **2009**, *325, 1510*.
(255) Scully, M.O. The QUASAR revisited: insights gleaned from analytical solutions to simple models. *Laser Physics* **2014**, *24,* 094014.
(256) Scully, M.O. A Simple Laser Linac. *App. Phys. B* **1990**, *51*, 238.
(257) Scully, M.O. Enhancement of the Index of Refraction via Quantum Coherence. *Phys. Rev. Lett* **1991**, *67,* 14.
(258) Schirber, M. Focus: Dark Physics Beats Light Limit. *Phys. Rev. Focus* **2008**, *21*, 6.
(259) Scully, M.O. Condensation of N Bosons and the Laser Phase Transition Analogy. *Phys. Rev. Lett* **1999**, *82*, 3927.
(260) Scully, M.O.; Svidzinsky, A.A. The Lamb Shift Yesterday, Today, and Tomorrow. Physical Review A **2010,** 81, 053821.
(261) Scully, M.O. Collective Lamb Shift in Single Photon Dicke Superradiance. *Phys. Rev. Lett.* **2009**, *102*, 143601.
(262) Baggott, J. *The Quantum Story: A History in 40 Moments*, Oxford University Press; p 328.
(263) McFadden, J.; Al-Khalili, J. *Life on the Edge: The Coming Age of Quantum Biology*; Crown Publishers: New York.
(264) Arrayas, M.; Bouwmeester, D.; Trueba, J.L. Knots in electromagnetism. *Phys. Rep.* **2017**, *667*, 1–61.
(265) Papasimakis, N.; Fedotov, V.A.; Savinov, V.; Raybould, T.A.; Zheludev, N.I. Electromagnetic toroidal excitations in matter and free space. *Nature Materials* **2016**, *15* (3), 263–271.
(266) Hellwarth, R.W.; Nouchi, P. Focused one-cycle electromagnectic pulses. *Phys. Rev. E* **1996**, *54*, 889.
(267) Ziolkowski, R.W. Localised transmission of electromagnetic energy. *Phys. Rev. A* **1989**, *39* (4), 2005.
(268) Raybould, T.; Papasimakis, N.; Fedotov, V.A.; Youngs, I.; Zheludev, N.I. Generation of flying electromagnetic doughnuts via spatiotemporal conversion of transverse electromagnetic pulses. In *Nanometa Conference*, Austria; Seefeld, **2017**.
(269) Raybould, T.; Fedotov, V.; Papasimakis, N.; Youngs, I.; Zheludev, N. Focused electromagnetic doughnut pulses and their interaction with interfaces and nanostructures. *Optics Express* **2016**, *24* (4), 3150–3161.
(270) Raybould, T.A.; et al. Toroidal circular dichroism. *Phys. Rev. B* **2016**, *94* (3), 035119.
(271) Savinov, V.; Fedotov, V.A.; Zheludev, N.I. Toroidal dipolar excitation and macroscopic electromagnetic properties of metamaterials. *Phys. Rev. B* **2014**, *89* (20).
(272) Fedotov, V.A.; Rogacheva, A.V.; Savinov, V.; Tsai, D.P.; Zheludev, N.I. Resonant Transparency Non-Trivial Non-Radiating Excitations in Toroidal Metamaterials. *Scientific Reports* **2013**, *3*, 2967.
(273) Thouless, D.J.; Kohmoto, M.; Nightingale, M.P.; den Nijs, M. Quantized Hall Conductance in a Two-Dimensional Periodic Potential, *Phys. Rev. Lett.* **1982**, *49*, 405.
(274) Haldane, F. D. (1988). Model for a Quantum Hall Effect without Landau Levels: Condensed-Matter Realization of the "Parity Anomaly". Phys. Rev. Lett. **1988**, 61 (18), 2015—2018.
(275) Berry, M.V. Quantal phase factors accompanying adiabatic changes. *Proc, R. Soc. London, Ser. A* **1984**, *392*.
(276) Su, W.P.; Schrieffer, J.R.; Heeger, A.J. Solitons in polyacetylene. *Phys. Rev. Lett.* **1979**, *42*, 1698.
(277) Raghu, S.; Haldane, F. Analogs of quantum-Hall-effect edge states in photonic crystals. *Physical Review A* **2008**, *78* (3), 033834.
(278) Wang, Z.; Chong, Y.; Joannopoulos, J.; et al. Reflection-free one-way edge modes in a gyromagnetic photonic crystal. *Phys. Rev. Lett.* **2008**, *100*, 13905.
(279) Wang, Z.; Chong, Y.; Joannopoulos, J.D.; Soljacic, M. Observation of unidirectional backscattering-immune topological electromagnetic states. *Nature* **2009**, *461*, 772.
(280) Hafezi, M.; Demler, E.A.; Lukin, M.D.; Taylor, J.M. Robust optical delay lines with topological protection. *Nature Physics* **2011**, 7, 907–912.
(281) Hafezi, M.; Mittal, S.; Fan, J.; Migdall, A.; Taylor, J.M. Imaging topological edge states in silicon photonics. *Nature Photonics* **2013**, *7*, 1001.
(282) Rechtsman, M.C.; et al. Photonic Floquet topological insulators. *Nature* **2013**,

*496* (7444), 196–200.
(283) Khanikaev, A.B.; S., H.M.; Tse, W.K.; Kargarian, M.; MacDonald, A.H.; Shvets, G. Photonic topological insulators. *Nature Materials* **2013**, *12* (3), 233.
(284) Chen, W.J.; et al. Experimental realization of photonic topological insulator in a uniaxial metacrystal waveguide. *Nat. Commun.* **2014**, *5*, 5782.
(285) Gao, W.; et al. Topological Photonic Phase in Chiral Hyperbolic Metamaterials. *Phys. Rev. Lett.* **2015**, *114*, 037402.
(286) Liu, F.; Li, J. Gauge Field Optics with Anisotropic Media. *Phys. Rev. Lett.* **2015**, *114*, 103902.
(287) Ma, T.; Khanikaev, A.B.; Mousavi, S.H.; Shvets, G. Guiding Electromagnetic Waves around Sharp Corners: Topologically Protected Photonic Transport in Metawaveguides. *Phys. Rev. Lett.* **2015**, *114* (12), 127401.
(288) Wu, L.H.; Hu, X. Scheme for Achieving a Topological Photonic Crystal by Using Dielectric Material. *Phys. Rev. Lett.* **2015**, *114*, 223901.
(289) Ma, T.; Shvets, G. All-Si valley-Hall photonic topological insulator. *New Journal of Physics* **2016**, *18*, 025012.
(290) Lai, K.; Ma, T.; Bo, X.; Anlage, S.; Shvets, G. Experimental Realization of a Reflections-Free Compact Delay Line Based on a Photonic Topological Insulator. *Scientific Reports* **2016**, *6*, 28453.
(291) Cheng, X.; Jouvaud, C.; Mousavi, S.H.; Ni, X.; Genack, A.Z.; Khanikaev, A.B. Robust reconfigurable electromagnetic pathways within a photonic topological insulator. *Nature Mat.* **2016**, *15*, 542.
(292) Ma, T.; Shvets, G. Scattering-free edge states between heterogeneous photonic topological insulators. *Phys. Rev. B* **2017**, *95*, 165102.
(293) Hilbert, D. Mathematical problems. *Bull. Am. Math. Soc.* **1902**, *8*, 437.
(294) Ginzburg, V.L. What problems of physics and astrophysics seem now to be especially important and interesting thirty years later, already on the verge of XXI century? *Physics Uspekhi* **1999**, *42, 353*.
(295) Allen, R.; Lidström, S. Life, the Universe and everything – 42 fundamental questions. *Phys. Scr.* **2017**, *92* (1), 012501.
(296) Coley, A.A. Open problems in mathematics. *Phys. Scr.* **2017**, *92*, 093003.